\pdfoutput=1

\documentclass[11pt, manuscript]{aastex}
\usepackage[]{hyperref}
\usepackage{float}
\usepackage{color}
\usepackage{graphicx}
\usepackage{natbib}

\begin{document}

\title{M Dwarf Flare Continuum Variations on One-Second Timescales:  Calibrating and Modeling of ULTRACAM Flare Color Indices\footnote{Based on observations obtained with the Apache Point Observatory 3.5 m
telescope, which is owned and operated by the Astrophysical Research
Consortium, based on observations made with the William Herschel Telescope operated on the island of La Palma by the Isaac Newton Group in the Spanish Observatorio del Roque de los Muchachos of the Instituto de Astrofísica de Canarias, and observations, and based on observations made with ESO Telescopes at the La Silla Paranal Observatory under programme ID 085.D-0501(A). }}
\author{Adam~F.~Kowalski}
\affil{Department of Astronomy, University of Maryland, College Park, MD 20742, USA.}
\affil{NASA/Goddard Space Flight Center, Code 671, Greenbelt, MD 20771.}
\email{adam.f.kowalski@nasa.gov\\} 
\and
\author{Mihalis Mathioudakis} 
\affil{Astrophysics Research Centre, School of Mathematics \& Physics, Queen's University Belfast, Belfast, BT7 1NN, UK}
\and
\author{Suzanne L. Hawley} 
\affil{Department of Astronomy, University of Washington, Box 351580, Seattle, WA 98195, USA.}
\and
\author{John P. Wisniewski}
\affil{HL Dodge Department of Physics \& Astronomy, University of Oklahoma, 440 W Brooks Street, Norman, OK 73019, USA}
\and
\author{Vik S. Dhillon}
\affil{Department of Physics and Astronomy, University of Sheffield, Sheffield S3 7RH, UK}
\affil{Instituto de Astrofisica de Canarias, 38205 La Laguna, Santa Cruz de Tenerife, Spain}
\and
\author{Tom R. Marsh}
\affil{Department of Physics, Gibbet Hill Road, University of Warwick, Coventry CV4 7AL, UK}
\and
\author{Eric J. Hilton}
\affil{Department of Astronomy, University of Washington, Box 351580, Seattle, WA 98195, USA.}
\and
\author{Benjamin P. Brown}
\affil{Laboratory for Atmospheric and Space Physics and Department of Astrophysical \& Planetary Sciences, University of Colorado, Boulder, Colorado 80309, USA.}

\begin{abstract}
We present a large dataset of high cadence dMe flare light curves obtained with 
custom continuum filters on the triple-beam, high-speed camera system ULTRACAM.  The 
measurements provide constraints for models of the NUV and optical 
continuum spectral evolution on timescales of $\approx$1 second. We provide a robust interpretation 
 of the flare emission in the ULTRACAM filters
using simultaneously-obtained low-resolution spectra during two moderate-sized flares in the dM4.5e star YZ CMi.  
By avoiding the spectral complexity within the broadband Johnson filters,
the ULTRACAM filters are shown to characterize bona-fide continuum emission in the NUV, blue, and red wavelength regimes.
  The NUV/blue flux ratio in flares is equivalent to a Balmer jump ratio, and the blue/red
flux ratio provides an estimate for the color temperature of the optical continuum emission.  We present a new ``color-color'' relationship for these continuum flux ratios at the peaks of the flares. 
 Using the RADYN and RH codes, we interpret the ULTRACAM filter emission using the dominant emission processes from a radiative-hydrodynamic flare model with a high nonthermal electron beam flux, which explains a hot, $T\approx10^4$ K, color temperature
at blue-to-red optical wavelengths and a small Balmer jump ratio as are observed in moderate-sized and large flares alike. 
We also discuss the high time-resolution, high signal-to-noise continuum color variations observed in YZ CMi during a giant flare, which 
increased the NUV flux from this star by over a factor of 100.  
\end{abstract}


\section{Introduction}
The white-light emission in chromospherically active M dwarf (dMe) flares
spans a large wavelength range, from the ultraviolet through the optical, and thus 
comprises a large fraction of the radiated flare energy \citep{Hawley1995, Osten2015}.  
Constraining the nature of the emission processes 
provides insight into the heights and densities of flare heating that result in white-light emission \citep{CramWoods1982, Houdebine1992, Christian2003}.  An optically-thin hydrogen recombination spectrum consisting of a relatively large amount of Balmer 
continuum emission and a large change in flux across the wavelength region, $\lambda\approx3600-3800$ \AA, of the Balmer jump
indicates heating of low-to-moderate densities in
the mid-to-upper chromosphere ($n_{\rm{H}} \approx 10^{13} - 10^{14}$ cm$^{-3}$), whereas a $T\approx10^4$ K blackbody-like spectrum and relatively
smaller change in continuum flux over the Balmer jump region
indicates heating at much higher densities \citep[$n_{\rm{H}} \ge 10^{15}$ cm$^{-3}$;][]{Kowalski2015}.

The coarse spectral energy distribution of the white-light emission has traditionally been investigated 
with broadband colors \citep{Kunkel1970, HawleyFisher1992, Hawley1995, Hawley2003, Zhilyaev2007, Davenport2012, Lovkaya2013} and is generally consistent with a blackbody with $T\approx9000$ K (or greater), implying a large heating rate at high densities in large and moderate-sized flares alike.
Although the $BVR$ broadband filters are dominated by continuum emission during dMe flares \citep[at most $\sim17$\% emission line contribution in the $B$-band in the gradual phase;][]{HawleyPettersen1991}, the interpretation of the Johnson $U$-band is more ambiguous.  The $U$-band includes emission at $\lambda<3500$ \AA\ that has rarely been characterized in detail with spectra \citep{Schmitt2008, Fuhrmeister2008}.  The $U$-band also straddles the Balmer jump wavelength ($\lambda=3646$ \AA) and includes 
 the complex spectral region with the higher order Balmer lines that broaden and merge producing additional continuum between $\lambda=3646-3800$ \AA\ \citep{Zarro1985, Doyle1988, DonatiFalchi1985, HawleyPettersen1991}, which varies in a complicated way as a function of wavelength from the changing ambient proton densities throughout the flare \citep[][hereafter K15]{Kowalski2015}.  \cite{Allred2006} concluded that there is a striking degeneracy for constraining emission mechanisms using broadband filters,
since an optically thin hydrogen recombination continuum with emission lines convolved with the broadband filters\footnote{In the \cite{Allred2006} models, 
the $B$-band contains H$\delta$, H$\gamma$, and H$\beta$ line emission, and the $U$-band integrates over a large Balmer discontinuity in the model spectrum.  We have recently shown that a Balmer discontinuity, however, is not expected when including Landau-Zener b-f and b-b opacities near the Balmer edge (K15).} exhibits the same general shape
as a hot blackbody with $T=9000$ K.

Spectral observations are necessary to break the degeneracy of emission mechanisms that contribute to the 
observed white-light spectrum.  
Recent spectral analyses around the Balmer jump ($\lambda=3646$ \AA) using near-ultraviolet (NUV), blue, and optical
spectra \citep{Kowalski2010, Kowalski2013}, reveal a more complex continuum distribution that requires several components 
 to fit the data.  In the recent, homogeneous analysis of a 
sample of twenty dMe flares \citep[][hereafter K13]{Kowalski2013}, the spectra were phenomenologically explained with the following
four continuum components:
A Balmer continuum (in emission or absorption) at $\lambda < 3646$\AA, a pseudo-continuum
of merged, Stark-broadened high-order Balmer lines from $\lambda = 3646-3700$\AA\ and in the region of lower order Balmer lines between $3700-3800$\AA, 
a hot ($T\approx10\,000-12\,000$ K) blackbody-like
component that dominates the flux at $\lambda > 4000$\AA, and a cool ($T\approx5000 $ K) blackbody continuum at $\lambda \gtrsim 5000$\AA.
Each of these components vary on different timescales, with the hot blackbody component the most 
rapid to brighten and then decay and apparently cool (to $T\approx8000$ K), while the Balmer continuum and the cooler (redder) blackbody continuum components decrease on longer timescales yet faster than the Balmer emission lines.  
 If related to two-ribbon spatial features observed in solar flares, 
the hot blackbody component may originate from compact, newly-heated kernels whereas the 
 Balmer continuum emission and cooler blackbody component in the gradual decay phase may
originate from spatially extended, previously heated ribbons \citep{Kowalski2012}.

Although spectra provide a more detailed view of the continuum components, most spectrographs are  
limited to relatively long readout times of $\gtrsim10$~s and comparably long or much longer\footnote{High speed spectroscopy is now possible with a few modern spectrographs:  the visiting instrument ULTRASPEC \citep{Ultraspec}, the Intermediate Dispersion Spectrograph and Imaging System/QUCAM \citep{qucam} on the William Herschel Telescope and the Robert Stobie Spectrograph (RSS) on the South African Large Telescope (SALT).} exposure times ($10-300$ s).
These timescales are greater than the evolution timescales of white-light emission which are as fast as $\approx$10~s in the impulsive phase \citep{Moffett1974}.  Furthermore, these timescales are far longer than the timescales over which atmospheric
properties change in 
radiative-hydrodynamic flare models \citep[][K15]{Allred2006}.
Thus, spectra may average over important temporal changes of each continuum component, preventing an accurate comparison to models
and possibly leading to an ambiguous interpretation of the emission.  

We have completed a large flare-monitoring campaign of dMe stars using ULTRACAM in order to 
probe short timescales that are currently inaccessible with spectral measurements.   We employed custom filters to
observe pre-selected continuum wavelength regions away from the complicated Balmer jump region, yet at wavelengths that are accessible to 
ground-based spectra.  
   In this paper, we present
 the properties of twenty flares observed with ULTRACAM. In our analysis, we calibrate the high-time resolution ULTRACAM filter ratios to
 the broader spectral energy distribution in simultaneous low-resolution spectra, which we use to interpret the emission with 
 new radiative-hydrodynamic (RHD) model predictions.

The paper is organized as follows.  In Section 2, we describe the ULTRACAM observations.  In Section 3, we describe how we calculate the
 absolute flare flux from the relative photometry.
  In Section 4, we analyze two moderate-sized 
flares with simultaneous low-resolution spectra and interpret the ULTRACAM filter emission.
 In Section 5, 
we present new properties of the flare continuum revealed at high time resolution for a sample of 20 flares.
 In Section 6, we discuss the largest flare in the sample compared to the YZ CMi Megaflare \citep{Kowalski2010}.
 In Section 7, we summarize five main results from the study and discuss future modeling directions.
 
\section{Observations and Data Reduction}

\subsection{High Speed ULTRACAM Photometry}\label{sec:ucred}
The observing log is given in Table \ref{table:obslog}.  We monitored five dMe stars with ULTRACAM \citep{Dhillon2007} at the 4.2 m William
Herschel Telescope (WHT) on La Palma and the 3.6 m New Technology Telescope (NTT) at the ESO La Silla Observatory.  
ULTRACAM employs two dichroic beamsplitters to allow data to be obtained in three filters simultaneously. ULTRACAM uses a frame transfer CCD in each of the three arms, resulting in very short (24 millisecond) readout time between exposures, and thus providing continuous observations of the flares. 
In the NUV and blue arms of ULTRACAM, we used the narrow-band filters NBF3500 
 (FWHM 100\AA, $\lambda_c=3500$\AA) and NBF4170 (FWHM 50\AA, $\lambda_c=4170$\AA), respectively.
These two filters were specifically designed 
to characterize the Balmer jump ratio in dMe flares, and a twin set of these filters is available with the ROSA instrument \citep{Jess2010} at the Dunn Solar Telescope.  
For the red arm, we used the Red Continuum \#1 filter (FWHM 120\AA,
$\lambda_c=6010$\AA; hereafter RC\#1) 
to provide a color temperature of the flux at wavelengths longer than the Balmer jump.  Our analysis of the ULTRACAM
data of the first three flares observed on EQ Peg A in 2008 was reported in \cite{Kowalski2011}. For those observations, the red filter employed was a narrow (FWHM 50 \AA) H$\alpha$ filter.  

The data were reduced using the ULTRACAM reduction pipeline\footnote{\url{http://deneb.astro.warwick.ac.uk/phsaap/software/ultracam/html/}} developed by T. Marsh,
and the photometry was measured relative to a nearby comparison star\footnote{To get sufficient
counts in the blue, the brightest comparison star was sometimes saturated in the red.  In these cases, the second 
brightest comparison star in the field was used for the red comparison measurement.}.  
A fixed aperture size was determined for each filter based on the size that gave the lowest standard deviation 
during non-flaring times, while also not affecting the photometry during flare periods compared to a very large 17-pixel radius aperture.  The best aperture radii for the NUV and blue arms were typically $6$ pixels and 
for the red arm $10$ pixels.  
The relative photometry was normalized to quiescence, giving the count flux enhancement in each filter (designated as $I_{f,\lambda}+1$; Section \ref{sec:analysis}).  
The standard deviations of the relative photometry in quiescence ranged between $1.5-6$\% for NBF3500, $1-3$\% for NBF4170, and was $\approx0.5$\% for RC\#1.
For almost all observations, the exposure time in the NUV arm
was twice the exposure time of the blue and red arms;  the exposure times in the blue and red arms were always equal.
The sky conditions were mostly clear during the observations.  During the observations after 22:45 UT on 2012 Jan 13, 
there were intermittent thin clouds that affected the relative photometry; we removed the bad observations from the 
light curves by assessing the variations in the nearby comparison stars.  The seeing was poor and variable during the photometry
 of Prox Cen on 2010 May 21-22; longer exposure times and larger aperture sizes were used to obtain reliable measurements.

\subsubsection{The ULTRACAM Flare Sample}
In Table \ref{table:obslog}, we give with the monitoring time, exposure times,
 and an approximate number of flare events in each night.  
From these $\approx$104 flare events, we select twenty flares with the
highest signal-to-noise in the photometry at peak time to analyze in detail (Section \ref{sec:sample}).
We obtained low-resolution spectra from the Apache Point Observatory (APO; Section \ref{sec:specdata}) during four of these events on YZ CMi (flare events IF4, IF8, IF11, GF2);  two of these events (IF4, IF11)
are analyzed in detail in Section \ref{sec:spec}.  Spectra with the RSS and SALT were obtained during the IF1 and IF3 events on YZ CMi \citep{BrownAAS}; a detailed analysis of these spectra will be presented in a future paper.  The ULTRACAM data for the IF1 and IF3 events are discussed in Section \ref{sec:ultraflare}.
The reduced and flux-calibrated ULTRACAM and spectral data from this paper are available in FITS format through Zenodo \citep{kowalski_2016_zenodo}.

\subsection{Spectral Data} \label{sec:specdata}
Low-resolution spectral monitoring data ($\lambda=3420-7400$\AA, R$\approx400$ or about $10-11$ \AA\ FWHM) 
of YZ CMi were obtained on 2012 Jan 14 from
2:43 UT to 5:16 UT with the Astrophysical Research Consortium (ARC) 3.5 m Telescope and the Dual-Imaging Spectrograph (DIS)
at the APO.  We obtained 225 spectra with exposure times ranging between 25\,--\,30~s.
The conditions were clear, and we used a wide 5\arcsec\ slit to facilitate absolute flux calibration of the spectra.
The data were reduced using standard IRAF procedures as described in K13\footnote{The correction 
to the quiescent $V$ and $B$ magnitudes was not applied as in K13 because the observations took place in the late decay phase of a large flare; 
we estimate that the correction would have been very small.}. Four flare events (IF4, IF8, IF11, and GF2) were observed over this time.
The procedure in Appendix A of K13 was used to determine a multiplicative scale-factor that minimizes the subtraction residuals in
the red molecular bands and results in a spectrum with flare-only emission.  Due to a small increase in the red flux
during these flare events, we adjusted the multiplicative scale factor in increments of 0.005, compared to increments of 0.01 in K13.
The increment of 0.005 is equal to the uncertainty of the ULTRACAM photometry in RC\#1 and to the 
 standard deviation of the residual (flare-only) flux in the spectra at the wavelengths of RC\#1.
The scaling of the spectra using this algorithm produces a robust estimate of the flare increase,
which was checked against the value of $I_f+1$ in RC\#1 from the ULTRACAM data binned to the integration time of the spectra.
 All of the flare-only emission quantities described in K13 were calculated from the
spectra.  Quiescent spectra from APO/DIS of YZ CMi, AD Leo, and EQ Peg A have been previously reported in K13; these quiescent spectra are used to calibrate the ULTRACAM photometry to flux density values (see Section \ref{sec:analysis}).

The wide slit width was significantly
larger than the seeing, causing radial velocity uncertainty due to underfilling of the slit by the stellar PSF.  
As a result, the FWHM resolution determined from the arc lines is larger than obtained from point sources through
 the wide slit.  The spectral resolution as determined from quiescent emission line profiles of YZ CMi
varies due to seeing fluctuations, which manifests as 0.4 \AA\ (root-mean-square) variations in the line widths during quiescence.  The line-integrated fluxes of the emission lines 
are not affected, but users of the spectral data\footnote{These data are available online through Zenodo in the same format as the spectral data in K13.}
should exercise caution with the detailed line profiles in these spectra in the early impulsive phase when the 
increased line flux from the flare is relatively low.  

Small errors in telescope pointing produced wavelength-independent shifts in the wavelength calibration by at most one pixel or 2.3 \AA.  
 These wobbles resulted
from shifts in the position of the star behind the wide slit (the faint stellar PSF wings that spill over the slit are used for telescope guiding).  
Thus, we coaligned the wavelength scale of all spectra using the center of the H$\alpha$ emission line profile before producing the flare-only emission spectra\footnote{These spectra
are unsuitable for characterizing emission line centroid shifts from mass motions.  We checked that the alignment of the molecular band heads in the blue and red arm DIS spectra agree with the wavelength shift derived from the H$\alpha$ line.}.

\subsection{Broadband Photometry Data} \label{sec:1m}
We obtained Johnson $U$-band data of YZ CMi with the New Mexico State University 1 m telescope \citep{Holtzman}
on 2012 Jan 14 beginning at 3:40 UT. The cadence of the photometry was $\approx$19~s, and the exposure time was 10~s.  
 The data were reduced with an automatic pipeline using standard IRAF procedures, giving 
$I_f+1$ in the $U$-band.  Two small flares occurred at 4:03 and 4:09 UT, and two moderate-sized
flares occurred at 3:30 and 4:32 UT (although the former was not observed with the 1 m).  The APO/DIS spectra
and ULTRACAM photometry of the two moderate-sized
flares have suitable signal-to-noise for a detailed analysis, which is presented in Section \ref{sec:spec}.

Outside of these flares, the $U$-band is elevated over its nominal level for this star ($I_{f}+1 \approx 1.4$ prior to the flare at 4:32 UT).  We attribute the additional emission in quiescence to the decay phase of a large flare peaking at 22:44:31 on 2012 Jan 13, which we discuss in Section \ref{sec:ultraflare}.
For the flare peaking at 4:32 UT on 2012 Jan 14, we calculate a $U$-band energy of $E_U=1.6\times10^{31}$ ergs and an equivalent duration \citep{Gershberg1972} of 400~s.
The smaller, more gradual flare event at 
$\approx$4:09 UT has a $U$-band energy of $E_U = 3\times10^{30}$ ergs and an equivalent duration of 75~s.
This corresponds to an average flare energy for YZ CMi \citep{Lacy1976} and is one of the lowest amplitude ($I_{f,U}\approx0.4$, $I_{f,\rm{NBF3500}}\approx0.6$)
and slowest evolving flare 
that we consider in our analysis.   The flare at 4:03 UT has a $U$-band energy of $E_U<10^{30}$ erg.  Later, another moderate energy flare ($1.6\times10^{31}$ erg) and a larger 
flare ($1.3\times10^{32}$ erg) occurred; the cloud coverage became much worse at the WHT after the flare at 4:32 UT, and so 
these two flares were not observed with ULTRACAM.  SDSS $u$, $g$, and $r$-band photometry of YZ CMi was obtained with the ARCSAT 0.5-m telescope at APO 
on 2012 Jan 14 from 2:54-5:20;  the analysis of these data is outside the scope of this paper, but the light curves are available upon request.

The $U$-band data were obtained to relate the NBF3500 filter characteristics to decades of previous observations in 
the $U$-band.  These filters have similar central wavelengths (3500 \AA\ for NBF3500 compared to $\approx3700$ \AA\ for the $U$-band), but the $U$-band is $\approx7$ times wider than the NBF3500 filter.  
The flare energies in the NBF3500 are $5-7$ times lower than in the $U$-band, which is consistent
with the ratio of the filter widths.  Thus, one can obtain an estimate of the $U$-band energy of a flare by multiplying the NBF3500 energy
by this factor.

\section{Flare Color Indices} \label{sec:analysis}
Using quiescent spectra of the target stars obtained from K13, 
we converted the relative photometry in each ULTRACAM filter to a ratio 
of flux densities at Earth.  Proxima Centauri is not included in K13 because it is in the 
Southern hemisphere.  We derive the quiescent flux for this star by comparing
the count rates in quiescence (during stable and clear conditions on MJD 55340) with a large aperture (radius of 17 pixels) to 
the count rates of a spectrophotometric standard DA white dwarf LTT 3218 obtained on the same night (MJD 55340 22:36-22:44). 
 An airmass correction is applied using the La Silla atmospheric extinction curve.
  Gl 644 AB is also not included in K13;  for this star, 
we use the quiescent spectrum for AD Leo, which is the same spectral type.  The quiescent spectrum described in \cite{Kowalski2011} is
used for EQ Peg A. 

Using the definition of the mean flux density per unit wavelength in a bandpass \citep{Sirianni2005}, the quiescent spectra were convolved with the ULTRACAM filter transmittance (\url{http://www.vikdhillon.staff.shef.ac.uk/ultracam/filters/filters.html}) and the CCD response curves 
 giving the quiescent flux density ratios, $R_{Q, 3500/4170}$ (QcolorB) and $R_{Q, 4170/6010}$ (QcolorR).
The quiescent flux densities in each continuum filter and the values
of QcolorB and QcolorR are given in Table \ref{table:stars}.
  Using the 
formula in \cite{Kowalski2011}, the flare color indices, FcolorB and FcolorR,
are determined as follows:

\begin{equation}\label{eq:fceq1}
\mathrm{FcolorB}(t)=\mathrm{QcolorB} \times \frac{I_{f,3500}(t)-I_{pre,3500}}{I_{f,4170}(t)-I_{pre,4170}}
\end{equation}

\begin{equation}\label{eq:fceq2}
\mathrm{FcolorR}(t)=\mathrm{QcolorR} \times \frac{I_{f,4170}(t)-I_{pre,4170}}{I_{f,6010}(t)-I_{pre,6010}}
\end{equation}

 $I_{\rm{f,\lambda}}(t)$ is the excess count flux during the flare in units of the quiescent count flux, $I_{f,o,\lambda}$ \citep{Gershberg1972, Hilton2011}.
$I_{f,\lambda}(t)$ is obtained directly from the normalized relative photometry light curve given by $I_{f,\lambda}(t)+1$ (Section \ref{sec:ucred}).
In Equations \ref{eq:fceq1} - \ref{eq:fceq2}, $I_{\rm{pre},\lambda}$ is the pre-flare value of $I_{f,\lambda}$
and is subtracted to account for elevated emission from a previously decaying flare event.  

The general formula for relating photometry to the intrinsic flare spectrum is given in \cite{Hawley1995}.  If a fraction, $X$, of the visible stellar hemisphere is flaring
 with a surface flux spectral energy distribution given by $S_{\rm{Flare}, \lambda}$, then the flux at Earth (above the atmosphere) from the star is given
by $(S_{\rm{o},\lambda} - XS_{\rm{o},\lambda} + XS_{\rm{Flare},\lambda})\times\frac{R^2}{d^2}$ (where $S_{\rm{o},\lambda}$ is the quiescent surface flux of the star, $R$ is the stellar radius, and $d$ is the distance to the star).  Incorporating the total system response ($T(\lambda)$), subtracting the quiescent star count flux, and dividing by the quiescent star count flux
then gives the value of $I_{\rm{f,\lambda}}(t)$:

\begin{equation} \label{eq:ifeq}
\frac{\int T(\lambda)\lambda(X S_{\rm{Flare},\lambda}- X S_{\rm{o},\lambda})\frac{R^2}{d^2} d\lambda}{ \int T(\lambda)\lambda S_{\rm{o},\lambda} \frac{R^2}{d^2} d\lambda}
\end{equation}
 
\noindent where $X$ and $S_{\rm{Flare},\lambda}$ vary as a function of time.
For example, $I_{f,3500}(t) \propto (XS_{\rm{Flare},3500}-XS_{\rm{o},3500}) \times \frac{R^2}{d^2}$, and thus the \emph{measured} flare color indices given by Equations \ref{eq:fceq1} and \ref{eq:fceq2} are directly related to the continuum ratios of the \emph{excess} surface flux density caused by the flare\footnote{Assuming for simplicity that all of the flare area at a given time is approximated by a single spectral energy distribution given by $S_{\rm{Flare,\lambda}}$.}.  
    The flare color indices are equal to the surface flux ratios of the intrinsic flare spectral energy distribution if $S_{\rm{Flare},\lambda}$ consists
of only optically thin emission above an unaffected photospheric radiation field or if 
  $S_{\rm{Flare},\lambda}>>S_{\rm{o},\lambda}$, which is the case at the blue and NUV wavelengths for an M dwarf (see also the recent discussions about spatially resolved spectral observations of solar flares in \citet{Kerr2014} and \citet{Kleint2016}).  At red wavelengths, this is generally a valid approximation
  because the model surface flux values must be large to be 
  consistent with the observed Balmer jump ratios\footnote{Very large Balmer jump ratios
  correspond to models with $S_{\rm{Flare},\lambda>5000}\approx S_{\rm{o},\lambda>5000}$ \citep{Allred2006};  small Balmer jump ratios are observed, which are characteristics 
   of model spectra with very high surface flux values at all optical wavelengths (e.g., see Figure 1 of K15).}. 
   Aside from these considerations,
  the inferred values of $X$ are usually very small ($<<0.01$) and subtracting $S_{\rm{o},\lambda}$ and  $XS_{\rm{o},\lambda}$ from $XS_{\rm{Flare},\lambda}$ to obtain $I_f(t)$ in Equation \ref{eq:ifeq} is nearly  
   equivalent to subtracting only $S_{\rm{o},\lambda}$. 
   
   FcolorB is a Balmer jump ratio, which is the ratio of the 
   flux in the NBF3500 filter (blueward of the Balmer jump) to the flux in the NBF4170 filter (redward of the Balmer jump at wavelengths between H$\delta$ and H$\gamma$).  FcolorR 
   gives a proxy of the color temperature of the blue-to-red optical wavelength continuum.  FcolorB is similar 
   to $\chi_{\rm{flare}}$, which is the ratio of continuum fluxes C3615/C4170 used in K13.  FcolorR is similar to the continuum flux ratio C4170/C6010 used in K13.
   The interpretation of the flare color indices and comparison to model spectra is discussed further in Section \ref{sec:interp} and Section \ref{sec:rhd}.

We estimate the $1\sigma$ systematic uncertainty of QcolorB and
QcolorR to be $\approx$5\%, obtained 
 from the systematic uncertainty of the flux calibration for the quiescent
 spectra in the NUV, blue, and red (Appendix A of K13).  The statistical errors of FcolorB and FcolorR are calculated
using standard error propagation of the count rate errors of the target star 
and comparison star returned by the ULTRACAM pipeline, including an uncertainty in the pre-flare value of $I_f$ which we take as the error in the mean
of $I_f$.  When comparing flare color variations through the evolution of the flares on the same star (on the same night), the systematic 
errors are not necessary to include in the error propagation.  Because we seek to compare flare color indices on different stars and to model predictions, 
we add the systematic and statistical errors in quadrature. 
 Since the NBF3500 filter is often obtained with twice the exposure time of the NBF4170 
and RC\#1 filters, we bin the NBF4170 exposures for the
calculation of FcolorB.  In the light curves, the values of FcolorR are shown at the original
cadence (usually twice the NUV arm), but in the tables and throughout the text, the FcolorR indices are averaged (using a weighted mean)
at the flare peak to increase the signal-to-noise.  
We consider only values of FcolorB and FcolorR with a 
significance greater than 4$\sigma$.

\section{The Relationship between ULTRACAM Flare Color Indices and Simultaneous Spectra for Two Moderate-Sized Flares} \label{sec:spec}

Obtaining simultaneous spectra and ULTRACAM data was necessary to understand 
how the coarse continuum distribution evolution given by the FcolorB and FcolorR indices relate to the detailed continuum evolution 
and the bright emission lines which lie outside of these narrow filters.  

We obtained simultaneous low-resolution spectra and ULTRACAM photometry during two moderate-sized flares on YZ CMi 
on 2012 Jan 14 (Section \ref{sec:specdata}).  The light curves of the NBF3500 photometry, the H$\gamma$ line flux, the
Ca \textsc{ii} K line flux, and the $U$-band for the two flares (which we denote as ``F1'' and ``F2'') are shown in Figure \ref{fig:summary1}. 
The NBF3500 light curve peak for the F1 event occurs at 2012 Jan 14 2:59:26, and for the F2 event at 2012 Jan 14 4:32:01.  Several smaller flares occur within this time interval at 3:10, 4:03, and 4:09 UT which produce insignificant enhancements of the Ca \textsc{ii} K emission line;  the flare events at 4:03 and 4:09 UT are included in the peak flare color analysis of Section \ref{sec:sample}, but the flares are too small to give useable signal-to-noise
 in the spectra.   As discussed in Section \ref{sec:1m}, the F1 and F2 events occurred 
in the decay phase 4.2 hours and 5.8 hours, respectively, after a much larger flare event, which is discussed in
Section \ref{sec:ultraflare}. 
 In Figure \ref{fig:summary2}, we show the ULTRACAM data of F1 and F2 for a narrow time range around the peaks.  We also show the values of FcolorB and FcolorR, with spectral
integration windows as gray bars in the top panel.  

\subsection{Description of the Flares} \label{sec:desc}

 The energy of the F2 event in NBF3500 is four times greater
than the energy of the F1 event, and the peak amplitude is 1.5 times larger ($I_f+1=4.7$ for F2 and 3.4 for F1, calculated with the pre-flare enhanced emission levels subtracted).
 The two flares have different temporal morphology in the impulsive and gradual phases.  To compare the timescales of the impulsive phases, we 
use the $t_{1/2}$ value, which is the FWHM of the light curve (K13).
These values are given in Table \ref{table:thalf}, where the values in the
NBF4170 and RC\#1 filters are calculated from the light curves that have
been binned to the NBF3500 exposure times.  In NBF3500, $t_{1/2}$ is 34~s and 14~s for F1 and F2, respectively.
The F2 event becomes brighter at peak and is a faster flare than F1; thus the impulsiveness index, $\mathcal{I}$ (where $\mathcal{I}=I_f/t_{1/2}$ and $t_{1/2}$ is 
expressed in minutes; K13)
for the F2 event is about a factor of four larger than F1.  
From the light curves in Figure \ref{fig:summary2}, there are several high-time resolution effects
 that generate a larger $t_{1/2}$ and smaller
value of $\mathcal{I}$ for F1.  The rise and peak phases of F1 consist of  
several smaller events, or ``bursts'', occuring in the impulsive phase and  
appear to sustain the
impulsive phase for a longer time but at a lower level than for F2.  In contrast, the F2 light curve has 
 a steep slope in the rise phase with no evidence of several bursts.  The flare event F1 has a total rise time of 13~s (due to the succession of several bursts), whereas F2
 has a rise time of 8~s.

The initiation of the gradual phase emission begins at a shorter 
time after the main peak in F1, and the time range corresponding to $t_{1/2}$ includes some of this gradual emission (see also discussion in Section \ref{sec:thalfs}).
  After the impulsive phase of F2, there is a gradual, low amplitude event which peaks at 4:34 UT (Figure \ref{fig:summary1}), 
$\approx130$~s after the main peak. At high time-resolution, several very small events are evident during this gradual event.
  This gradual flare event within the F2 event corresponds
to the peak of the hydrogen Balmer lines and Ca \textsc{ii} K line emission in Figure \ref{fig:summary1}. 
The peak of the Balmer lines and Ca \textsc{ii} K line emission during F1 corresponds to the 
beginning of the gradual decay phase of the NBF3500 light curve, just after a secondary gradual peak in NBF3500 occuring 16~s after the main peak.  The continuum flux (C3615 and C4170) light curves calculated from the spectra also peak before the peaks of the emission line light curves, indicating that the time delays between the peaks in the light curves of the ULTRACAM data and the emission lines are not a result of the higher temporal resolution of the ULTRACAM continuum data.  These time delays are similar to those observed in the flare events from K13, indicating a change in the heating (or cooling) phases in these dMe flares, and have yet to be explained.

\subsection{Comparison to Simultaneous Spectra}
In Figure \ref{fig:spectra}, we show the APO/DIS spectra (Section \ref{sec:specdata}) for the impulsive and gradual phases for these two events.  The 
flare-only emission is calculated by subtracting a pre-flare spectrum (shown in gray and scaled by 1/10 in the top panel).
Averaged over the 30~s integration time of the spectra, the flare-only emission is a small fraction of the quiescent emission,
especially at redder wavelengths\footnote{The H$\alpha$ profiles show
a central dip that attains negative values.  This effect is likely due to seeing fluctuations behind the wide slit, described in Section \ref{sec:specdata}.}.
In Figure \ref{fig:spectra}, we show the impulsive phase of F1 (i, top panel),
the first gradual phase of F1 (ii, second panel),
the impulsive phase of F2 (i, third panel), and the coadded gradual decay phase (or a secondary gradual event) spectrum of F2 (ii-v, bottom panel).  
 The ULTRACAM filter curves are shown as dotted lines, which confirm
that these filters sample only continuum emission outside of any major or minor emission lines during the impulsive and gradual phases. 

We calculate synthetic values of FcolorB and FcolorR from these spectra, and we find they are quite consistent with the values of FcolorB and FcolorR
 obtained at flare peak in the ULTRACAM data.  In Table \ref{table:calib}, 
we give the synthesized values (columns 4 and 5) and the ULTRACAM values (columns 6 and 7).  The agreement between the spectral
and ULTRACAM flux calibration methods is remarkable for 
the peaks/impulsive phases of both flares.  The high time-resolution evolution of FcolorB and FcolorR in Figure \ref{fig:summary2} in the impulsive phases shows that at a high cadence of $1-2$~s,
 FcolorB does not reach significantly lower values and  
FcolorR does not reach significantly higher values than calculated from the longer integration (30~s) spectra.  
 The flux calibration for the coadded gradual phase spectra for F2 is
also consistent with the ULTRACAM color indices.  We also give the values of $\chi_{\mathrm{flare}}$, where $\chi_{\mathrm{flare}}$ is C3615/C4170, and C4170/C6010 as defined in K13
(columns 2 and 3) obtained from the spectra; as expected, the value of  $\chi_{\mathrm{flare}}$ is similar to FcolorB and
the value of C4170/C6010 is similar to FcolorR.  Even with the low signal-to-noise at the bluest
wavelengths, the spectra give the correct relative level of flare emission averaged over the ULTRACAM NBF3500 filter curve.  
The flux calibration thus appears consistent between the ULTRACAM high time-resolution color indices and the broad wavelength 
coverage of the low time-resolution spectra.  The flare color indices of individual flares can be related to the continuum characteristics, which
we describe in detail in Section \ref{sec:interp} and Section \ref{sec:rhd}.

\subsection{A Phenomenological Interpretation of the ULTRACAM Filter Emission} \label{sec:interp}
The spectra in Figure \ref{fig:spectra} illustrate how the broader wavelength continuum evolution leads to the changes in the values 
of FcolorB and FcolorR between the peak and gradual phases. 
For example, it is apparent how the value of FcolorB becomes larger in the gradual phase, whereas FcolorR decreases.  
FcolorB is a ratio of NUV to blue continuum fluxes, and is a measure of the Balmer jump ratio, similar  
to $\chi_{\mathrm{flare}}$ used in K13 (FcolorB samples the flare flux at $\lambda \approx 3450-3550$ \AA, whereas $\chi_{\mathrm{flare}}$ averages the flux between $3600-3630$ \AA).  FcolorR is a measure of the color of the continuum emission at wavelengths longward of the 
Balmer jump, and we will use it to provide an estimate of the color temperature, $T_{\rm{FcolorR}}$, for the blue-to-red ($\lambda=4170 - 6010$ \AA) optical wavelength regime.  

K13 found that 
the color temperature of the continuum regions in the blue-optical wavelength regime ($\lambda=4000-4800$\AA) could be fit by a single blackbody function with temperature, $T_{\rm{BB}}$.  
A value of $T_{\rm{BB}}\approx8500$ K or greater indicates a large optical depth in the flare plasma at $T_{\rm{e}}\gtrsim 10\,000$ K (see K15 and Section \ref{sec:rhd}),
Thus, high-time resolution information on this hot blackbody, or blackbody-like, emission helps constrain future radiative-hydrodynamic models.  
Since the flux calibration of the ULTRACAM data and spectra are consistent, we are able to determine a 
relationship between $T_{\rm{BB}}$ and $T_{\rm{FcolorR}}$ using the simultaneous spectra and ULTRACAM photometry.
Following K13, we fit the continuum emission in the blue wavelength regions, designated as BW (``blue windows'') which are given in Table 4 of K13, with a blackbody function to obtain $T_{\rm{BB}}$; these 
fits are shown as light blue curves in Figure \ref{fig:spectra}, and the best-fit parameters of $T_{\rm{BB}}$ and the areal filling 
factor of emission \citep[$X_{\rm{BB}}$;][]{Hawley2003} are given in Table \ref{table:bbtable}. 
The values of $T_{\rm{BB}}$ for F1 and F2 are 12,100 K in the impulsive phases (F1(i) and F2(i)) and $8600-9300$ K
in the gradual phases (F1(ii) and F2(ii-iv)) of the flares.  
The spectra in Figure \ref{fig:spectra} illustrate
that NBF4170 and RC\#1 filters are dominated by hot blackbody-like continuum emission during the peak and gradual phases 
during both F1 and F2 events.
 
The statistical errors on $T_{\rm{BB}}$ are $ \approx 600-1200$ K. 
However, these flares produce relatively faint emission in the
 blue-to-red optical wavelength regimes, which can be seen by comparing 
 the flare-only emission to the quiescent spectra scaled by 0.1 in Figure \ref{fig:spectra}.
 Thus, the systematic errors of the color temperature from the spectra can
 be comparable or larger.  
If we exclude the blue-most continuum windows (BW1 and BW2; K13 Table 4) from the blackbody fits, the values of $T_{\rm{BB}}$ become $\approx11,000$ K for F1(i) and F2(i) and are consistent
with the systematic errors of 1200 K quoted in K13.  An important step in the flux calibration of these spectra includes scaling each spectrum by a multiplicative factor in order to minimize the subtraction residuals in the red when isolating the flare-only emission (see Section \ref{sec:specdata} and Appendix A of K13).  For the low levels of emission produced in these flares, the measured color temperature varies significantly if we adjust this scale factor by a small amount around the value returned by the scaling algorithm.
The resulting temperature range if we adjust the scale factor by $\pm$0.01 is $9,600-18,100$ K  for F1(i) and $10,200-15,500$ K for F2(i);  adjusting the scale factor by 
$\pm0.005$ (the $1\sigma$ relative uncertainty in the flare increase), the range is $10,600-14,300$ K for F1(i) and  $11,000-13,500$ K for F2(i). 
The uncertainties from the scaling algorithm and the blackbody fitting procedure give approximately the same
lower limit to the value of $T_{\rm{BB}}$, and thus the detection of hot blackbody-like emission at the optical wavelengths in the impulsive phase of these
flares is robust.

The value of FcolorR can be used as another measure of the color temperature of the continuum at wavelengths longer than, and significantly far away from, the complicated Balmer jump spectral region.  
According to Table \ref{table:calib}, the FcolorR values determined from the spectra are consistent with the ratios determined from the ULTRACAM data.
However, the ULTRACAM data have higher signal-to-noise and higher cadence than the spectra, and thus the FcolorR indices give values for the color temperatures for the F1 and F2 events that are more directly comparable to model predictions.
We fit a blackbody to the synthesized spectral values of FcolorR (column 5 of Table \ref{table:calib}), resulting in a best-fit color temperature of the blue-to-red optical wavelength regime, $T_{\mathrm{FcolorR}}\approx7000-10\,400$ K.  These values are also given in Table \ref{table:bbtable} and are shown in Figure \ref{fig:spectra} as red-dashed lines.  Compared to the values of $T_{\rm{BB}}$ in Table \ref{table:bbtable}, $T_{\mathrm{FcolorR}}$ from the spectra are systematically lower by $\approx1400-2400$ K, thus giving an approximate relationship between these parameters:

\begin{equation} \label{eq:calibeq}
T_{\rm{BB}} \approx T_{\rm{FcolorR}} + 1900
\end{equation}

The color temperatures corresponding to FcolorR from the spectra are fully
consistent with $T\gtrsim10^4$ K blackbody (or blackbody-like)
emission in the peak/impulsive phases of F1 and F2, as found for most flares in the spectroscopic sample of K13. 
The systematic uncertainty on the values of $T_{\rm{BB}}$ at peak is large ($\approx 1200$ K),
and the values of $T_{\mathrm{FcolorR}}$ fall within the lower range of the possible values of $T_{\rm{BB}}$ temperatures.

\begin{figure}
\begin{center}
\includegraphics[scale=0.6]{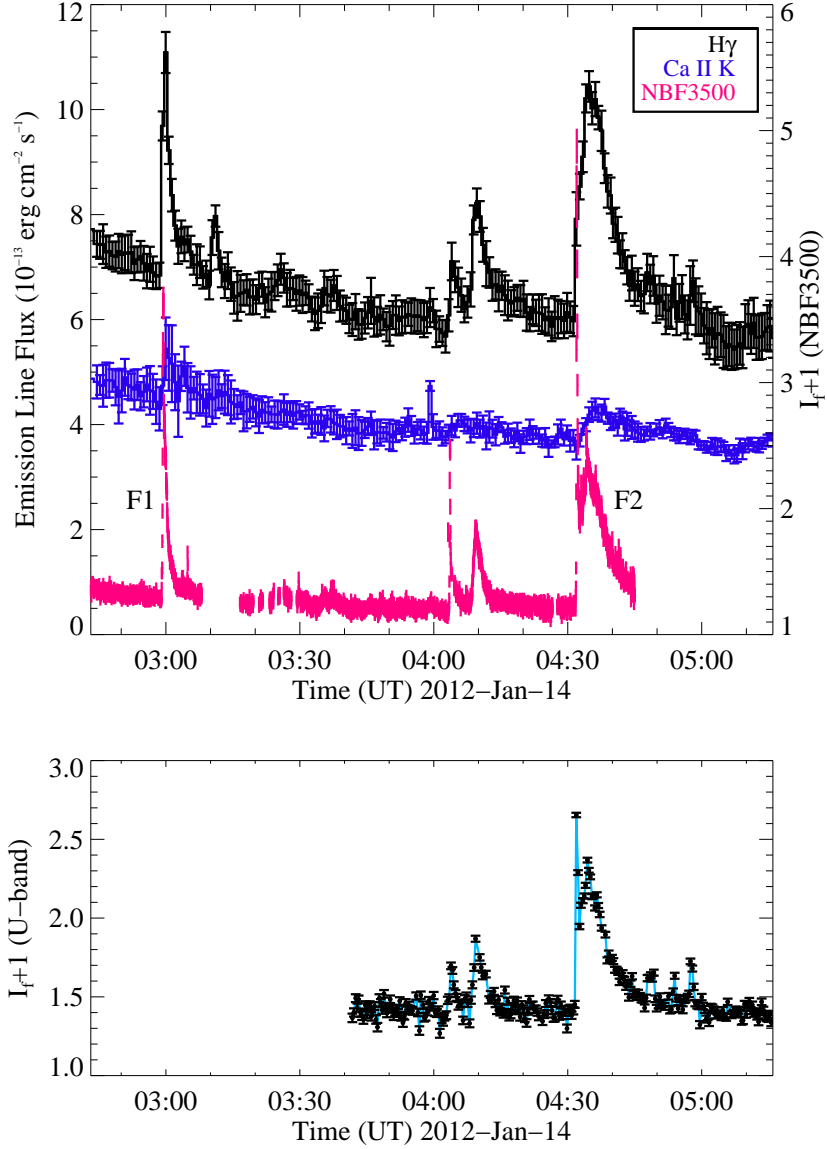}
\vspace{-10mm}
\caption{Simultaneous data on YZ CMi obtained from the NMSU 1m and the ARC 3.5m telescopes at APO (H$\gamma$, Ca \textsc{ii} K, $U$-band) and from the WHT (NBF3500).  The 
line flux values give the total emission from the star as a function of time (includes non-flaring regions, newly flaring regions, and previously decaying flare regions).  Flares discussed in the text 
include F1 (IF11) at 2:59, IF8 at 4:03, GF2 at 4:09, and F2 (IF4) at 4:32.  Time intervals with transparency variations at the WHT have been removed from the NBF3500 light curve; the flare times are unaffected.  } \label{fig:summary1}
\end{center}
\end{figure}

The value of FcolorB gives a measure of the Balmer jump ratio, similar to the $\chi_{\mathrm{flare}}$ value employed
in K13.  As can be seen in Figure \ref{fig:spectra}, there is excess flare-only emission above the blue and red-dashed line (blackbody curve)
extrapolations to the shortest wavelengths in the range of the NBF3500 filter.  Thus, the blackbody continuum component that determines the value
of FcolorR cannot account for the emission in all three ULTRACAM filters. 
This is the case for both impulsive and gradual
phase spectra of F1 and F2;  the excess emission above a blackbody extrapolation was given as the ``BaC3615'' measure in K13, or
an excess Balmer continuum emission component.  
Both flares illustrate that the NBF3500 filter has a large
contribution (60\,--\,75\%) from blackbody emission in the impulsive phase, 
but the excess Balmer continuum emission becomes larger (both relatively and absolutely)
in the decay phases as the blackbody component decreases in (apparent) temperature while the emission lines become both relatively and absolutely stronger (Figure \ref{fig:summary1}).  
These effects were observed in K13 in other (larger) flares, and the excess flux at $\lambda < 3646$ \AA\ was interpreted as evidence of Balmer recombination radiation and a Balmer jump in emission.
The value of FcolorB thus follows the relative importance of the excess Balmer continuum emission to that of the blackbody emission in NBF4170 filter.  

 The hydrogen Balmer emission
lines are small in the impulsive phase of F1 and F2 and become stronger in the gradual phase consistent with the line flux evolution in Figure \ref{fig:summary1}.
 In Figure \ref{fig:summary2} (top panels), we show that the FcolorB trend is similar to the evolution of H$\gamma$/C4170 (the H$\gamma$ equivalent width relative to the $\lambda=4170$ \AA\ continuum), which is a measure used by K13 and K15 to compare to models.
 In the impulsive phases of F1 and F2, the value of H$\gamma$/C4170
 (where C4170 is similar to NBF4170) is 40 and 20, respectively, and the value for F2 is comparable to the smallest in K13 (Figures $11-12$ of K13).
 In the gradual phases of the flares, the 
Balmer jump ratio becomes larger at the same time as the hydrogen Balmer emission lines
become stronger relative to the continuum (and according to the absolute line flux; see Figure \ref{fig:summary1} and Section \ref{sec:timeevol}). 
  Thus, FcolorB also gauges the relative importance of Balmer line
  emission to the blue-optical continuum emission in NBF4170.
In the two-ribbon solar flare analogy of the gradual phase, the excess Balmer continuum (and line) emission in the NBF3500 filter would originate from previously heated ribbons
and the blue-optical continuum emission would originate from cooling kernels or ``hot spots'' \citep{Kowalski2012}.  
An attempt at more advanced modeling was made in K13 to explain the impulsive phase emission (using hot-star Kurucz models to represent the hot blackbody component and to account 
for ``missing BaC'' emission);  in Section \ref{sec:rhd} we improve on this phenomenological interpretation of FcolorB and FcolorR using recent results from radiative-hydrodynamic model flare spectra produced with the RADYN and RH codes. 

\subsection{Time-Evolution of FcolorB and FcolorR in F1 and F2} \label{sec:timeevol}
Now that the values of FcolorB and FcolorR can be readily related to the broader continuum and emission line properties, the ULTRACAM
 data can set new constraints on the high time-resolution evolution of the white-light emission for comparison to radiative-hydrodynamic flare models.
In this section, we describe the detailed evolution of FcolorB and FcolorR for the F1 and F2 events; in Section \ref{sec:ultraflare}, we describe the evolution
 during the largest flare in the sample.

\begin{figure}
\includegraphics[scale=0.65]{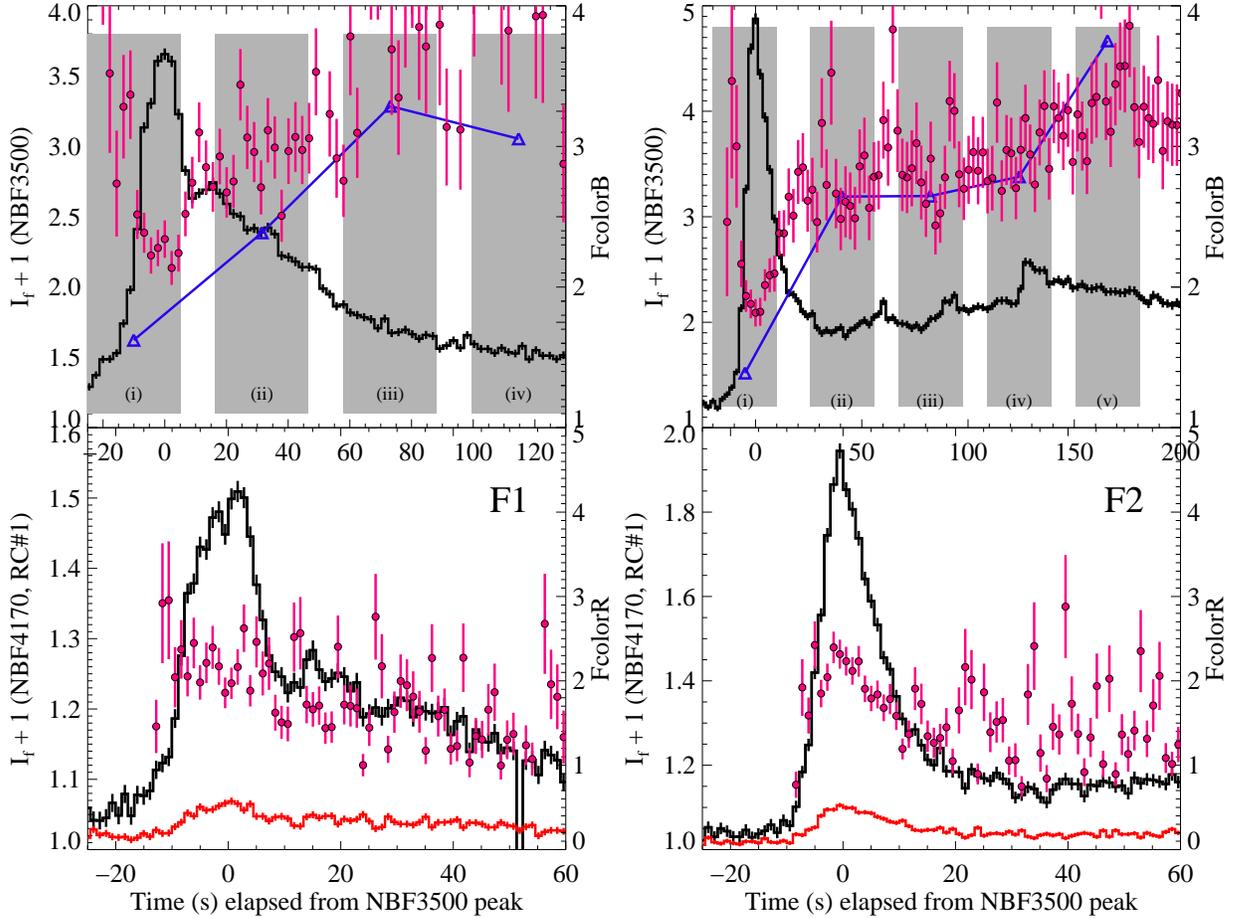}
\vspace{-10mm}
\caption{ Light curves (black) and flare color (pink with error bars) for F1 and F2 on YZ CMi.
  The RC\#1 filter data are also shown as the red light curves in the bottom panels.  Spectral
  integration times are highlighted as grey bars in the top panels, and the Roman numerals correspond to the 
  spectra in Figure \ref{fig:spectra}. Note different time axes on the top and bottom panels.  Note, the F2 event includes a gradual event with a second maximum in NBF3500 
  at $t-t_{\rm{peak}}=130-150$~s.
  The blue triangles are the values of the H$\gamma$/C4170 ratios (see text), plotted on a linear scale ranging from 
   $0-200$ in the top left panel and $0-150$ in the top right panel.  The pre-flare emission levels have not been subtracted in these figures.}  \label{fig:summary2}
\end{figure}

\begin{figure}
\begin{center}
\includegraphics[scale=0.5]{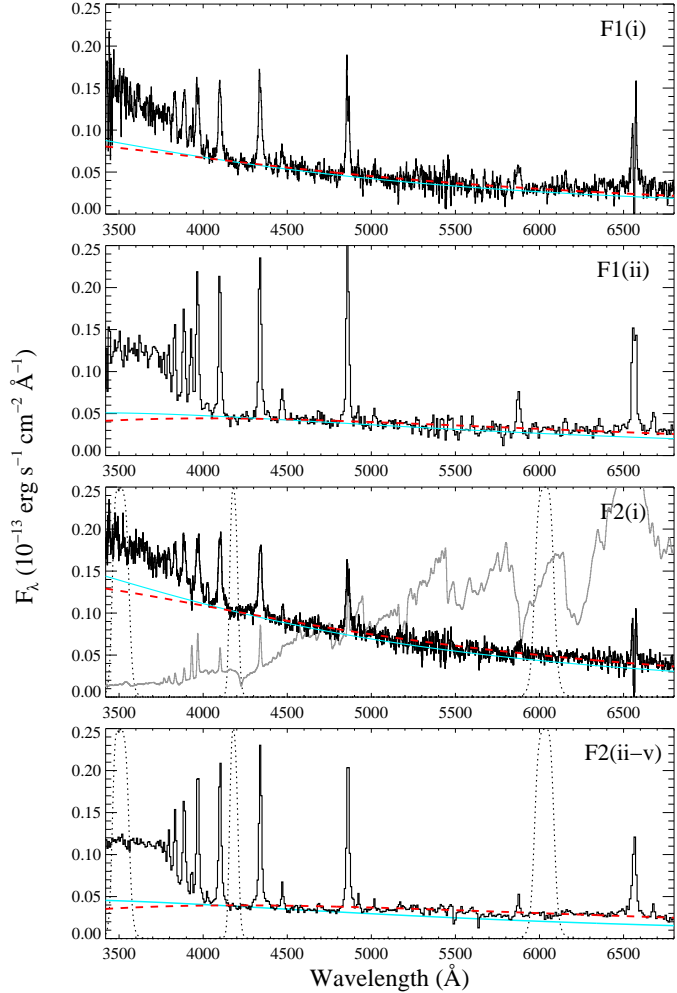}
\caption{APO/DIS spectra corresponding to the times of the gray shaded bars (designated by Roman numerals) in Figure \ref{fig:summary2}. The gray spectrum in the panel  
that is third from top is a pre-flare spectrum of YZ CMi scaled by 0.1.  The light blue curves show the blackbody 
fit to the blue continuum wavelength regions (the ``BW'' ranges from Table 4 of K13) between $\lambda=4000-4800$ \AA, and the red-dashed curves
show the blackbody fits to the synthesized fluxes in the NBF4170 and RC\#1 ULTRACAM filters.  The gradual
phase spectra (second and fourth panels) have been binned in the dispersion direction by 4 pixels to increase
the signal-to-noise.  The gradual phase spectra F2(ii), F2(iii), F2(iv), and F2(v) were averaged to 
obtain F2(ii-v).  Note, the dichroic of DIS is known to affect the flux calibration from $\lambda=5200-5800$ \AA, but the 
general trend seems to be preserved here. } \label{fig:spectra}
\end{center}
\end{figure}

The F1 event shows a sustained peak phase and slower decay compared to the F2 event, which is evident in the bottom panels of Figure \ref{fig:summary2}.
For F1, we calculate a weighted average of the ULTRACAM FcolorR values over times corresponding to the fast rise phase, 
middle peak phase, maximum peak phase, and fast decay phase,
and we find that FcolorR is approximately constant ($\approx$2.1\,--\,2.2) through the impulsive phase. 
These values of FcolorR for F1 correspond to color temperature values
of $T_{\rm{FcolorR}}=10,500-11,000$ K, consistent with the value of $T_{\rm{FcolorR}}=10,400$ K obtained 
from the F1(i) spectrum (Figure \ref{fig:spectra}, Table \ref{table:bbtable}).
 The inferred
areal coverage of this emission from the ULTRACAM data is $1-1.2\times10^{17}$ cm$^{2}$ at the peak.  The area from the
spectrum using the synthesized FcolorR is $A_{\rm{FcolorR}} \approx 6\times10^{16}$ cm$^{2}$, which makes sense because the 
spectrum has a 30~s integration time and is averaged over the impulsive
phase amplitude.  During the initial gradual decay phase emission (corresponding to 
 the spectrum F1(ii) in Figure \ref{fig:spectra}),
the weighted average value of FcolorR from the ULTRACAM filters gives $T_{\rm{FcolorR}}\approx8000$ K (compare to synthesized spectral value of 7200 K; Table \ref{table:bbtable}) and an emitting area of
$\approx1.4\times10^{17}$ cm$^{2}$.  A lower color temperature by $\approx2000$ K is characteristic of the late fast decay and early gradual decay
phase spectra (K13), but a comparable or larger inferred area at this
time is at odds with the time-evolution of the hot blackbody-emitting area that was found to decrease significantly in the gradual phase 
for the IF3 event in K13 (see Figure 33 of K13 and Figure 6.17 of \cite{KowalskiThesis}).
Perhaps better models (e.g., Section \ref{sec:rhd} here or Appendix F of K13) are required to properly infer the areas in
the gradual decay phase.

The unbinned FcolorR values have higher signal-to-noise in F2, and its evolution is also shown in Figure \ref{fig:summary2} (bottom panel).
The evolution of FcolorR differs slightly between the impulsive phases of F1 and F2.  Whereas
FcolorR in F1 is approximately constant at $\approx 2.1-2.2(\pm0.1)$ through the rise, peak, and fast decay phases, 
FcolorR in F2 varies between $1.8- 2.4$ (corresponding to $T_{\rm{FcolorR}} \approx 8900 - 12300$ K), is maximum around the time of maximum emission, and
 decreases steadily from $\approx$2.3 to $\approx$1.5 in the fast decay phase.  
 
On the other hand, the time evolution of FcolorB is quite similar between the two flares, with 
a decrease in the rise phase, a minimum value at the peak, and an increase in the fast decay phase, followed
by a slower increase in the gradual decay phase.  This anti-correlation between FcolorB and $I_f$
 is common to dMe flares and is due to the relative importance of the blackbody emission compared to the Balmer continuum emission (see Chapter 6 of \cite{KowalskiThesis}).  
As discussed in Section \ref{sec:interp}, FcolorB also follows the trend of the ratio of Balmer line flux to the $\lambda=4170$ \AA\ continuum.  
However, the FcolorB value attained at the peak of F1 is 
slightly larger (2.3) than for F2 ($1.8-1.9$).  Propagating the errors on the color indices, we find that 
this difference corresponds to a confidence level of 3$\sigma$.
The impulsiveness index ($\mathcal{I}$) of the NBF4170 emission for F2 is about 2.8 times that for F1. 
This difference is consistent with larger Balmer jump ratios appearing in less impulsive flares, as observed for the flare sample in K13.  Note
that F2 (the more impulsive flare) is also the flare with more variation in the FcolorR index in the impulsive phase.  The F1 impulsive phase event has an extended peak phase due to several smaller bursts, whereas the F2 event is comprised of one larger burst.  We speculate that these differences generate
the stronger variations in FcolorR and lower value of FcolorB in the F2 impulsive phase.
The differences in FcolorB between F1 and F2 may also be related to the different decay phase evolution in the two flares:
 F1 has a larger amount of gradual emission at the end of the fast decay whereas the F2 event has a more extended fast decay phase followed by a delayed gradual event ($t=30-200$~s in Figure \ref{fig:summary2}). 
Therefore, these spatially unresolved observations may include emission from regions with gradual emission (large values of the Balmer jump ratio)
that are spatially superimposed with impulsive emission (small values of the Balmer jump) to give the total flare spectrum.

These F1 and F2 flare events are among the smallest amplitude flares ($\Delta$ mag $=-1.4,-1.7$, respectively, in NBF3500 at the peaks) yet observed that exhibit convincing evidence 
for hot, $T_{\rm{BB}}\approx10^4$ K blackbody-like emission in the impulsive phase spectra.  In \cite{Hawley2003}, similarly small filling factors
were found for $\approx$9000 K blackbody emission using broadband photometry of a $\Delta U=-1.8$ mag event on AD Leo, and \cite{Lovkaya2013}
 also used broadband colors to deduce $T\sim 14\,000$ K blackbody components at the peaks of two moderate-amplitude events, $\Delta U=-1.65$ and $-1.76$ mags, also on AD Leo.  The simultaneous information from our spectra allow us to assess the 
excess Balmer continuum emission and obtain a more detailed view of the continuum during moderate-sized flares. 
 According to \cite{Lacy1976}, the average flare energy on YZ CMi 
is comparable to the F1 event (and the F2 event is about four times more energetic).  Our spectral observations (Figure \ref{fig:spectra})
thus confirm that an energetically dominant, hot ($10^4$ K) blackbody-like emission component is present during the impulsive phase of small and large
flares, and that this continuum emission can be characterized at high-time resolution with ULTRACAM data.

We do not find unambiguous evidence of a cooler continuum component in the red (referred to as
the ``Conundruum'' continuum component in K13) in the spectra of F1 and F2.
  The continuum at all wavelengths $> 4000$\AA\ can be adequately explained 
by the blackbody curve fits to the optical regime in Figure \ref{fig:spectra}.  K13 noted
that the amount of Conundruum continuum emission varies from flare to flare without any apparent preference for the flare temporal morphology.  
This continuum component was also found
to contribute relatively more in the late gradual phase.  For the F1 and F2 events, the late gradual phase emission is at such a low level
that a significant detection, as for much larger flares (e.g., Figure
31 of K13), is not possible.  If the Conundruum component were 
present, it would give rise to a flattening of the flare-only emission in the reddest wavelengths around RC\#1. The spectrum F2(ii-v) 
 may exhibit some emission
 in this filter that is not accounted for by the light blue curve ($T_{\rm{BB}}$), and the spectrum F1(ii) is suggestive of a flattening/reddening in the flare-only
emission around this filter, which may indeed be evidence of some contribution from this 
continuum component in the gradual phases of these two events.  A signature of the
Conundruum component in the impulsive phase may be the  
difference between $T_{\rm{FcolorR}}$ and $T_{\rm{BB}}$ calculated in Section \ref{sec:interp}, which
would indicate that the larger Conundruum contribution 
in the RC\#1 filter produces an apparently multithermal continuum over
the entire blue-to-red optical wavelength regime.  Alternatively, there
may not be any Conundruum emission in the impulsive phases of these flares,
since the red continuum flux can be fully accounted for by a new radiative-hydrodynamic flare model (Section \ref{sec:rhd}).

\subsection{The ``F13 Interpretation'' of the Impulsive Phase ULTRACAM Flare Emission} \label{sec:rhd}

A recent radiative-hydrodynamic (RHD) simulation from
K15 \citep[see also][]{Kowalski2015IAUS} using a high flux of nonthermal electrons
provides new insight into the 
 atomic processes that generate the flare continuum emission in the ULTRACAM
 filters.  K15 used the RADYN \citep{Carlsson1992, Carlsson1994, Carlsson1995, Carlsson1997, Carlsson2002, Abbett1999, Allred2005, Allred2006} and RH \citep{Uitenbroek2001} codes to 
 model the atmospheric response to a beam of nonthermal electrons with an energy flux of $10^{13}$ erg cm$^{-2}$ s$^{-1}$ (F13).
  The electron beam followed a double power-law distribution with $\delta_{\rm{upper}} =
  4$ above $E=105$ keV and $\delta_{\rm{lower}} = 3$ below 105 keV.  Just after 2~s of flare heating,
  a dense ($n_{\rm{H,max}} = 7.6\times10^{15}$ cm$^{-3}$), heated
  ($T=12,000-13\,500$ K) compressed region of the atmosphere, or a chromospheric ``condensation'', had formed due to
  thermal pressure in a downward moving shock front.  
  This chromospheric condensation resulted in strong continuum emission with a color temperature of $T_{\rm{BB}}=10^4$ K and a small
  Balmer jump ratio ($\chi_{\rm{flare}}=2.0$).  
  
\begin{figure}
\includegraphics[scale=0.75]{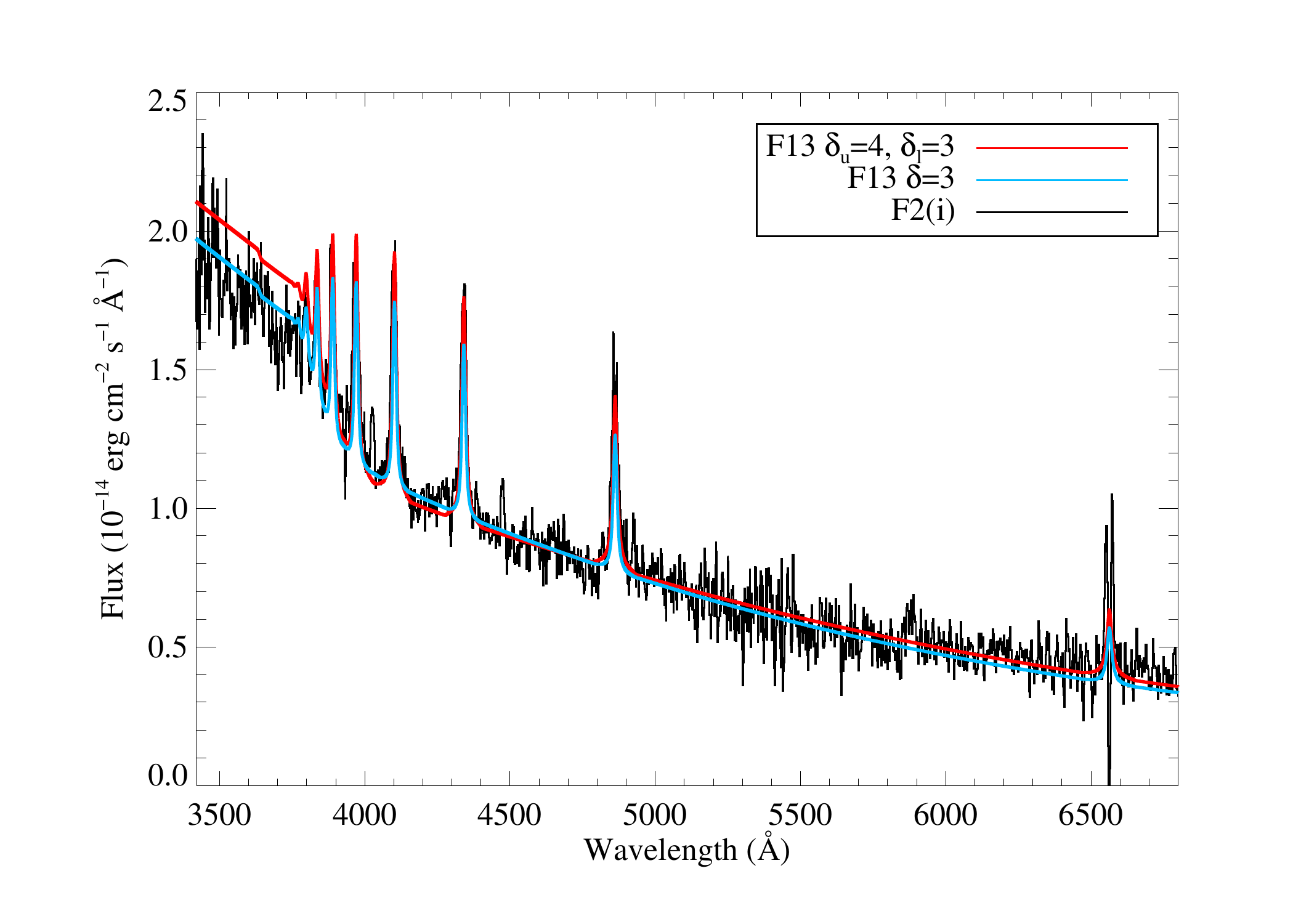}
\caption{ The F2(i) spectrum from Figure \ref{fig:spectra} is shown in black and model spectra produced by the RADYN and RH codes are shown for the single and double power-law 
electron heating distributions as the blue and red, respectively.  The spectra have been scaled to the flux at $\lambda=4700$ \AA\ in order to show that the continuum
ratios from the models are generally consistent with the observations. } \label{fig:f13}
\end{figure}

  To demonstrate the applicability of the (instantaneous) F13
  chromospheric condensation model to the impulsive phase emission of
  the F2 flare event in our ULTRACAM sample, we compare the F13 model flux spectrum ($S_{\rm{Flare},\lambda}$; Section \ref{sec:analysis}) at $t=2.2$~s (thick red line)
  to the F2(i) (impulsive phase) spectrum in Figure \ref{fig:f13}.
   There is general agreement between the model and 
observation, but the Balmer jump ratio (FcolorB) of the model spectrum is larger (2.1) and FcolorR is smaller
  (1.9) compared to the synthesized values of FcolorB (1.8) and FcolorR (2.1) in Table \ref{table:calib}.  
  The values of FcolorB and FcolorR at the high-time resolution of ULTRACAM are similarly discrepant, and the value of FcolorR 
  is even higher (2.3).

We have computed a large grid of radiative-hydrodynamic flare models of M dwarf flares
covering the parameter space of nonthermal electron distribution properties, varying $\delta$, the low-energy cutoff, and the total energy flux.
This grid employs the test-particle, non-relativistic treatment of
energy deposition \citep{Emslie1978} used in K15 and will be updated
with a grid of models using the fully relativistic Fokker-Planck solution \citep{Allred2015} and presented in a future paper (Kowalski et al 2016a, in prep).
Here, we discuss one variation of the F13 model from the grid, and 
we find that changing the ratio of high to low energy electrons in the nonthermal electron
distribution produces a flux spectrum that is more
consistent with the spectral and ULTRACAM observations of the F2 event than the double-power law simulation from K15.

 We use the RADYN code to simulate a single power-law distribution of
 nonthermal electrons with a power-law index $\delta=3$ and heating
 duration of $t=2.3$~s.  We use the \cite{Emslie1978} prescription for nonthermal electron energy distribution, and all other details of the simulation
are also kept the same as in K15 in order to facilitate direct comparison to the double power-law simulation.  
  Compared to the double power-law model in K15, the
   $\delta=3$ electron distribution has the same total heating flux but
  a larger average electron energy 
because the distribution is harder:  the ratio of the number of $E=200$ keV electrons to the number of $E=40$ keV electrons 
is a factor of two larger in the single power-law $\delta=3$ distribution\footnote{The absolute number of lower energy electrons  with $E < 125$ keV is smaller in the single power-law $\delta=3$ simulation.}.  The
resulting atmospheric evolution for the $\delta=3$ simulation is 
very similar to the atmospheric evolution for the double power-law analyzed in K15.  Both simulations produce a dense ($n_{\rm{H}}>10^{15}$ cm$^{-3}$), 
heated chromospheric condensation, although the maximum density 
 in the double power-law simulation is about 30\% larger.
The $\delta=3$ simulation also significantly heats dense, stationary layers just below the chromospheric condensation, as in the double power-law simulation
 of K15.  K15 found that $E\approx60-200$ keV electrons heat the chromospheric condensation whereas only the $E\gtrsim200$ keV electrons can reach
 the stationary flare layers without losing all of their energy.

In both simulations, at $t=2.2$~s, the emergent flux spectrum
achieves a blue-optical color temperature of $T_{\rm{BB}} \gtrsim 9000$ K. However, in the $\delta=3$ simulation,
  the Balmer jump ratio is lower (FcolorB$=1.8$), which is more
  consistent with the peak phase values of F2 (Table \ref{table:calib}).
 The value of FcolorR ($=2.2$) is also higher in the $\delta=3$ simulation at
 $t=2.2$~s, and thus the blue-to-red optical color temperature is
 higher and closer to the observations (FcolorR$=$2.3).   Although the Balmer jump ratio is better 
reproduced by the harder nonthermal electron spectrum, the value of H$\gamma$/C4170 is 20 in the observed spectrum and only 10 in the $t=2.2$~s, $\delta=3$, F13 model; the F13 double power-law model at $t=2.2$~s is more consistent in this respect. 
 
 In Figure \ref{fig:f13}, we show the single power-law $\delta=3$ F13 model (thick blue line) and the K15 double power-law F13 model (red line) emergent flux spectra ($S_{\rm{Flare},\lambda}$; Section \ref{sec:analysis}) at $t=2.2$~s compared
to the impulsive phase spectrum of F2.  We use a projected areal coverage of $A_{\rm{flare}}\approx 5\times10^{16}$ cm$^2$ and the distance to YZ CMi (5.97 pc) to scale the model surface flux spectra to the observed flux at $\lambda=4690-4710$ \AA\ in order to facilitate comparison of the spectral energy distribution. 
 As in K15,  the RH code was used to produce this model spectrum from a snapshot of the atmospheric parameters in the dynamical
simulation at $t=2.2$~s.   
The RH calculation includes higher levels of hydrogen in addition to the bound-free and bound-bound opacity modifications \citep{Dappen1987, HM88, Tremblay2009}
that are due to Landau-Zener transitions of electrons 
between dissolved upper levels of hydrogen.  As a result of the Landau-Zener transitions, Balmer continuum 
emission is produced at $\lambda > 3646$ \AA\ and the model spectrum exhibits a continuous transition 
from the Balmer continuum at $\lambda<3646$ \AA\ to the higher order
Balmer lines (see K15 and \cite{Kowalski2015IAUS}), as in the spectral observation
of F2 in Figure \ref{fig:f13}.  We note that no additional
continuum component (the ``Conundruum" continuum component discussed in Section \ref{sec:timeevol}) is required to
explain the red continuum emission for F2 in
the impulsive phase.

The smaller Balmer jump ratio in the single power-law simulation is due to 
more high-energy ($E\gtrsim200$ keV) electrons and fewer low-energy electrons than in the double power-law simulation.  
 The larger number of high-energy electrons 
can heat the stationary layers of the atmosphere to $T\approx10^4$ K 
just below the chromospheric condensation (K15),
causing an increased ionization fraction (e.g., $X_{\rm{ion}}=80$\% at $z=200$ km) and hence higher electron
density in these layers than in the double power-law simulation (e.g., $X_{\rm{ion}}=60$\% at $z=200$ km).
The resulting hydrogen b-f (Paschen) recombination emissivity from
the stationary lower depths is therefore larger.  Furthermore, this increased emission can escape because the optical depth in the chromospheric condensation 
at $\lambda=4300$ \AA\ (the closest wavelength calculated in these models to the NBF4170 filter) has only attained a value of $\tau_{4300} \approx0.6$ (at $\mu=0.95$) in the chromospheric
condensation.  In fact, the
optical depth in the chromospheric condensation of the single-power
law is even less than in the double power-law (where $\tau_{4300} \approx 0.8-0.9$; see Figure 5 of K15)
 because the density, and thus the hydrogen bound-free (b-f) opacity, in the chromospheric condensation is not as high as in the double power-law simulation at $t=2.2$~s. \emph{The $\delta=3$ single power-law electron beam effectively penetrates further into the 
 atmosphere, resulting in a smaller $n_{\rm{H}}$ (and $n_e$) in the chromospheric condensation but larger $n_e$ just below in the stationary flare layers.}
  We use the contribution function analysis described in K15 \citep[see also][]{Kowalski2015IAUS}
to calculate that $\approx45$\% of the emergent intensity for
wavelengths near the NBF4170 filter ($\lambda=4300$ \AA)
escape from the dense, heated, partially ionized layers below the
chromospheric condensation for the $\delta=3$ F13 model.  In the double power-law F13 simulation,
 only $20-25$\% of the emergent intensity at this wavelength 
can escape from heights below the chromospheric condensation
(K15) due to the larger optical depth in the
chromospheric condensation.  

Therefore, the larger emergent intensity of blue-optical light in the
single power-law simulation produces a
smaller value of FcolorB and a larger value of FcolorR.
The emission in NBF3500 and RC\#1 experiences a large optical depth ($\tau_{3550,\mu=0.95} \approx 3.5$ and $\tau_{5790,\mu=0.95} \approx 1.4$; $\lambda=5790$ \AA\ is the 
closest model wavelength to the RC\#1 filter)
 in the chromospheric condensation, so light at these wavelengths
 can only escape from the uppermost layers in both simulations.  Thus,
 the increased hydrogen (predominantly Balmer and Paschen recombination) emissivity at lower heights caused by the $E>200$ keV
 electrons does not escape.  
 The emergent intensity at these wavelengths does not change as much
 as the emission in NBF4170 in the single power-law simulation\footnote{The emergent intensity ($\mu=0.95$) at $t=2.2$~s 
 in the $\delta=3$, F13 simulation is 5\%, 23\%, and 7\% larger at $\lambda=3550$, 4300, and 5790 \AA, respectively,
 than in the double power-law F13 simulation.}.

 We have demonstrated that an F13 heating flux model
 can reproduce the impulsive phase continuum emission for our ULTRACAM flare F2, which provides an
 interpretation of the ULTRACAM filter emission using the emission mechanisms
 considered in K15. 
 The continuum emissivity was found to predominantly result from hydrogen recombination (bound-free)
  radiation in the dense regions of atmosphere where the optical depth
  causes wavelength-dependent attenuation\footnote{There is also
    opacity and emissivity contributions from hydrogen free-free processes; see K15.}.  
  Thus, the phenomenological hot ($T\approx10^4$ K) blackbody emission and the excess
  Balmer continuum emission components (Section \ref{sec:interp}) both result from the combination of Paschen and Balmer hydrogen
  recombination radiation
   originating from different physical depth ranges ($\Delta z$) due
   to wavelength-dependent optical depth in the flare layers that are heated to $T=10\,000-13\,500$ K.  Assuming the ULTRACAM flare emission results from the formation
  of a dense, hot chromospheric condensation, the NBF3500 filter samples primarily Balmer recombination radiation originating from a small physical depth range within the chromospheric
  condensation, and the NBF4170 filter (attributed to blackbody radiation in Section \ref{sec:interp}) samples primarily Paschen recombination radiation
  originating over a larger physical depth range extending from the dense chromospheric condensation through
  the deeper layers of the flare atmosphere that are heated by the $E\gtrsim200$ keV nonthermal electrons.  The RC\#1 filter
  also contains Paschen recombination radiation, but the red wavelengths corresponding to this filter are more optically thick than the blue wavelengths of the NBF4170 filter, and therefore the red wavelengths originate from a 
  smaller physical depth range (confined to the chromospheric condensation,
  as in NBF3500).  
  A small value of the FcolorB index ($\approx2$) in the ULTRACAM data
  therefore is produced by a large Balmer bound-free optical depth ($\tau>>1$) in a
  chromospheric condensation and also significant heating at lower
  depths where blue light (NBF4170) can still escape due to the smaller optical depth.
  A large value of FcolorR ($\approx2$) is produced by a moderate
  Paschen bound-free optical depth ($\tau\approx0.5-1$) at blue
  wavelengths, and a larger physical depth range of emitting material for
  blue light compared to red light.  The combination of these effects generates a color
  temperature of $T_{\rm{BB}}\approx10^4$ K.  
  As noted by K15, the color temperature of the emergent Paschen continuum flux is due to the wavelength-dependent variation of the 
physical depth range ($\Delta z$) of emitting material, and therefore
the value of $T_{\rm{BB}}$ gives only an \emph{apparent} temperature
(however, material must be heated to $T\approx10,000$ K to produce
significant hydrogen b-f opacities).  Using this interpretation, we infer that
a large value of FcolorB ($>2.5$) and a small value of FcolorR
($<1.5$) signifies lower hydrogen Balmer and Paschen bound-free
optical depths and less varying physical depth
ranges of emitting material as a function of wavelength.  Our grid of
M dwarf flare models (Kowalski et al 2016a) will elucidate which heating parameters 
produce these conditions and will allow us to interpret other flares in the sample (Section \ref{sec:sample}) in addition to the 
 gradual phase emission in the F1 and F2 events (Figure \ref{fig:summary2}).

\section{Flare Sample Properties} \label{sec:sample}
We have observed many flares with ULTRACAM at high time-resolution (Table \ref{table:obslog}).    As we have demonstrated in Section \ref{sec:interp}, 
FcolorB and FcolorR can be used to characterize the Balmer jump ratio and the
color temperature of the blue-to-red optical continuum emission.  These data provide an opportunity 
to double the number of flares in the previous spectroscopic sample of K13 while revealing new global trends of the white-light
emission at high time-resolution.  
From the observing log in Table \ref{table:obslog}, we homogeneously analyze the peak properties of the twenty largest amplitude flares, 
which give the highest signal-to-noise constraints on the flare color indices.  The NBF3500 light curves of the twenty flares are presented in Figures \ref{fig:lcsample1_3500}, \ref{fig:lcsample2_3500}, \ref{fig:lcsample3_3500}, and the NBF4170 light curves of the twenty flares are presented in Figures
\ref{fig:lcsample}, \ref{fig:lcsample2}, \ref{fig:lcsample3}.  
We use the impulsiveness index $\mathcal{I}$ from K13, where $\mathcal{I}=I_{f,\rm{peak}}/t_{1/2}$, to classify and order the flares into three categories according to the time-evolution: 
``impulsive flares''
(IF events; $\mathcal{I} \gtrsim 2$), ``gradual flares'' (GF events;
$\mathcal{I} \lesssim 0.6$), and ``hybrid flares'' (HF events;
$\mathcal{I}\approx 0.6-1.8$).  Where K13 used $U$-band photometry, and we use the NBF3500 photometry for the ULTRACAM sample,
giving 13 IF events, 5 HF events, and 2 GF events.  The flare events F1 and F2 (Section \ref{sec:desc}) are IF11 and IF4, respectively.  
Here, we do not attempt to analyze a similar number of flares in each flare category, but instead have selected only the flares that can give the highest signal-to-noise values for the FcolorR and FcolorB indices.
These tend to be the IF events because they exhibit the largest amplitudes. 

Of these twenty flares,
we only provide information on FcolorR for the flares that have a significant enhancement in RC\#1 ($I_f > 0.025$),
which typically corresponds to a $5\sigma$ enhancement.
This excludes a flare
on AD Leo (HF4), a flare on Gl 644AB (HF1)\footnote{We also obtained 
low-resolution spectra and broadband photometry from APO during the flares on Gl 644 AB on 2010-May-25.  The
 analysis is outside the scope of this paper since the flares were too weak in the red to 
provide a full wavelength coverage comparison as for the flares F1/IF11 and F2/IF4 on YZ CMi discussed in Section \ref{sec:spec}.  However, the data are available upon request.}, and a flare on YZ CMi (IF13); the flares on EQ Peg A from 2008-Aug-10 (IF6, HF3, HF5) were observed in an H$\alpha$ filter and are also not
included in the FcolorR analysis.
The GF2 event on YZ CMi is included in the FcolorR analysis because the RC\#1 data could be coadded to increase the signal-to-noise.

 Properties of these twenty flares in order of the value of $\mathcal{I}$ are given Table \ref{table:flare_properties}:
(1) Flare ID, (2) Star name (3) peak time, (4-6) peak amplitude in the
three ULTRACAM filters, (7) FcolorB at peak, (8) FcolorR at peak, (9) $t_{1/2}$ in NBF3500, 
 (10) the impulsiveness index in NBF3500 ($\mathcal{I}$), (11) the rise phase duration, (12) the fast decay phase duration
  and (13) an
 adjusted impulsiveness index (discussed in Section \ref{sec:bursty}).  
  
The peak amplitude and the impulsive phase time-evolution ($t_{1/2}$) do not strongly correlate with the spectral energy 
distribution given by either FcolorB or FcolorR; a simple measure from the light curve would function as a useful 
proxy for the nonthermal energy deposition flux or another physical quantity (e.g., the hardness of the electron distribution; Section \ref{sec:rhd}).
 It was found that in the spectroscopic sample of K13,
flares with $\mathcal{I} > 2$ (IF events) all exhibit  values of $\chi_{\rm{flare, peak}}$ (similar to FcolorB at peak) that are $\lesssim 2.2$
and flares with $\mathcal{I} < 2$ all exhibit values of $\chi_{\rm{flare, peak}}$ at peak $\gtrsim$2.3 \citep[cf. Figure 10 of][]{Kowalski2013}, suggesting 
that the Balmer jump ratio was correlated with the time-evolution of the impulsive phase.  
Of the five most impulsive flares ($\mathcal{I} > 10$) in the ULTRACAM sample, only three have FcolorB $<$2.2 at peak;  the others have 
 higher values of 2.5 and 3.2.  Flares classified as IF events
 are not expected to have such large Balmer jump ratios according to the
 relationship in K13 (Section 4.3.1 and Figure 10).
  Note especially that the IF10 event on Prox Cen has a very large value of FcolorB, $\approx7$.  In the sample from K13, most of the IF events exhibited $\chi_{\rm{flare, peak}}$ between 1.6 and 1.8, and two IF events had $\chi_{\rm{flare, peak}} \approx 2.2$, which were the largest values at peak among IF events.  
The increased sample size when including the ULTRACAM data shows that IF events can exhibit Balmer jump ratios (FcolorB) at the flare peak that are significantly larger than 2.
Only four (out of thirteen) IF events have such low values of the Balmer jump ratio (FcolorB $\lesssim$2.2) at peak.  The smallest values (FcolorB$<2$) do indeed occur at the peaks of the flares with high values of $\mathcal{I}$, but a large value of $\mathcal{I}$ does not 
necessarily correspond to a very small Balmer jump ratio. 

 The high-time resolution of the ULTRACAM data results in larger values of the impulsiveness index than would 
 be calculated with the $U$-band data from K13, which has longer integration times and significant readout times ($\approx10$~s) between exposures.  
 From the $U$-band data obtained during F2 (IF4; Figure \ref{fig:summary2}), we calculate a value of $\mathcal{I}\approx1.5$ which is significantly lower
 than $\mathcal{I}\approx16$ obtained from the NBF3500 data.  
 The difference in the calculated values of $\mathcal{I}$ is due to the low time-resolution of the $U$-band data ($\approx$20~s) compared to the duration of the impulsive phase.  
  $U$-band data were not obtained during F1 (IF11), but averaging the NBF3500 filter data over 10~s gives a value of  $\mathcal{I}\approx3$, which is an upper limit to the 
   value of $\mathcal{I}$ that would be obtained with low cadence photometry since we have not taken into account the dead-time between exposures.
  Thus, the values of $\mathcal{I}$ are affected by the time-resolution of the data, and the ULTRACAM IF event
 sample includes a larger sample of flares than would be categorized as IF events with lower resolution photometry. 
 \emph{Therefore, some impulsive flare events have large Balmer jump ratios.}
  In Section \ref{sec:bursty}, we explore the role of a second parameter that accounts for
   some of the variation between FcolorB and $\mathcal{I}$ among the IF and HF events at high-time resolution. 
   
Although the ULTRACAM IF event sample includes some flares that are not consistent with the Balmer jump ratio properties of the IF events in the K13 spectroscopic sample, the
HF and GF events in both samples share the property of having larger Balmer jump ratios of FcolorB$\approx2.2-3.7$ (the lowest is
the FcolorB value of $2.2$ for the HF3 event in EQ Peg A).
The HF events in the ULTRACAM sample have a particularly striking
similarity to the HF events in K13 (in particular to HF1 in K13). The
HF events in both samples are typically medium-amplitude ($I_f+1
\approx 2-4$) flares with multiple, temporally-resolved peaks in the
impulsive phase, although the peaks in the ULTRACAM sample have much better temporal coverage than the lower-cadence $U$-band photometry in K13.
The (ULTRACAM) GF1 event is more similar to the (ULTRACAM) HF events in having several well-defined bursts in 
the impulsive phase (Figures \ref{fig:lcsample2} - \ref{fig:lcsample3}).
The evolution of (ULTRACAM) GF2 in YZ CMi is similar to the evolution of GF5 in K13; both events are clearly gradually evolving (e.g., Figure \ref{fig:summary1})
and have relatively large values of FcolorB.  
If the gradual event following the impulsive phase of IF4/F2 ($t\gtrsim30$~s in Figure \ref{fig:lcsample}) is considered a separate flare event,
it would be classified as a GF event with large Balmer jump ratio of
FcolorB$\approx 3$ at the peak (Table \ref{table:calib}).  

\subsection{The Relationship Between Rise Phase Bursts and the Balmer Jump Ratio} \label{sec:bursty}
What aspect of the flare energy release determines the spectral energy distribution properties at peak time during a flare?  Is there a measured 
quantity from a flare light curve that can be used as a proxy for both the flare energy release and the spectral energy distribution?
K13 related the value of $\chi_{\rm{flare, peak}}$ (similar to FcolorB) with the impulsive phase time-evolution:  smaller Balmer jump
ratios are observed during more impulsive flare events, and larger Balmer jump ratios are observed during more
gradual flare events.  In Section \ref{sec:sample}, we found that the ULTRACAM sample does not reveal such a clear
relationship within the IF events; the HF and GF events do follow the trend. 
A more complicated relationship involving a second parameter is needed to explain the full range of behavior.

We employ the high time-resolution of the ULTRACAM data and investigate a modified impulsiveness index,
$\mathcal{I}^{\prime}$.  This modified index divides $\mathcal{I}$ by the factor $t_{\rm{rise}}/t_{\rm{fast-decay}}$, which measures the degree of asymmetry of the 
impulsive phase.  In many flares, the rise phase can consist of several ``bursts'' leading up to the peak time;  the rise time from these 
bursts in included in the calculation of the total rise time if the burst reaches 20\% of the peak flux.  The fast
decay phase is measured from the peak time to the end of the impulsive phase.  The start of the rise and the end of the fast decay phase are determined from the NBF4170 photometry (the highest cadence) when the signal-to-noise is sufficient; otherwise these times are determined from NBF3500 photometry.  The start of the rise and the end of the fast decay phase 
are denoted by vertical dotted lines for each flare in Figures \ref{fig:lcsample1_3500}-\ref{fig:lcsample3_3500} and \ref{fig:lcsample}-\ref{fig:lcsample3}.
 For the IF events, it is generally easy to distinguish 
the end of the fast decay when the gradual decay phase begins.  However for the HF and GF events, the change from fast to slow emission can be 
ambiguous.  In some flares (IF3, HF3, HF5, GF1) with low-level bursts after the peak time, the end of 
these occurs with the onset of gradual emission and is taken as the end of the fast decay phase.  For IF3, there are many secondary events in the decay phase of this very large flare, but there is a 
notable break to much slower emission that begins either at 530~s or 1050~s after the peak.  We use $t-t_{\rm{peak}}=1050$~s to denote the start of the gradual phase to compute $\mathcal{I}^{\prime}$ for this flare (the flare color indices also notably reach a constant value at this time; see Section \ref{sec:ultraflare}). 
The values of $\mathcal{I}^{\prime}$ are given as the last column in Table \ref{table:flare_properties}.
We find that flares with $t_{\rm{rise}}/t_{\rm{fast-decay}} < 1$ are those with typically lower Balmer jump ratios,
whereas flares with  $t_{\rm{rise}}/t_{\rm{fast-decay}} \ge 1$ have a relatively slower rise phase that is usually related to the presence of light curve variations in the rise phase which are indicative of several, temporally superimposed events (Figures \ref{fig:lcsample}-\ref{fig:lcsample3}) giving a relatively high value of FcolorB at the peak.

Incorporating this extra parameter can qualitatively account for the flare-to-flare variations of the peak values of FcolorB within the IF event category.  For example, IF5 and IF12 are
 fast flares in YZ CMi observed on 2012 Jan 12 and have large values of FcolorB (3.2 and 4, respectively).  The values of $\mathcal{I}$ for these flares are not largely affected by the 
 time-resolution of the data (Section \ref{sec:sample}).
 However, adjusting the  impulsiveness indices  by $t_{\rm{rise}}/t_{\rm{fast-decay}}$ decreases these values from $\mathcal{I}=13$ to $\mathcal{I}^{\prime}=9$ (for IF5) and from $\mathcal{I}=3$ to $\mathcal{I}^{\prime}=1$ (for IF12).
 In both cases, the high time-resolution of the NBF4170 light curve reveals asymmetries in the impulsive phases that are due to a series 
of bursts in the rise phase and/or substructure at the peak that is indicative of a series of bursts (Figure \ref{fig:lcsample2}).
 We discussed the detailed morphological differences between IF4 (F2) and IF11 (F1) in Section \ref{sec:desc}.  
F1 exhibits a bursty impulsive phase and a value of $t_{\rm{rise}}/t_{\rm{fast-decay}}=1.6$ whereas F2 exhibits a smooth rise phase
and a largely asymmetric impulsive phase with $t_{\rm{rise}}/t_{\rm{fast-decay}}=0.5$.  They are also different in their 
values of FcolorB with F2 at peak having $1.8-1.9$ and F1 at peak having 2.2.  Although not obvious
from the spectra (Figure \ref{fig:spectra}), the value of FcolorB from ULTRACAM is higher signal-to-noise, and the 
difference between the two is significant by $3\sigma$.  Thus the asymmetry differences for the impulsive phases of these two flares 
are consistent with having different values of FcolorB.
Also, consider the flares in EQ Peg A discussed in \cite{Kowalski2011}:
IF6, HF3, and HF5.  These flares were designated by eye as ``impulsive'', ``traditional'', and ``gradual'', respectively.  
IF6 and HF3 have different values of the impulsiveness index,  $\mathcal{I}$, yet have a very similar value of FcolorB ($\approx2.2$) at peak.  
Although very fast, IF6 has a bursty rise phase, whereas HF3 has a more sudden rise phase but a small value of $t_{\rm{rise}}/t_{\rm{fast-decay}}\sim0.15$.  Thus, the 
value of  $\mathcal{I}^{\prime} \approx 6-8$ for both flares (note, smaller bursts occur during the HF3 event, but after the main peak in Figure \ref{fig:lcsample2}; we take the end of these series of bursts to be the end of the fast decay phase).  HF5
consists of three bursts in the rise phase and has a higher value of FcolorB at peak.
Although the modified impulsiveness index explains some of the variation in FcolorB among the IF and HF events, there is some variation in FcolorB from flare to flare (e.g., IF5 compared to IF6) at peak times that is not related to the time-evolution given by either impulsiveness indices.
 Notably, the largest flare in the sample (Section \ref{sec:ultraflare})
has  a small value of FcolorB at peak, an impulsive phase that is largely asymmetric with a significantly shorter rise time (270~s) compared to the
fast-decay time (1050~s; and thus a large value of $\mathcal{I}^{\prime} \approx 72$), but there are at least six bursts in the rise phase of this event.

In Section \ref{sec:thalfs}, we explain the general trend between relatively larger amounts of Balmer emission (larger Balmer jump ratios) at peak and rise phase evolution consisting of several bursts.

\subsection{Timescales of the Continuum Emission} \label{sec:thalfs}
Using the high cadence ULTRACAM sampling in NBF3500 and NBF4170, we can accurately compare the differences in the timescales
of the continuum emission on either side of the Balmer jump. 
 In Figure \ref{fig:thalf}, we show the ratio of $t_{1/2, \rm{NBF3500}}/t_{1/2, \rm{NBF4170}}$ vs $t_{1/2, \rm{NBF4170}}$.  
 The median ratio is 1.3, and most of the flares spanning nearly 1.5 dex of $t_{1/2, \rm{NBF4170}}$ fall 
within the interquartile range (0.36), which is indicated by the shaded region.
A $t_{1/2, \rm{NBF3500}}/t_{1/2, \rm{NBF4170}}$ ratio of 1.3 means that the NBF4170 and NBF3500 emission generally evolve similarly in the impulsive phase,
 but bright NBF4170 emission is maintained for a shorter 
 amount of time (shorter impulsive phase) compared to bright NBF3500 emission.  
 
The flare events with very large values of $t_{1/2, \rm{NBF3500}}/t_{1/2, \rm{NBF4170}} >> 1.3$ are more ``bursty'' events with longer
impulsive phases.
The flares that have $t_{1/2, \rm{NBF3500}}/t_{1/2, \rm{NBF4170}}$ ratios greater than three are outliers in Figure \ref{fig:thalf} and are the four 
least impulsive flare events:  HF4, HF5, GF1, and GF2.  Three of these   
flares exhibit multiple, time-resolved bursts each separated by $\Delta t \approx 50-100$~s (Figures \ref{fig:lcsample2}-\ref{fig:lcsample3}).  HF3 has five 
bursts in the impulsive phase from $t=0-100$ s, but each of these are separated by only $\Delta t \approx20$~s and 
are relatively small in amplitude compared to the peak emission.
In the insets of Figure \ref{fig:thalf}, we show the normalized light curves for HF5 (bottom left) and GF1 (bottom right)
to illustrate how the $t_{1/2}$ values are measured for NBF3500 and NBF4170 emission.  Both flares have a peak corresponding to a
bright, nearly symmetric burst (at $t=0$~s in the light curves of Figures \ref{fig:lcsample2}-\ref{fig:lcsample3}).  In this main burst, the NBF4170 emission
 rises and decays quicker than the NBF3500 (consistent with the
 $t_{1/2, \rm{NBF3500}}/t_{1/2, \rm{NBF4170}}$ value of the typical flares having ratios of $\approx$1.3), and thus the $t_{1/2}$ for NBF3500 spans a larger
 fraction of the flare event. 
 Another outlier in Figure \ref{fig:thalf} is
 the F1 (IF11) event (Section \ref{sec:desc}), which has a $t_{1/2}$ ratio of
 $\approx2.2$ (Table \ref{table:thalf}).  This high ratio is due to
 the more prominent decay phase emission relative to peak in NBF3500;
 thus the measurement of $t_{1/2, \rm{NBF3500}}$ includes impulsive and decay phase
 emission whereas the measurement of $t_{1/2, \rm{NBF4170}}$ only
 includes the impulsive phase evolution.

 The relationship between a rise phase with significant emission in bursts and a higher value of FcolorB at peak (Section \ref{sec:bursty}) can be
explained by the different timescales of NBF3500 and NBF4170.  Suppose that each fundamental burst contributing to the total spatially integrated emission
exhibits a spectral evolution like that of spectrum F2(i) in Figure
\ref{fig:spectra}: a larger Balmer jump ratio in the decay than in the
impulsive phase and a ratio of $t_{1/2, \rm{NBF3500}}/t_{1/2, \rm{NBF4170}}\approx1.1-1.2$ (Table \ref{table:thalf}).  It follows that a flare 
with several partially-resolved heating episodes will be composed of the sum of the individual rise and decay emission from each episode, as in the complex flare decomposition of white-light flares from Kepler data \citep{Davenport2014}.
Then the total emission during each successive heating event would include decay emission from the previous heating event.  The peak emission following
a series of bursts would thus have a larger Balmer jump ratio as a result of
the temporal superposition of several decaying bursts, each of which
has a significantly larger Balmer jump ratio due to the slower decay of emission in NBF3500.  For flares that don't show evidence of several heating bursts in the rise (e.g., F2), the value of FcolorB at peak can attain a lower value because there are fewer decaying bursts at peak.  If all heating bursts are identical over a flare, then a superposition of bursts in the rise phase of F2
would produce a smaller Balmer jump ratio very early in the rise than at peak since the NBF4170 emission in the early times would not have had time to decay significantly.  However, the smallest Balmer jump ratios are not observed in the early rise (Figure \ref{fig:summary2}) suggesting that either the F2 event consists of a superposition of bursts with heating properties that change over the course of the event or this event actually represents a fundamental heating episode.  In Section \ref{sec:concl}, we discuss future modeling directions to determine the likely heating scenario in events like F2.

The shorter duration of bright emission in NBF4170 explains the anti-correlation between
FcolorB and the flare flux in the impulsive phase \citep[see Figure \ref{fig:summary2} and Section \ref{sec:timeevol}; the anti-correlation was also discussed in][]{Kowalski2011}.  If the emission in NBF4170 and NBF3500 evolved with 
the same impulsive phase timescale, FcolorB would be constant.
However, FcolorB decreases to a minimum at the flare peak and increases again by the start of the gradual phase.  

\subsection{Peak Flare Color Indices:  A Relationship Among All dMe Flares}
The values of FcolorB and FcolorR at the peak times (from Table \ref{table:flare_properties}) are shown in Figure \ref{fig:fcolor_relation} and Figure \ref{fig:fcolor_relation2}, where Figure \ref{fig:fcolor_relation} shows the data and Figure \ref{fig:fcolor_relation2} shows the data compared to 
model predictions and interpretation from Sections \ref{sec:interp} and \ref{sec:rhd}.  
Note, only fourteen flares are plotted because six of the twenty events have been excluded from the FcolorR analysis (see Section \ref{sec:sample}).  
For the majority of the flares, the FcolorB indices range between 1.6\,--\,4 and the FcolorR indices range between 1.3\,--\,2.3.  In
general, a lower value of FcolorB corresponds
to a larger value of FcolorR.  According to the spectral interpretation in Section \ref{sec:interp}, as the Balmer
  jump becomes relatively smaller,
the blue-to-red optical continuum (from $\lambda=4170$\AA\ to $\lambda=6010$\AA) becomes bluer.  
To guide the eye, we show two unweighted linear fits (gray lines) to all the data in Figure \ref{fig:fcolor_relation2}:  $y=12.2-5.9x$ for $x=$FcolorR$<1.7$.
  and $y=5.3-1.6x$ for $x=$FcolorR$>1.3$.  The values of the color temperature obtained from FcolorR  ($T_{\mathrm{FcolorR}}$) from the ULTRACAM sample are shown on the 
top x-axis of Figure \ref{fig:fcolor_relation2}.
A bluer continuum gives a larger value of the color temperature ($T_{\rm{FcolorR}}$), and thus the continuum is apparently hotter for flares with a smaller Balmer jump ratio in the peak phase.
 The vertical dotted line in Figure \ref{fig:fcolor_relation2} at FcolorR of 1.7 corresponds to a value of
$T_{\mathrm{FcolorR}} \approx 8500$ K.  Flares in the ULTRACAM sample
with FcolorR indices greater than this value of $T_{\mathrm{FcolorR}}$ exhibit bona-fide hot
blackbody-like emission at their peaks. Only about 1/3 of
the sample are in this regime.  Interestingly, these flares are not exclusively the
largest flares but span a range of peak amplitude enhancements (from $I_{f,\rm{NBF3500}}+1 \approx 3.4$ for IF11 to $I_{f,\rm{NBF3500}}+1 \approx 106$ for IF3).

The spectral calibration of the ULTRACAM filter ratios in Section \ref{sec:interp} indicate that the values of the color temperature derived from
FcolorR can be $1400-2400$ K lower than the color temperature ($T_{\rm{BB}}$) calculated from
detailed fitting of blackbody spectra to line-free wavelength windows 
between $\lambda=4000-4800$\AA\ (Section \ref{sec:interp}, Equation \ref{eq:calibeq}), due to a
multithermal continuum and to systematically increased emission in the
bluemost continuum fitting windows (BW1 and BW2) between H$\epsilon$ and H$\delta$.  The calibrated values of the blue-optical color temperature
$T_{\rm{BB}}$ from Equation \ref{eq:calibeq} are also given as a top axis in Figure \ref{fig:fcolor_relation2}, which provides a guess for the possible value of $T_{\rm{BB}}$ for these flares.  
The vertical dashed line in Figure \ref{fig:fcolor_relation2} indicates the adjusted hot blackbody threshold ($T_{\rm{BB}} \approx 8500$ K corresponding to $T_{\rm{FcolorR}} \approx 6600$ K) after
applying this empirical correction to $T_{\mathrm{FcolorR}}$.
Even after applying this empirical correction, we note several (definitely two, five more within uncertainties) flare events in our sample 
 do not show conclusive evidence of hot blackbody emission at the
 peak.  This significantly increases the number of dMe flare events in the
 literature that do not show an energetically dominant, bona-fide hot color temperature
 at blue-optical wavelengths.  
These flare events also produce larger values of FcolorB.
Interestingly, all flare events on YZ CMi from the night of 2012 Jan 12-13 (green circles) exhibit
low values of FcolorR and high values of FcolorB, whereas most flare events
 on YZ CMi on the night of 2012 Jan 13-14 (black circles) exhibit the highest values of
FcolorR and the lowest values of FcolorB.  
Also, none of the four flare
events on Proxima Centauri have low values of FcolorB ($<2.5$), even though the second most impulsive ($\mathcal{I}$) flare event is among these four flares.

The peak flare color indices for several flares from the flare spectral
  atlas of K13 are shown as black squares  in Figure \ref{fig:fcolor_relation} for comparison.
The following flare events are included:
IF2, IF3, IF5, IF7, IF9, HF1, HF2, HF3, and GF1.  All spectra do not have sufficient signal-to-noise at $\lambda=3500$\AA, and we show here C4170/C6010 (x-axis) which is similar to FcolorR and 
  $\chi_{\mathrm{flare}}$ (y-axis) which is similar to FcolorB (Table \ref{table:calib}).
$\chi_{\mathrm{flare}}$ and C4170/C6010 clearly follow the same trend
as the ULTRACAM FcolorB and FcolorR indices.  The values of $T_{\rm{BB}}$
for these flare events (Table 7 of K13) are systematically higher by 
$800-3000$ K compared to the values of  $T_{\mathrm{FcolorR}}$ (top
axis) that correspond to the C4170/C6010 ratios, which is consistent with the relation between $T_{\rm{BB}}$ and $T_{\mathrm{FcolorR}}$ calculated in Section \ref{sec:interp} for IF4 and IF11.

 In addition, we show the values (blue filled circles) from the rise and peak of the MDSF2
(``megaflare decay secondary flare \#2'')
from Figure 56 of K13 (see also Section \ref{sec:ultraflare} here).  This secondary flare produced Vega-like spectra
with the Balmer continuum in absorption, a phenomenon that was only directly detected with spectra during MDSF2.  The purple filled circles 
are estimates of the flare color indices for the newly formed emission during three secondary flares in the decay phase of IF3 (from the ULTRACAM sample), which are discussed
in Section \ref{sec:ultraflare}.
The relation in Figures \ref{fig:fcolor_relation} - \ref{fig:fcolor_relation2}
extends to flares that exhibit spectra similar to a hot star's photospheric spectrum and
to flares that have large Balmer jump ratios and apparently cooler blue-to-red
optical continua.  Thus, we suggest that the trend in Figures \ref{fig:fcolor_relation} - \ref{fig:fcolor_relation2} represents a ubiquitous
relation for impulsive phase white-light flare emission in dMe stars; a flare ``color-color'' diagram that can be used
to readily determine the physical properties of any flare through photometry alone.
Straightforward comparisons to radiative-hydrodynamic model predictions can also be achieved with this diagram;
we show the model predictions (Table \ref{table:calib}) at $t=2.2$~s from Section \ref{sec:rhd} and from K15 in Figure \ref{fig:fcolor_relation2}.
Because the observed flare color indices are proportional to the excess flare spectrum (Equation \ref{eq:ifeq}, Section \ref{sec:analysis}), we subtract the pre-flare
spectrum before calculating the model flare color indices, which significantly changes only the value of FcolorB from $\sim9$ to 11 for the F11 model.  Subtracting the pre-flare spectrum $S_{o,\lambda}$ has a very small effect on the higher flux model spectra since $S_{\rm{Flare},\lambda} >> S_{o,\lambda}$.
These model predictions lie along the general trend indicated by the two linear relationships, but only the two F13 models
give reasonable values of the flare color indices for the flares that exhibit hot, blackbody-like continua in the optical.

\section{The ``Ultraflare''} \label{sec:ultraflare}
In Figure \ref{fig:ultraflare}, we show the ULTRACAM light curves and time-evolution of FcolorB and FcolorR for the largest flare event
(hereafter, ``Ultraflare'') in our sample\footnote{Observations of this flare event were obtained with exposure times of 2.2~s in NBF3500 and 1.1~s in NBF4170 and RC\#1 until the main peak time, at which point the exposure time in the NBF3500 was reduced to 1.1~s until the late decay phase.}.
This flare occurred in the dM4.5e star YZ CMi with a large peak at 22:33:54 UT on 2012 Jan 13 (IF1, $I_f + 1=30$  in NBF3500) and a second (main) peak 
at 22:44:33 UT (IF3, $I_f + 1 = 105$ in NBF3500).  The NBF4170 light curve is shown with a linear flux scale in Figure \ref{fig:lcsample}, 
which indicates a change\footnote{The secondary flares in the gradual phase (discussed in Section \ref{sec:secondary}) make it difficult to precisely identify the break from fast to slow decay phase emission in the IF3 event.  A break may occur near $t-t_{\rm{peak}}\approx550$~s, but we choose the break at $t-t_{\rm{peak}}\approx1050$~s because this also occurs at time when the flare color indices reach a relatively constant value (discussed in Section \ref{sec:secondary}).} from fast decay emission to a gradual decay emission at $t-t_{\rm{peak}}\approx1050$~s.
   The total NBF3500 energy 
for this flare event is 6.2$\times10^{32}$ erg; from this, we estimate a $U$-band energy of $\approx4\times10^{33}$ erg, which is a factor of 
five times less energetic than the ``Megaflare'' that occurred on this star on 2009 Jan 16 \citep{Kowalski2010}, which we use for comparison throughout this section.
The $t_{1/2}$ value is $\approx$5.5 minutes, compared to 7.7 minutes
for the Megaflare, and the peak of the Ultraflare is half the peak
amplitude of the Megaflare. The IF11, IF8, GF2, and IF4 events occurred $t-t_{\rm{peak}}>4$ hr after the
main (IF3) peak, when the gradual decay emission from the Ultraflare was still
present.  

The flare color indices in Figure \ref{fig:ultraflare} are among
the highest time-resolution, highest signal-to-noise constraints on the white-light emission
 yet obtained during a dMe flare, and we envision that these data will be invaluable for guiding future modeling efforts.
The Ultraflare event exhibits a complex light curve after the peak of IF1, through the rise of IF3 and the decay of IF3. 
We refer to the sub-peaks in the rising phase of emission as ``bursts'' and the sub-peaks in the decaying phase of emission
as ``secondary flares''.
The anti-correlation between FcolorB and the flare flux is present through almost the entire event (top panel), whereas the variations 
of FcolorR are correlated with the changes in the flare flux (bottom panel).  Note especially the long timescale rise of FcolorR and the response
to the $\approx6-7$ bursts between $t-t_{\rm{peak}}=-350$~s and $t-t_{\rm{peak}}=0$~s.  
The values of FcolorB experience a faster change after the secondary flares (e.g., at $t-t_{\rm{peak}}=-500$~s to $-400$~s and at $t-t_{\rm{peak}}\approx500$~s) than in the 
 gradual decay phase of IF3 ($t-t_{\rm{peak}}>1050$~s).  Interestingly, the value of FcolorR attains an approximately constant value of $\approx1$ at the 
beginning of the gradual decay phase of IF3 ($t-t_{\rm{peak}}\approx1050$~s), whereas the value of FcolorB 
begins a slow increase.  The values of FcolorB (FcolorR) during the bursts and secondary flares do not attain values 
as low (high) as the values during the two large peaks, 
which is probably due to newly formed emission (with small areal coverage) superimposed on the previously decaying
emission from the larger peaks (with large areal coverage) and possibly also to intrinsic variations in the flare heating during the sub-peaks.  
 
In the rise phase of the IF1 event, the FcolorB and $I_f$ variations are anti-correlated, but after the peak of the IF1 event the value of FcolorB declines from 2.0 to 1.8 at $t-t_{\rm{peak}}\approx10-14$~s, which is an instance of when the anti-correlation breaks down. 
 The maximum of $I_f$ and the minimum 
of FcolorB are indicated by vertical dotted and dashed lines, respectively, in Figure \ref{fig:ultraflare};  the minimum of FcolorB also corresponds to a break in the fast decay phase of $I_f$.
 Note, IF1 has the lowest value of $t_{1/2, \rm{NBF3500}}/t_{1/2,
   \rm{NBF4170}}\approx1.05$ in the sample (Figure \ref{fig:thalf}),
 which is consistent with not exhibiting strict anti-correlated
 variations between the light curve and flare color as explained in Section \ref{sec:thalfs}.
 A time lag between the maximum of $I_f$ and a minimum of $\chi_{\rm{flare}}$ (similar to FcolorB) was also found
during the secondary flare MDSF2 during the YZ CMi Megaflare (K13; see also Section \ref{sec:secondary}). The break down of the anti-correlation provides
a critical, high-time resolution constraint on models of flare evolution.

\subsection{Properties of the Secondary Flares} \label{sec:secondary}

During the decay phase of the Ultraflare, there are several secondary flares 
that are similar to the secondary events observed in the Megaflare.
In Figure \ref{fig:ultraflare_megaflare}, we compare the NBF3500 and $U$-band lightcurves during times in the gradual decay phase of the Ultraflare and Megaflare,
when strikingly similar sequences of secondary flares occurred.  Although the general pattern of the NUV emission is similar to the eye,
the timescales of the secondary flare events in the Ultraflare (the rise time of the secondary flare at $t-t_{\rm{peak}}=320$~s is 10~s) are much faster than those in the Megaflare (the rise time of the secondary flare ``MDSF2'' at $t-t_{\rm{peak}}=6000$~s is nearly 5 min);  in the Ultraflare, this sequence also occurs much sooner after the peak (3 minutes compared to an hour after the Megaflare peak).

The newly formed flare emission during the secondary flares provide important constraints on the particle acceleration (or other heating mechanisms) that produces these events.  
For example, \cite{Ayres2015} observed a series of far-ultraviolet continuum flares which did not produce enhanced emission in the transition region lines during the decay of a large flare on the rapidly rotating G dwarf EK Draconis, which is suggestive that the heating was localized to deep layers. In the decay phase of a large solar flare from \cite{Warmuth2009}, a hard X-ray (nonthermal) event was observed 
without a simultaneous increase in the soft X-ray (thermal) emission;
the hard X-ray spectrum was consistent with a very high-energy ($E\gtrsim100$ keV) nonthermal electron component during this secondary event.  
During the secondary flare MDSF2 in the decay phase of the Megaflare, the optical flare spectrum was found to closely resemble the spectrum of Vega (K13), and the 
Balmer emission line flux from the flare decreased during
this secondary event \citep{Kowalski2010}.  Apparently, the decrease in Balmer emission
flux from the \emph{entire spatially unresolved flare} resulted from the formation of Balmer line absorption in the secondary flare.  The Vega-like spectrum
also implies significant heating at high densities resulting in
photospheric-like hydrogen opacities, as implied from a preliminary ``hot spot'' phenomenological modeling investigation \citep{Kowalski2011b}.

 The values of FcolorB and FcolorR are shown in Figure
 \ref{fig:ultraflare_megaflare} for a series of secondary flares within the Ultraflare decay (the time intervals are indicated by gray bars in Figure \ref{fig:ultraflare}).  At 150~s
 after the main peak ($t=0$~s), the Ultraflare FcolorR value is 1.7, consistent
 with a $T_{\rm{FcolorR}}$ of 8400 K; at the main peak, the value is
 2.2, consistent with a value of $T_{\rm{FcolorR}}$ of $11\,000$ K, implying that the blue-to-red optical continuum has become
apparently cooler, and may be explained by decreasing the Paschen b-f
optical depth in chromospheric condensations (Section \ref{sec:rhd}).  From $t=150-380$~s, the value of FcolorR decreases to 1.5, suggesting that the Paschen b-f opacity decreases further.
 The value of FcolorB is 1.8 at $t\approx150$~s, compared to 1.6 at peak.  From $t=150-380$~s, the value of FcolorB increases to 2.0.  Thus, the relative amount of excess Balmer continuum emission increases, suggesting that the Balmer b-f optical depth
decreases over the decay phase, allowing Balmer continuum emission to escape from a larger physical depth range in the atmosphere (Section \ref{sec:rhd}).

For the Megaflare,
 we calculated synthetic values of FcolorB and FcolorR by convolving the ULTRACAM filter transmittance and CCD response with the spectra from K13 \citep[see also][]{Kowalski2011}.  The time interval presented here for the Megaflare is at $3000-7000$~s after the peak\footnote{This time interval corresponds to the time over which spectra were obtained.}, much later than for the Ultraflare, and thus the 
the b-f opacities have had more time to decrease, leading to smaller values of FcolorR and larger values of FcolorB -- similar to the values in the Ultraflare at $t\approx700$~s after the main peak (Figure \ref{fig:ultraflare}).  During the secondary flares of the Megaflare,
the evolution of FcolorB is anti-correlated with the $U$-band whereas the evolution of FcolorR is correlated with the $U$-band.
During the Ultraflare secondary flares over this time interval, there is little response of the FcolorB and FcolorR values, except during the 
secondary flare at $t=320$~s when the value of FcolorR increases from
1.5 to 1.6, and there is also a slight decrease in FcolorB.  In
contrast, the value of FcolorR increases significantly from 1.0 to 1.5 for MDSF2 ($t=6000$~s).  At
$t=5800$~s when FcolorR is 1.0, the $\lambda=4900-9200$\AA\ spectrum
(spanning the ULTRACAM filters) is dominated by a significant red
continuum component in the Megaflare, which was referred to as the Conundruum in K13,
whereas the $\lambda=4000-4800$\AA\ spectrum exhibited a color
temperature of $\approx8000-8500$ K  (see Figure 31 of K13).  Thus, the value of 
FcolorR is determined by the relative contributions of these components.
During MDSF2 at $t=6000$~s, the newly formed flare emission is Vega-like, which
causes the change in FcolorR and FcolorB for the total emission that
we observe.  The secondary flare at $t=320$~s during the Ultraflare is much less energetic (shorter duration)
than MDSF2 in the Megaflare but it produces the
same change in peak amplitude as MDSF2, $\Delta I_f \approx 5-6$.
However, the relative change in NBF3500 in the Ultraflare is only 10\% ($I_f = 51$ to 57 at peak)
whereas in the Megaflare the $U$-band changes by 30\% ($I_f\approx17$
to 22 at peak).  Therefore, the change in the flare color indices (for
the total flare emission) over the times of the secondary flares is
less in the Ultraflare than in the Megaflare.

Does the Ultraflare produce secondary flare spectra that are Vega-like?
 If the newly formed flare emission is the result of a
Vega-like flare spectrum forming in the Ultraflare, then the emission
is not as easily seen against the much larger amount of previously
decaying emission present at $t\approx300$~s.  We can estimate the
values of FcolorB and FcolorR for the newly formed flare emission at
$t=320$~s by subtracting the emission at $t=300$~s:  this gives values
of 1.9 ($\pm0.1$) for FcolorB and $\approx$2.8 ($\pm0.3$) for FcolorR.  For comparison, 
synthetic values for the newly formed flare emission for MDSF2 of
$0.7-0.8$ (FcolorB) and $2.5-2.7$ (FcolorR) (see Figure
\ref{fig:fcolor_relation}).
The estimated values of FcolorB and FcolorR at the peaks of the newly formed emission in the secondary flare at $t\approx1500$~s are
1.5 and 2.5, respectively; at the peak of the secondary flare at $t\approx 2400$~s the values are 2.4 and 2.0, respectively.   The peak values for these
secondary flares are also shown as purple circles in the flare color-color diagrams of Figures \ref{fig:fcolor_relation}-\ref{fig:fcolor_relation2}.
  The newly formed flare emission during
the secondary flares is apparently very hot with color temperatures of 
$T_{\rm{FcolorR}}\approx15\,000-16\,000$ K (see also K13 Table 9), but the Balmer jump ratios are apparently larger in the 
Ultraflare secondary flares.
A blackbody function fit to the value of FcolorR (2.8) at the peak of the secondary flare
at $t\approx300$~s extrapolated to $\lambda=3500$ \AA\ would give
a value of FcolorB of 1.5; values of FcolorB $<1.5$ are thus required for a Balmer continuum in absorption.
  The peaks of the secondary flares do not conclusively show that a Vega-like spectrum with Balmer continuum
  in absorption is formed, but precisely isolating the newly formed emission in the Ultraflare could be affected by several factors.
  For the secondary flare at $t\sim300$~s, the peak increase is relatively small compared to the brighter
 (more obscuring) Balmer continuum emission from the previously decaying (main peak) emission.  The values of $\Delta I_f$ (NBF3500) at the peaks of 
 the secondary flares at $t\approx1500$~s and $t\approx2400$~s are 1.4 and 1.1, respectively, which implies 
  a much smaller areal coverage of the newly formed flare region in these events.

\section{Concluding Remarks}\label{sec:concl}

We present a high-time resolution dataset of dMe light curves obtained with the triple-beam ULTRACAM and
 custom narrow-band continuum filters.  In Sections \ref{sec:concl1}-\ref{sec:concl4}, we summarize and discuss four 
 aspects of dMe flares that are addressed by the ULTRACAM flare sample.  In Section \ref{sec:concl5}, we discuss future modeling directions with the ULTRACAM data.  
 
 \subsection{Flare Color-Color Diagrams}\label{sec:concl1}
The NBF3500 filter in the NUV avoids the ambiguity of the broad $U$-band, which integrates over the complex spectral
region where higher order hydrogen lines merge together and where there is a combination of hydrogen Balmer and Paschen\footnote{Which we also refer to as the hot blackbody-like emission component; Section \ref{sec:interp}.} continuum emission components.  
We found agreement in the ULTRACAM flux ratios and simultaneous spectra for two moderate amplitude flares 
in YZ CMi, demonstrating that the ULTRACAM filter ratios (FcolorB and FcolorR) give  
values of the Balmer jump ratio and the blue-to-red ($\lambda=4170-6010$ \AA) optical
continuum color temperature that can be compared directly to radiative-hydrodynamic flare model predictions.  From the spectra, we calibrated
the values of the color temperature ($T_{\rm{FcolorR}}$) calculated from FcolorR to values of the color temperature in the blue-optical 
wavelength regime, $T_{\rm{BB}}$, which were found to be $1400-2400$ K larger (Section \ref{sec:interp}).  The comparison to spectra
also demonstrated that FcolorB follows the evolution of emission line flux in H$\gamma$ relative to the continuum flux in NBF4170.

Drawing from new RHD modeling results, the ULTRACAM data were used to constrain properties of the heating mechanism (e.g., the nonthermal electron flux and hardness of the nonthermal electron distribution) and the origin of the M dwarf flare continuum emission as either optically thin or thick hydrogen recombination.
High time-resolution (0.1~s) broadband colorimetry ($U-B$ vs $B-V$) variations of moderate-amplitude dMe flares as a function of time evolve over ``spaces'' corresponding to optically thin Balmer continuum emission, optically thick
Balmer continuum emission, or pure blackbody emission \citep{Zhilyaev2007, Lovkaya2013}.  These colorimetry analyses are often consistent with pure blackbody emission with $T=14\,000-22\,000$ K at flare peak. 
Even some flare spectra at peak times can be approximated by a single, isothermal blackbody model across the NUV and optical \citep{HawleyPettersen1991}, 
 but calibrating the ULTRACAM flux ratios to simultaneous spectra (e.g., for IF4 and IF11) allow us to characterize important deviations from an isothermal blackbody in the NUV and/or in the red that appear in most other flares.  

Following the $U-B$ vs $B-V$ diagrams of \cite{Kunkel1970}, \cite{Zhilyaev2007}, \cite{Lovkaya2013}, we have introduced a ``flare color-color diagram'' using the ULTRACAM filter ratios FcolorB vs FcolorR (Figure \ref{fig:fcolor_relation}).  
We found that the flare color-color distribution at the peak of flares follows a general inverse relationship:  larger values of FcolorR correspond to smaller values of
 FcolorB, but that the statistically significant variation from flare to flare within this relationship indicates intrinsic variations in the heating mechanism. 
 Future models of the heating mechanism or particle acceleration
 evolution should aim to 
self-consistently explain the interflare trend between FcolorB and
FcolorR.  

We have synthesized the flare color index values from the spectra
from K13, and thus we know how the broad wavelength distribution changes in the different regions of this diagram.
For example, the coordinates in the flare color-color diagram of Figure \ref{fig:fcolor_relation} for the \emph{total} flare emission at the peak of MDSF2 in the YZ CMi Megaflare are from Figure \ref{fig:ultraflare_megaflare}, $(x,y)=(1.6,1.5)$.  The total 
optical flare emission is best fit with two (phenomenological) blackbody continuum components:  a hot blackbody emission component ($T_{\rm{BB}} \approx 10,800$ K) accounts for the flare flux at NUV and blue wavelengths \citep{Kowalski2012} and a
cooler blackbody continuum is necessary at red wavelengths (the ``Conundruum''; K13).  Thus a low Balmer jump ratio and a low FcolorR value indicate such a scenario where the newly formed emission in a secondary flare is located in the region around $(x,y)=(2.5,0.6)$ and exhibits Vega-like emission, but that 
 a significant red continuum dominates the spectral energy distribution of the total emission.
We propose the use  
of flare color-color diagrams in future studies to characterize both the interflare and intraflare continuum variations and to directly compare to RHD model predictions.

 \subsection{The Color Temperature of the Blue-to-Red Optical Continuum Emission}\label{sec:concl2}

The two medium-sized flares IF4 and IF11 in YZ CMi that were analyzed using simultaneous spectra and ULTRACAM data
 exhibit small Balmer jump ratios and a blue-optical continuum at peak that is apparently hot, similar to other flares in K13 and larger flares in the ULTRACAM sample.  
 The IF4 and IF11 events are among the smallest energy and lowest amplitude flares to conclusively produce 
a continuum that has $T_{\rm{BB}} > 10^4$ K.  

We showed that the measurements of the blue-optical color temperature from blackbody fitting of the spectra have large ranges ($\Delta T_{\rm{BB}} \ge 5000$ K) of possible values within the systematic uncertainty (of 2$\sigma$) for these moderate-sized flares (Section \ref{sec:interp}).  
The uncertainty results primarily from subtracting the background spectrum and to a lesser extent emission line contamination within the continuum-fit regions. 
 The uncertainties can be mitigated by higher signal-to-noise and higher time resolution ULTRACAM photometry using FcolorR, giving values of $T_{\rm{FcolorR}}$ that are also consistent with a hot color temperature for these flares.  However, significant deviations from a single blackbody function (such as due to the Conundruum continuum component in the red) are not possible to identify and characterize without spectra.  
 
 As shown in the models of K15 and the spectral observations of K13 \citep[Figures $8-9$; see also][]{Fuhrmeister2008}, the color temperature in the NUV is predicted and observed to be similar to (or apparently hotter than) the blue continuum color temperature, $T_{\rm{NUV}} \gtrsim T_{\rm{BB}}$.  High signal-to-noise observations of the spectral range at $\lambda < 3600$ \AA\ are generally difficult because of the atmospheric transmission cutoff, but this wavelength regime is not subject to the uncertainty in the subtraction of the dMe background spectrum.  Thus, we suggest that the Balmer jump ratio be combined with new measurements of continuum flux ratios in the NUV (giving $T_{\rm{NUV}}$) to provide the most robust constraints on future models of dMe and solar flares.  In Kowalski et al.\ 2016b (in preparation), we will present  improved spectral constraints from data obtained with the \emph{Hubble Space Telescope}, the Keck 1 Telescope, and the William Herschel Telescope in the NUV during two moderate-sized dMe flares and compare to the results for IF4 and IF11.

 \subsection{The Time-Evolution of the Continuum Emission}\label{sec:concl3}
Using the flare color-color diagram (Figure \ref{fig:fcolor_relation2}) we found that not all flares
produce an apparently hot optical continuum at peak with a color temperature of $\gtrsim$8500 K characterizing the blue-to-red optical wavelength regime.  
These flares also have larger Balmer jump ratios at their peaks, and
some have a (surprisingly) fast time-evolution in the impulsive phase
(e.g., IF2, IF5, IF7, IF8 and IF10 in this sample; Table
\ref{table:flare_properties}). 

  In the large spectroscopic and photometric flare sample of K13, a relation was found 
between the impulsiveness of a flare and the Balmer jump ratio at peak (Section 4.3.1 of K13).  With the ULTRACAM sample, we do not find such a clear 
relationship such that some IF events have large Balmer jump ratios.  
 Thus, we introduce a second parameter that qualitatively accounts for the variation between some flares.  The parameter $t_{\rm{rise}} / t_{\rm{fast-decay}}$ gives the asymmetry of the impulsive phase, and we found that flares with values greater than unity tend to have partially resolved 
bursts at significant flux levels in the rise phase.  The slower decaying Balmer continuum emission in the NBF3500 filter resulting from the superposition of each of these bursts could sensibly produce a larger Balmer jump ratio at the peak of the light curve.  A shorter fast decay phase also corresponds to the onset of gradual emission at a flux value that is relatively closer to the peak (e.g., IF4 compared to IF11; Figure \ref{fig:summary2});  if the impulsive and gradual emission are spatially distinct (either in atmospheric height or across the stellar surface), the total flux at flare peak would have a larger Balmer jump ratio in the flares where the onset of impulsive and gradual emission coincide more closely in time.  In addition, the high-time resolution of the ULTRACAM data gives more accurate values of the 
impulsiveness index compared to the lower time-resolution $U$-band photometry employed in K13.  Therefore some IF events would have a lower value of the impulsiveness index if observed at a lower cadence that doesn't resolve the impulsive phase. 

For the flares where the $t_{1/2}$ in NBF3500 and NBF4170 are well-measured (i.e., they measure the similar general timeframe over the flare, which tends not to be the case for flares that have several temporally resolved bursts following the peak; the insets in Figure \ref{fig:thalf}), 
the high-time resolution of the ULTRACAM data has revealed an
important new relationship among flares: the median ratio of the $t_{1/2}$ values
in NBF3500 to NBF4170 is $\approx1.3$.  The NBF4170 continuum emission is very bright for a slightly shorter amount of time than the bright continuum emission in NBF3500.  
The physical origin of $t_{1/2}$ ratios of $\approx1.1-1.5$ (the
interquartile range) should be investigated with
RHD flare models.  The 1D RHD models with the RADYN code include many of the important timescales self-consistently, such as the timescale of optical
depth changes, which can be related to the dynamic timescales of the chromospheric condensation evolution and the timescales of recombination
and statistical equilibrium.  Important timescales not currently incorporated into our 1D RHD flare modeling prescription include arcade 
development timescales (which may be helpful to sustain bright emission for longer periods of time), a timescale for energy-dependent time-of-flight of the nonthermal beam particles, a timescale for the 
evolution of the \emph{initial} flux and energy distribution of nonthermal beam particles, and a separate timescale for gradual heating (e.g., thermal conduction from direct reconnection heating in the corona, XEUV radiative-backwarming from nearby hot loops, and/or delayed nonthermal particle precipitation from magnetic mirroring).

\subsection{The Origin of the White-Light Continuum Emission in a Chromospheric Condensation} \label{sec:concl4}

\subsubsection{Intraflare Variation} \label{sec:concl4a}
The flare emission in the three ULTRACAM filters can self-consistently be interpreted with hydrogen recombination bound-free (and free-free) radiation from a flare atmosphere with a dense, heated ($T\approx12,000-13,500$ K) chromospheric
condensation with wavelength dependent optical depth effects (Section \ref{sec:rhd}).   
Using results from RHD flare models with a high energy flux of nonthermal electrons, $10^{13}$ erg cm$^{-2}$ s$^{-1}$ (F13).
If M dwarf flares produce dense, hot ($T\gtrsim10^4$ K) chromospheric
condensations in the impulsive phase, the emission in NBF3500 is dominated by optically thick Balmer continuum emission, and the
the emission in NBF4170 and RC\#1 by optically thick Paschen continuum emission. In this scenario,
the NBF3500 light escapes over the smallest physical depth range ($\Delta z$ along the line of sight) because Balmer bound-free opacity is the largest,
the NBF4170 light escapes over the largest physical depth range because Paschen bound-free opacity at this wavelength is the smallest, and the RC\#1 light escapes over a physical depth range that is intermediate between the NBF3500 and NBF4170 light.  The flare color index FcolorB is 
the ratio of the amount of Balmer continuum emission from the chromospheric condensation to the amount of Paschen continuum emission
at the most optically thin (bluest) wavelengths in the Paschen continuum.  
The flare color index FcolorR is a measure of the color temperature
of the Paschen continuum; a color temperature of $T\approx10^4$ K or greater is attained for flares
 that have a large amount of heating at high densities in a chromospheric condensation resulting from the impact of an F13 beam.
The relationship between flare evolution and the ULTRACAM filter emission can be understood as follows:  
 \begin{enumerate}
 \item Initial heating leads to Balmer continuum emission, thus giving NBF3500 emission and a moderately large Balmer jump ratio.
 
 \item A chromospheric condensation forms, resulting in both Balmer continuum (NBF3500) and Paschen continuum (NBF4170) emission.
 
 \item The relative amount of Balmer and Paschen continuum emission is modified by the increase in optical depths;  emission in NBF4170 becomes
 relatively stronger.  
 
 \item Nonthermal electron heating stops, the continuum optical depths decrease, leading to relatively more NBF3500 emission compared to NBF4170 emission.
 \end{enumerate}

\subsubsection{Interflare Variation} \label{sec:concl4b}

Variations in the hardness of an F13 nonthermal electron beam can explain some of the interflare variation of the ULTRACAM flare colors (Figure \ref{fig:fcolor_relation2}).  
The double power-law F13 distribution\footnote{A double power-law distribution is thought to result from return current energy loss \citep{Holman2012, Zharkova2006} or from non-uniform ionization \citep{Su2011}.} from K15 
was found to best represent  
 either the early rise or gradual phases (burst-averaged spectrum) or the mid-rise (instantaneous model spectrum) during a large flare (IF3 in K13).  
We compared the impulsive phase values of FcolorB for the IF4/F2 event to this double-power law simulation and determined that the instantaneous ($t=2.2$~s) model spectrum resulting from a harder ($\delta=3$)
F13 nonthermal electron beam could explain the residual discrepancy between the continuum distribution in this flare and the double power-law model (Table \ref{table:calib} and Figure \ref{fig:f13}).  The $\delta=3$ F13 electron beam
produces a similar atmospheric response as the double power-law nonthermal beam, but the harder beam distribution has more high-energy ($E>200$ keV) electrons
which heat and ionize the stationary layers of the atmosphere immediately below the chromospheric condensation to a larger extent. 
  This causes increased electron density and therefore a larger hydrogen
bound-free emissivity at the depths of the stationary flare layers.  The large optical depths for NUV ($\lambda=3500$ \AA) and red ($\lambda=6010$ \AA) light in these layers
prevent the photons from escaping, but the lower optical depth at $\lambda=4170$ \AA\ leads to a greater escape probability for the blue photons.  
Compared to the double power-law simulation in K15, the density in the chromospheric condensation in the $\delta=3$ power-law distribution is lower (at $t=2.2$~s), making it even more likely for blue light to escape from these deeper layers.  These effects produce a smaller value of FcolorB and a larger value of FcolorR in the harder ($\delta=3$) nonthermal electron simulation, which can explain some of the interflare variation in Figure \ref{fig:fcolor_relation2}. The interflare peak flare color index distribution in Figure \ref{fig:fcolor_relation} is not explained by the short bursts of lower beam fluxes between $10^{11}-10^{12}$ erg cm$^{-2}$ s$^{-1}$ with 
a moderate low-energy cutoff ($E_c=37$ keV).

\subsection{Future Modeling Directions} \label{sec:concl5}

  We intend to improve our RHD modeling approach in several important ways. 
   An F13 electron beam produces a large return current electric field, as discussed by
  K15.  In future work (Kowalski et al 2016c, in prep), we will model the atmospheric heating that results from
  a F13 nonthermal electron distribution that undergoes energy loss from a return current electric field and compare 
  to a model with lower nonthermal electron flux and a higher low-energy cutoff (thus mitigating return current effects).
  In these models, the Fokker-Planck solution to the electron distribution will be used to
  account for the relativistic energy loss effects that are important for the higher energy (E$>$200 keV) electrons \citep{Allred2015}.  
  In future work (Allred et al 2016a, in prep), we will also use the emission line constraints from the lower order Balmer lines (e.g., H$\gamma$) to critically test these models with an improved treatment of the broadening from the    
   Stark effect (as discussed in Section 7 of K15).
  
  Since increasing the hardness of an F13 nonthermal electron distribution has the effect of increasing FcolorR and decreasing FcolorB (Section \ref{sec:rhd}), we 
  speculate that the (intraflare) time-evolution of the color indices over a dMe flare impulsive phase (Figures \ref{fig:summary2}, \ref{fig:ultraflare}) is also related to the variation of the injected nonthermal electron 
 distribution.  In solar flares, a ``soft-hard-soft'' (SHS) evolution of the hard X-ray spectrum power-law index ($\gamma$)
  is observed during hard X-ray sub-peaks, which show a range of timescales ranging from $\lesssim10$~s to a minute \citep{Lin1987, Fletcher2002, Grigis2004, Holman2011}.   The SHS variation in hard X-ray flux is attributed to
 a corresponding increase and decrease of the hardness of the accelerated nonthermal electron distribution over an individual flare sub-peak, which 
 is consistent with the theory of stochastic acceleration of electrons \citep[e.g.,][]{Petrosian2004} resulting from a gradual increase and decrease in the turbulence in a magnetic loop \citep{LiuFletcher2009}.   In future models, we will incorporate the nonthermal flux and SHS spectral evolution predicted by the stochastic acceleration
theory of \cite{LiuFletcher2009} in order to understand how the intraflare variation of the FcolorR and FcolorB due to 
SHS variations compares to that from the development of a chromospheric condensation (optical depth variation; Section \ref{sec:concl4a}).  Flare color-color diagrams of individual events
will be utilized for guiding each of these modeling approaches.
  
  It will also be important to lengthen the timescales over which bright continuum emission persists in the models, since the F13 impulsive phase model timescale (2~s) is not consistent with the timescales of continuum emission observed with ULTRACAM ($t_{1/2}$ values of $\lesssim$ ten seconds to several hundred seconds; Figure \ref{fig:thalf}).
     The integration times of $1-2$~s with ULTRACAM would average 
  over the emission produced by the F13 heating burst considered in this work and in K15.  Although the average model for the $\delta=3$ 
  simulation is not considered here, we note that the burst-averaged F13
  double-power law model was found to exhibit lower values of FcolorR and higher values of FcolorB compared to the instantaneous spectrum at the time of the brightest continuum emission ($t=2.2$~s).  
  A more realistic superposition of impulsively heated and decaying bursts is outside the scope of this paper but will be necessary to directly compare to the
observations.  The spatial development of two-ribbon flare arcades on the Sun consists of a series of sequentially heated footpoints \citep{Kosovichev2001, Wang2009, Qiu2010, Tian2015, Graham2015}.  A method for modeling solar flare arcade development consisting of a superposition of $\sim10-30$~s bursts is currently being developed (Allred et al 2016b, in prep).
  
  We observe the Balmer jump ratio (FcolorB) to vary over the timescale of the impulsive phases of IF1 and IF4, both of which have flare fluxes that 
smoothly vary without evidence of partially resolved
sub-peaks\footnote{In the IF1 event, there is a gradual increase from
  $t-t_{\rm{peak}}\approx -60$~s to $-30$~s, which only attains
  relatively low flux level ($\lesssim10$\% of the peak flux).  There is also a change from fast to faster rise emission at $t-t_{\rm{peak}}\approx -20$~s.}
at a cadence of $\Delta t=1-2$ s.  Notable, the impulsive phase timescales of the IF4 and IF1 events are also similar to the timescales of some sub-peaks in solar flare hard X-ray light curves \citep{Lin1987}. 
   Do the timescales of these events ($t_{1/2} \approx$14~s for IF4 and 43~s for IF1) represent a fundamental heating timescale of an ensemble of flux tubes,
 or do many sub-resolution heating events occur within the rise phase of the flares (as for some of the flare events that show evidence of partially resolved 
  bursts in the rise phase in the NBF3500 light curves; Section \ref{sec:bursty})?   For the IF1 event, the value of FcolorR varies only over a narrow range ($1.8-2.0$) in the rise phase, and the  corresponding color temperature remains approximately constant ($T_{\rm{FcolorR}}=9000-10,000$ K).  Therefore, the observed flux increase
in this flare is predominantly due to the increase in the areal coverage of the continuum-emitting regions\footnote{We assume that a narrow range of color temperature variation corresponds to a narrow range of surface fluxes, thus giving an general proportionality between the observed flux at Earth and the flare area.}.
Using a grid of short duration RHD M dwarf flare model bursts with lower energy fluxes (Kowalski et al 2016a in prep), we will investigate whether
the impulsive phase evolution of the flare color for these events represent the fundamental timescale of a SHS heating burst of an ensemble of flux tubes, 
or whether such flares (and by extension some hard X-ray sub-peaks in solar flares) are composed of a series of high-flux (e.g., F13)
bursts staggered in time with increasing hardness in the nonthermal electron distribution from the rise to peak times.  From our RHD modeling here (Section \ref{sec:rhd}), we hypothesize that the impulsive phase of IF4 consists of a 
superposition of short F13 bursts with softer nonthermal electron beams (e.g., similar to the double power-law $\delta=3,4$ F13 beam) 
 in the rise phase and harder nonthermal electron beams (e.g., a $\delta=3$ F13 beam) during the peak phase, thus giving the observed flare color variations. 
We note that if the nonthermal electron spectrum varies between bursts (in addition to the variation over the duration of a single burst)
a mechanism would be required to explain a larger amount of turbulence in the flare loops reconnected at 
peak times (e.g., through the development of turbulent velocities, which have been observed in the wake of a coronal mass ejection; \cite{McKenzie2013}). 
In summary, improving the time-evolution prescription (either through a superposition of bursts or through optical depth changes that occur over a longer time)
 while incorporating a more realistic physical description of 
nonthermal electron beams (using the relativistic Fokker-Planck solution with SHS variations and return current effects) are necessary future directions for modeling the impulsive phase flare emission observed with ULTRACAM.

We also plan to continue work on the Ultraflare (Section \ref{sec:ultraflare}) with low-resolution spectra 
obtained with the RSS on the SALT \citep{BrownAAS}.  In future work (Brown et al 2016, in prep), the 
spectral properties of this flare will be presented and compared to RHD flare models.  In particular, we will compare the emission
line and continuum evolution to place constraints on the formation of photospheric ``hot spots'' \citep{Kowalski2012} in the secondary flares of the decay phase via the anti-correlation presented in \cite{Kowalski2010} for the secondary flares in the Megaflare.
 
 We seek to understand the nature of these secondary flares with a scenario that self-consistently explains 
 both the triggering and atmospheric heating.  The general timing of the Megaflare secondary events were shown to be consistent with magneto-hydrodynamical processes from oscillating loops \citep{Nakariakov2006, Anf2013} whereas the full range of timescales was found to be consistent with a Fourier power spectrum \citep{Inglis2015}.  The properties of the secondary flares
 in the Ultraflare will provide important additional constraints on theoretical predictions of flare energy release.
  \cite{Kowalski2012} envisioned an expanding disturbance originating from the site of primary energy release would ``trigger'' the secondary events in the YZ CMi Megaflare.  
 Turbulent velocities have been detected in the wake of coronal mass ejections in solar flares \citep{Hara2009, McKenzie2013}.  
 Since electron beam intensity can be attributed 
 to the amount of magnetized turbulence in a loop \citep{LiuFletcher2009}, we speculate that propagating shock waves from the main peaks of the Ultraflare and Megaflare events generated turbulent flows in nearby active regions with especially strong magnetic fields, thus leading to the secondary events.  Assuming these waves propagated at the same speed across the surface (the flares occurred on the same star), the different timing of the intriguing pattern of flares in Figure \ref{fig:ultraflare_megaflare} may be related to the distance to the triggered events from the main flare site.
Observations of nonthermal
velocities in the decay phase of stellar flares \citep[e.g.,][]{Osten2005, Ayres2015} may help connect the role of 
post-impulsive phase turbulence to the generation of accelerated particles in neighboring magnetic loops.  Spatial information from sympathetic flares following large flares on the Sun (such as during the X5.4 flare of 2012-Mar-07) and 3D magneto-hydrodynamic models coupled with particle acceleration mechanisms will be crucial for explaining these enigmatic post-peak events.  

\section*{Acknowledgments}
We thank an anonymous referee for a helpful review of this article.
 A.F.K. acknowledges the support from NSF grant
AST08-07205, the NASA Postdoctoral Program at the Goddard Space Flight Center, administered by Oak Ridge Associated Universities through a contract with NASA, 
and from UMCP GPHI Task 132.  
M. M. acknowledges support from the European Community's Seventh Framework Programme (FP7/2007-2013) under grant agreement No.\ 606862 (F-CHROMA).
T. R. M. acknowledges support from the STFC, ST/L00073.
A. F. K. acknowledges Dr. J. Holtzman for providing the $U$-band data from the NMSU 1 m telescope at the APO, Dr. S. J. Schmidt for supporting
observations with ARCSAT at the APO, Dr. C. Copperwheat and Dr. S. Littlefair for assistance with ULTRACAM observations at the NTT, Dr. J. Allred for helpful discussions about flare modeling and for assistance with the RADYN and RH codes, Dr. M. Carlsson for helpful assistance with the RADYN code, Dr. H. Uitenbroek for helpful assistance with the RH code, Dr. P. Yoachim for  discussions about spectra obtained with a wide slit, Dr. R. Ryans for installing the ULTRACAM data reduction pipeline, and Dr. L. Kleint for discussions about the interpretation of excess flare emission.
\clearpage

\begin{figure}
\includegraphics[scale=0.75]{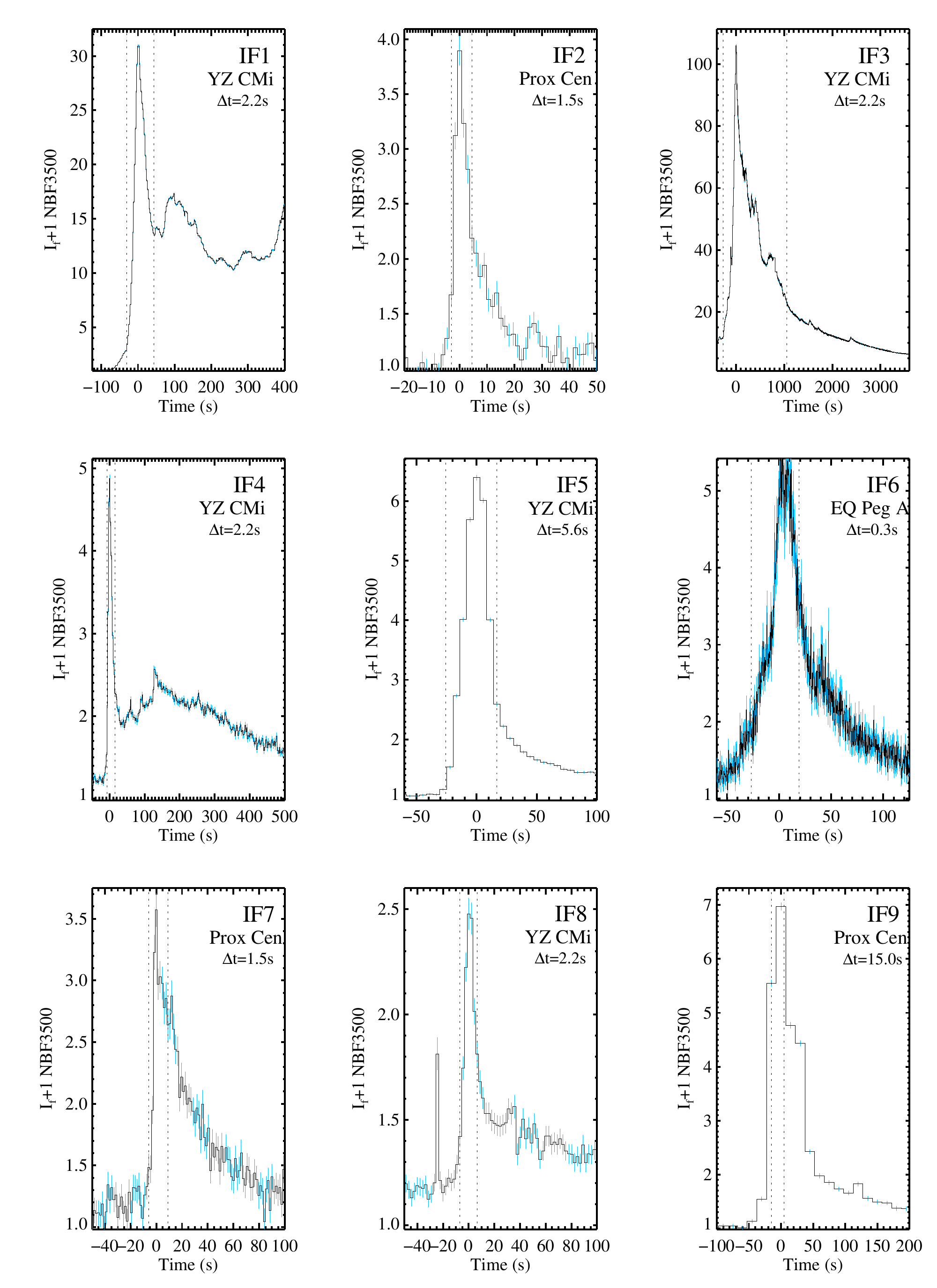}
\caption{Light curves in NBF3500 for the twenty flares in our sample.  The vertical dotted lines indicate the start of the rise phase and end of the fast decay phase. } \label{fig:lcsample1_3500}
\end{figure}

\begin{figure}
\includegraphics[scale=0.75]{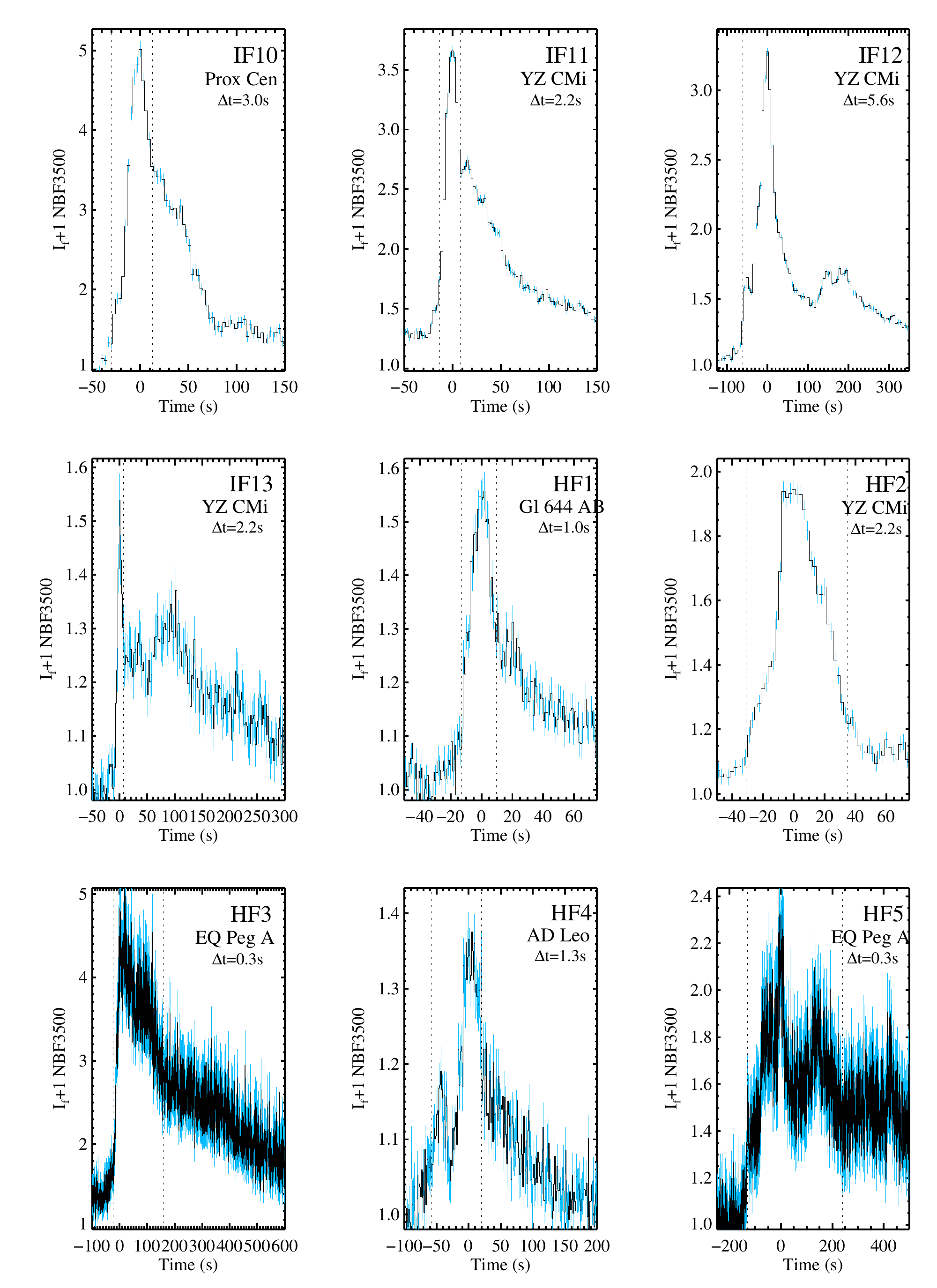}
\caption{(continued) Light curves in NBF3500 for the twenty flares in our sample. } \label{fig:lcsample2_3500}
\end{figure}

\begin{figure}
\includegraphics[scale=0.75]{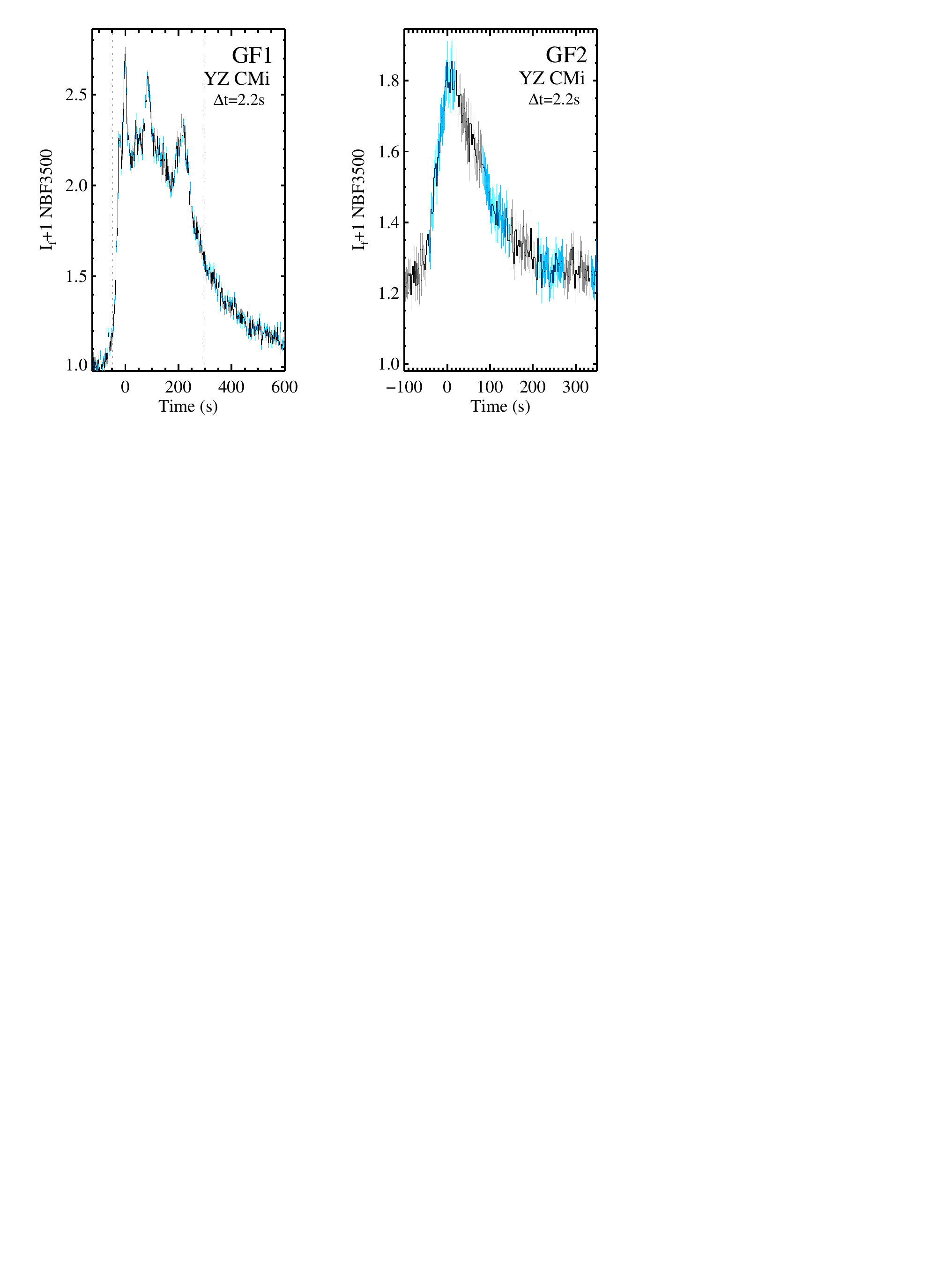}
\caption{(continued) Light curves in NBF3500 for the twenty flares in our sample. } \label{fig:lcsample3_3500}
\end{figure}

\begin{figure}
\begin{center}
\includegraphics[scale=0.65]{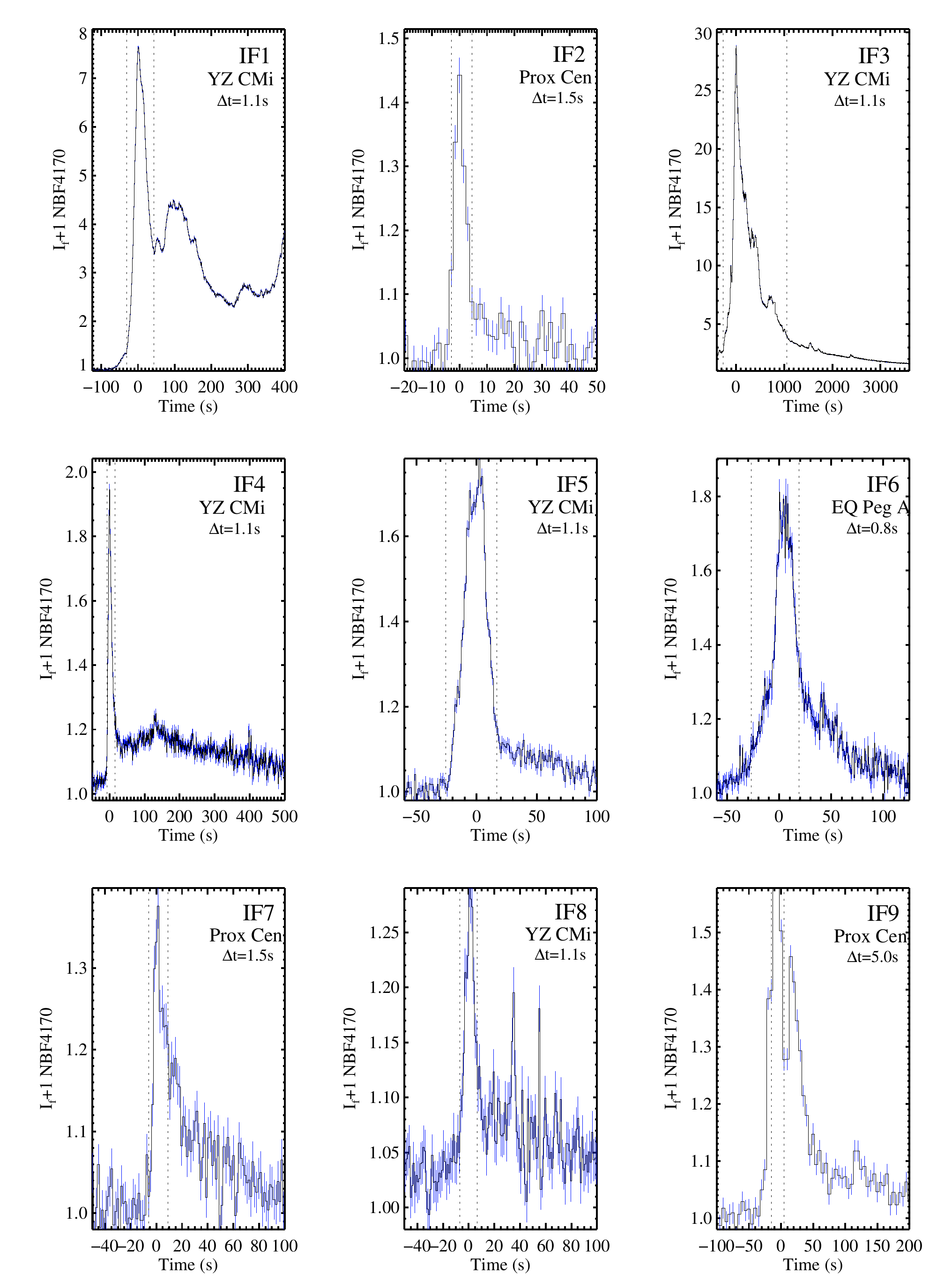}
\caption{Light curves in NBF4170 on approximately one-second timescales for the twenty flares in our sample. The data have been binned 
 for flares IF6, HF1, HF3, HF4, HF5, and GF2 to increase the signal-to-noise.  The vertical dotted lines indicate the start of the rise phase and the end of the fast decay phase. } \label{fig:lcsample}
 \end{center}
\end{figure}

\begin{figure}
\begin{center}
\includegraphics[scale=0.65]{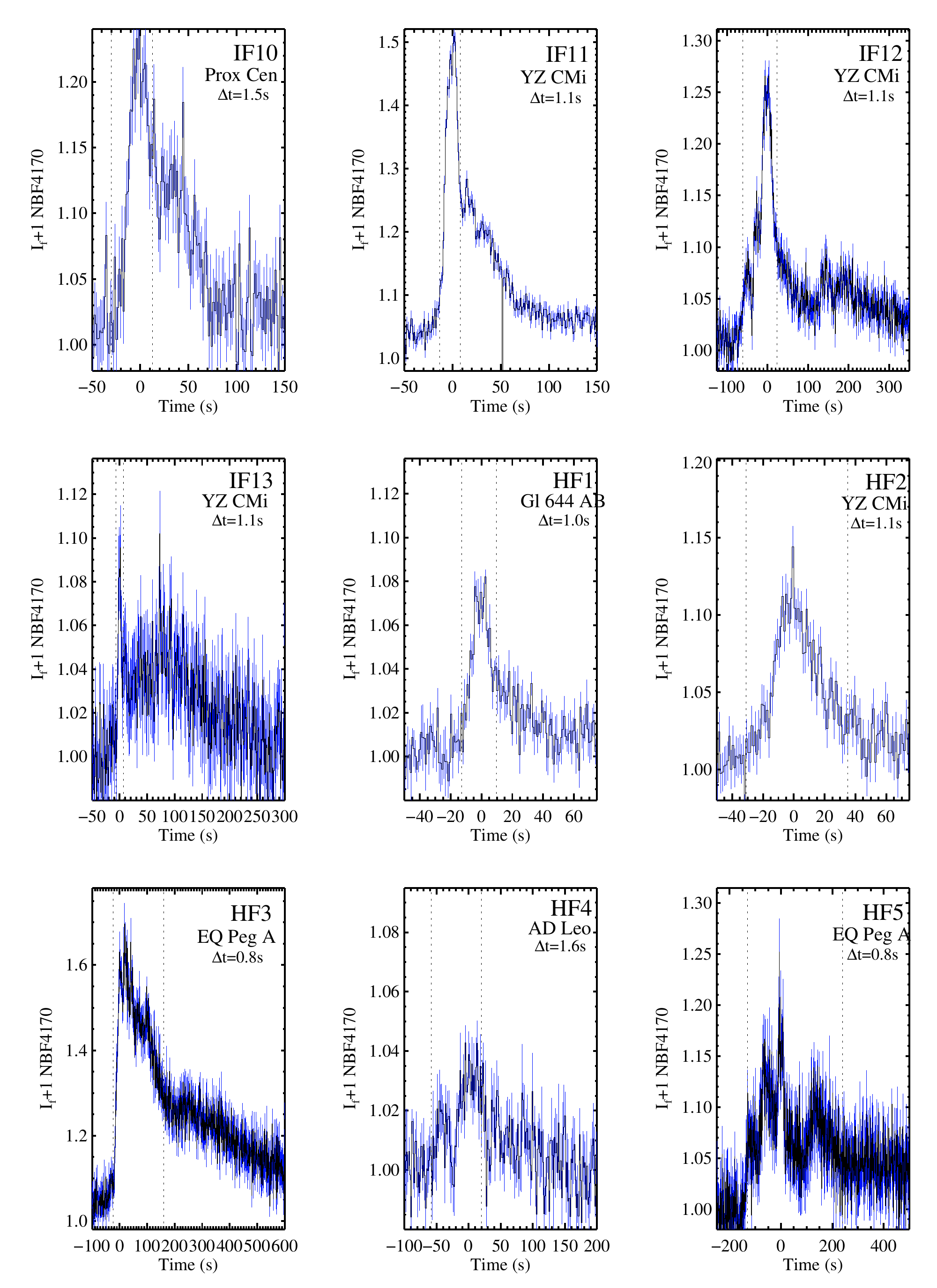}
\caption{(continued) Light curves in NBF4170 for the twenty flares in our sample. } \label{fig:lcsample2}
\end{center}
\end{figure}

\begin{figure}
\includegraphics[scale=0.65]{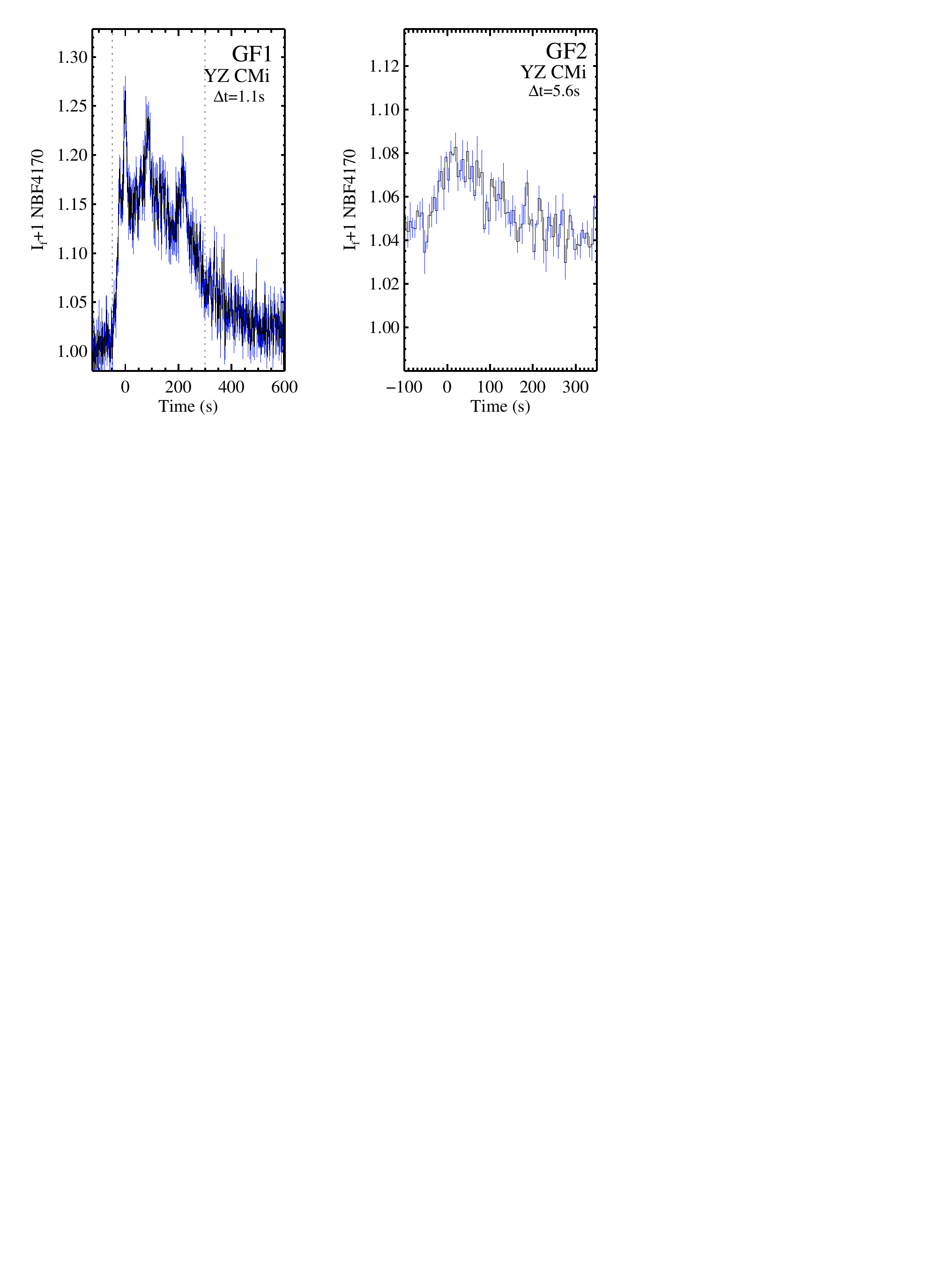}
\begin{center}
\caption{(continued) Light curves in NBF4170 for the twenty flares in our sample. } \label{fig:lcsample3}
\end{center}
\end{figure}

\clearpage
\begin{figure}
\begin{center}
\includegraphics[scale=0.75]{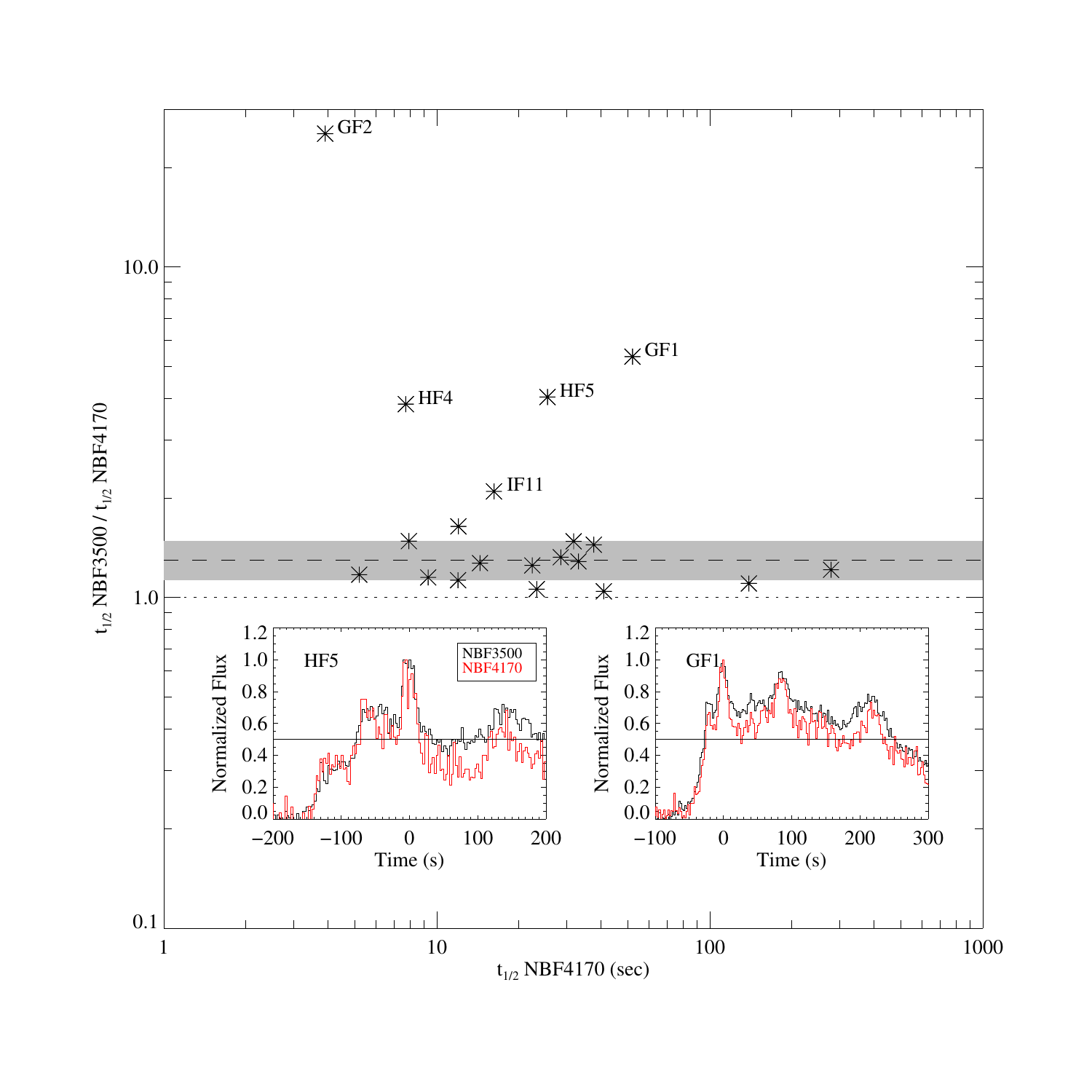}
\vspace{-10mm}

\caption{ The values of $t_{1/2, \rm{NBF3500}}/t_{1/2, \rm{NBF4170}}$ for the largest 20 flares in the ULTRACAM sample shown as a function 
of the $t_{1/2, \rm{NBF4170}}$ value. (top:) The dotted line gives a ratio of 1, the dashed line is the median ratio (1.3), and the shaded region
indicates the interquartile range (0.36).  Only five flares (IF11,
HF4, HF5, GF1, GF2) lie significantly above the shaded region; these
flares are labeled. The IF1 and IF5 events have the lowest ratios, $\approx1.05$. (bottom:)  The insets show the individual light curves for HF5 and GF1 in the NBF3500 (black) and NBF4170 (red) filters with the $t_{1/2}$ values indicated. } \label{fig:thalf}
\end{center}
\end{figure}

\begin{figure}
\begin{center}
\includegraphics[scale=0.75]{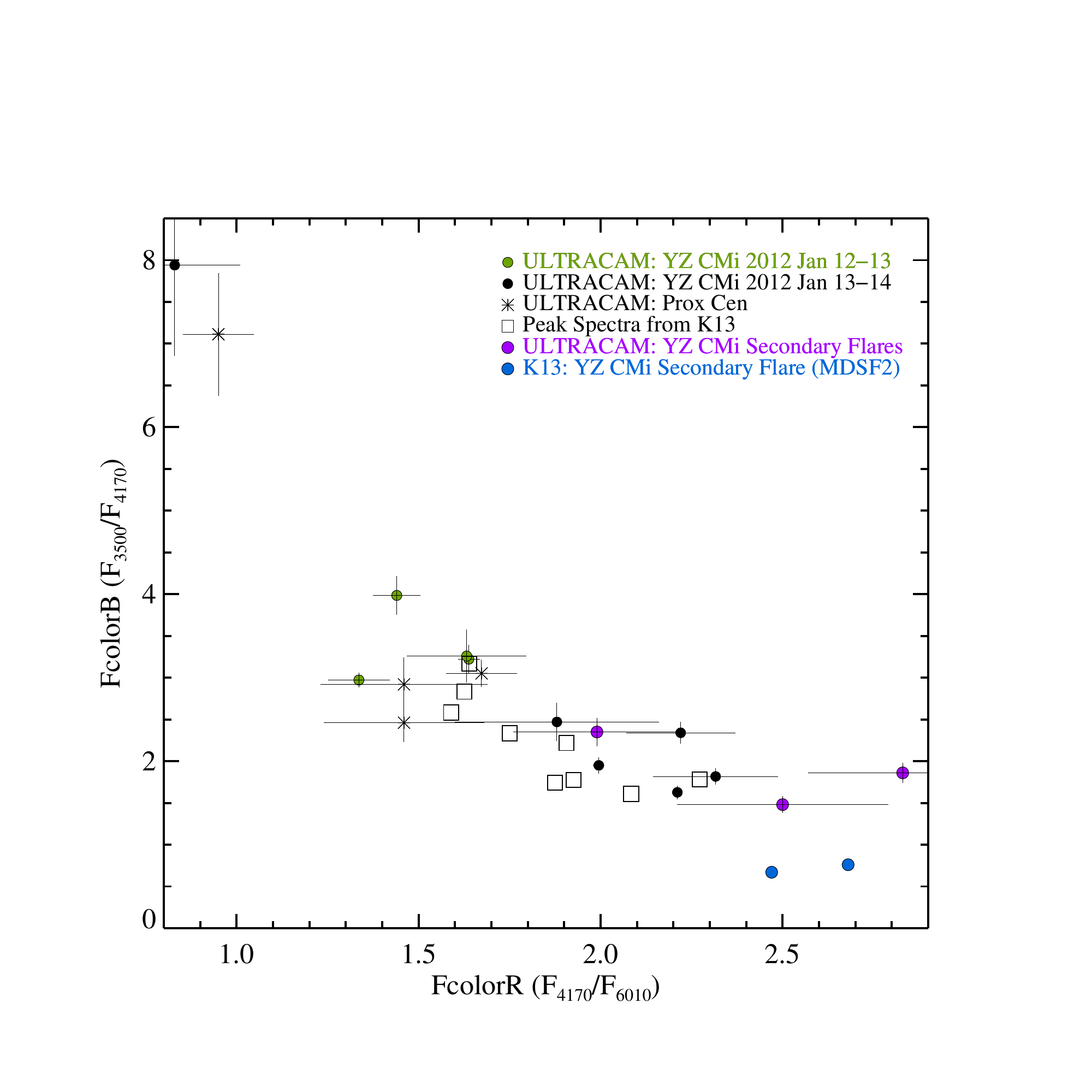}
\vspace{-10mm}
\caption{
  The values of FcolorB vs FcolorR for the ULTRACAM sample of fourteen flares
  in the dMe stars YZ CMi and Proxima Centauri.   The IF10 and GF2 flares have very small values of FcolorR and very large
  values of FcolorB at their peaks. The blue filled circles indicate
  the values for the
  Vega-like flare spectra from K13 (mid-rise and peak).  The purple filled circles show the flare color indices for the newly formed emission
  at the peaks of the secondary flares discussed in Section \ref{sec:secondary}.   See text for discussion. 
 } \label{fig:fcolor_relation}
 \end{center}
\end{figure}

\begin{figure}
\begin{center}
\includegraphics[scale=0.75]{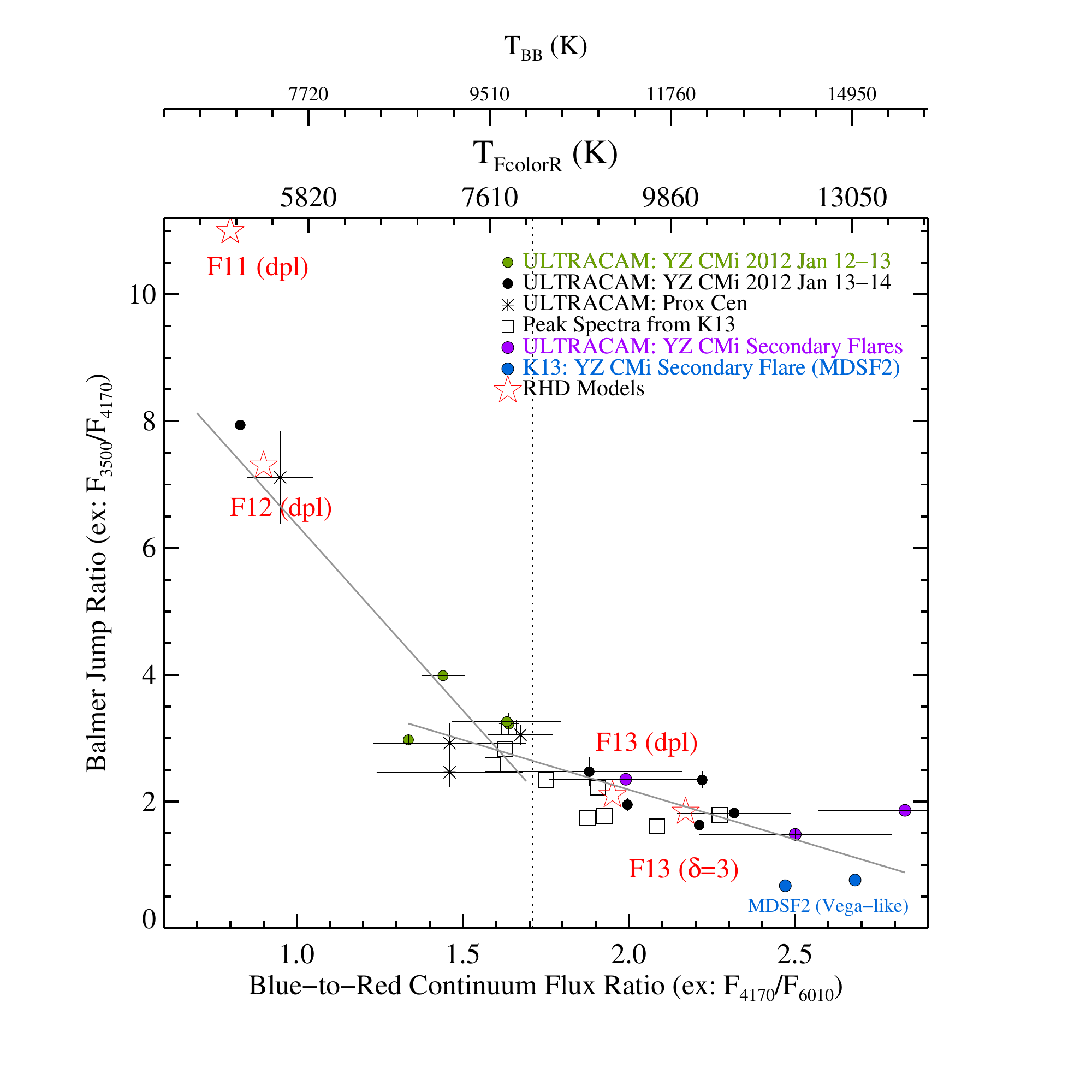}
\vspace{-10mm}
\caption{The data from Figure \ref{fig:fcolor_relation} are shown with an extended
  range for the x- and y-axes (FcolorR and FcolorB, respectively).  The values of $T_{\mathrm{FcolorR}}$ corresponding to the blue-to-red continuum flux ratio FcolorR and the calibrated values of $T_{\rm{BB}}$ are shown on the 
  top x-axes.  The RHD model predictions from Table \ref{table:calib}
  are shown as red star symbols, where ``dpl'' indicates the double
  power law simulations at $t=2.2$~s from K15.  A double linear
  (unweighted) fit (gray line) to the data is shown to guide the eye.   See text for discussion. } \label{fig:fcolor_relation2}
  \end{center}
\end{figure}

\begin{figure}
\begin{center}
\includegraphics[scale=0.6,angle=90]{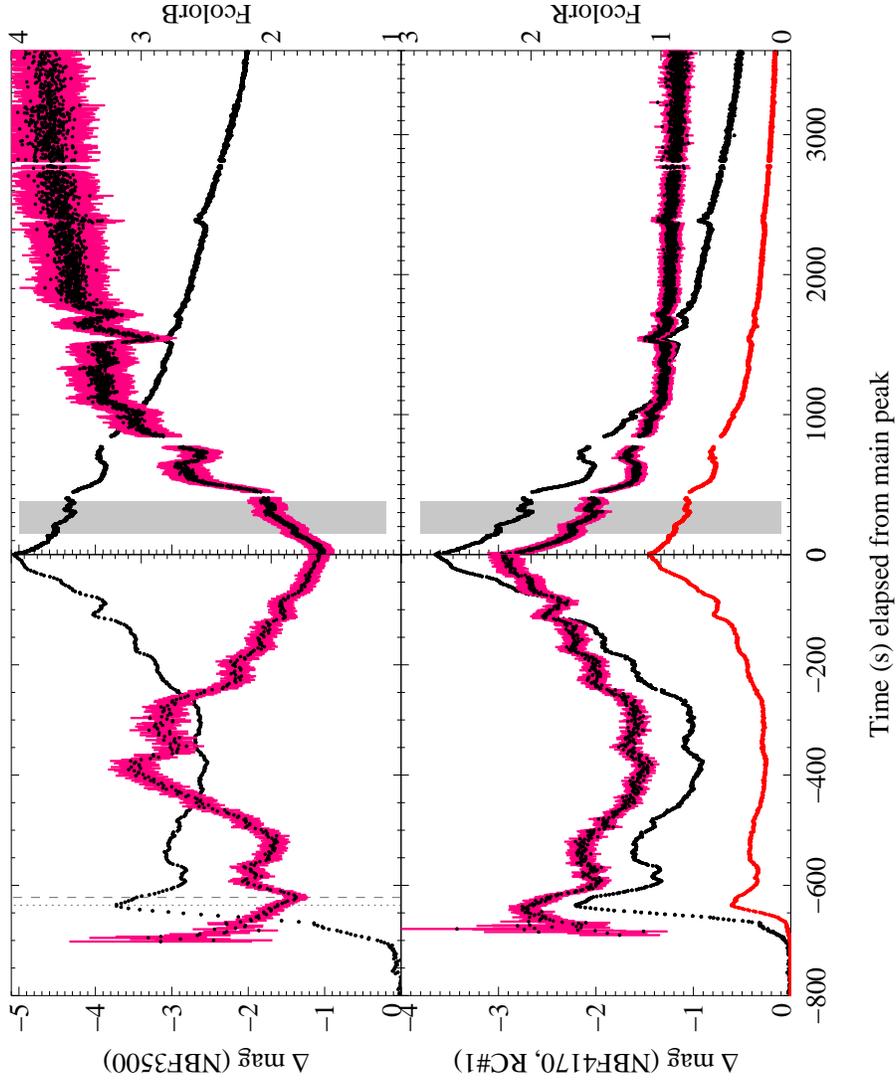}
\vspace{-13mm}
\caption{ Light curves for NBF3500 (top, black) and NBF4170 and RC\#1 (bottom, black and red, respectively) for the Ultraflare on the dMe star YZ CMi.  The values of FcolorB (top) and FcolorR (bottom) are shown as black points with pink error bars (right axis).  The left panel encompasses the IF1 event and the rise of IF3.  The vertical dotted line indicates the peak of IF1 and the  vertical dashed line indicates the minimum value of FcolorB (1.8) which occurs 14~s after this peak.  The right panel shows the first 3600~s of the decay of 
IF3; note the different time coverage in each panel.  The time corresponding to  $t=0$~s is the peak of IF3 at 22:44:33 UT on 2012 Jan 13.  The gray shaded bars indicate the times shown in Figure \ref{fig:ultraflare_megaflare}.  See text for discussion. }  \label{fig:ultraflare}
\end{center}
\end{figure}

\begin{figure}
\includegraphics[scale=0.7]{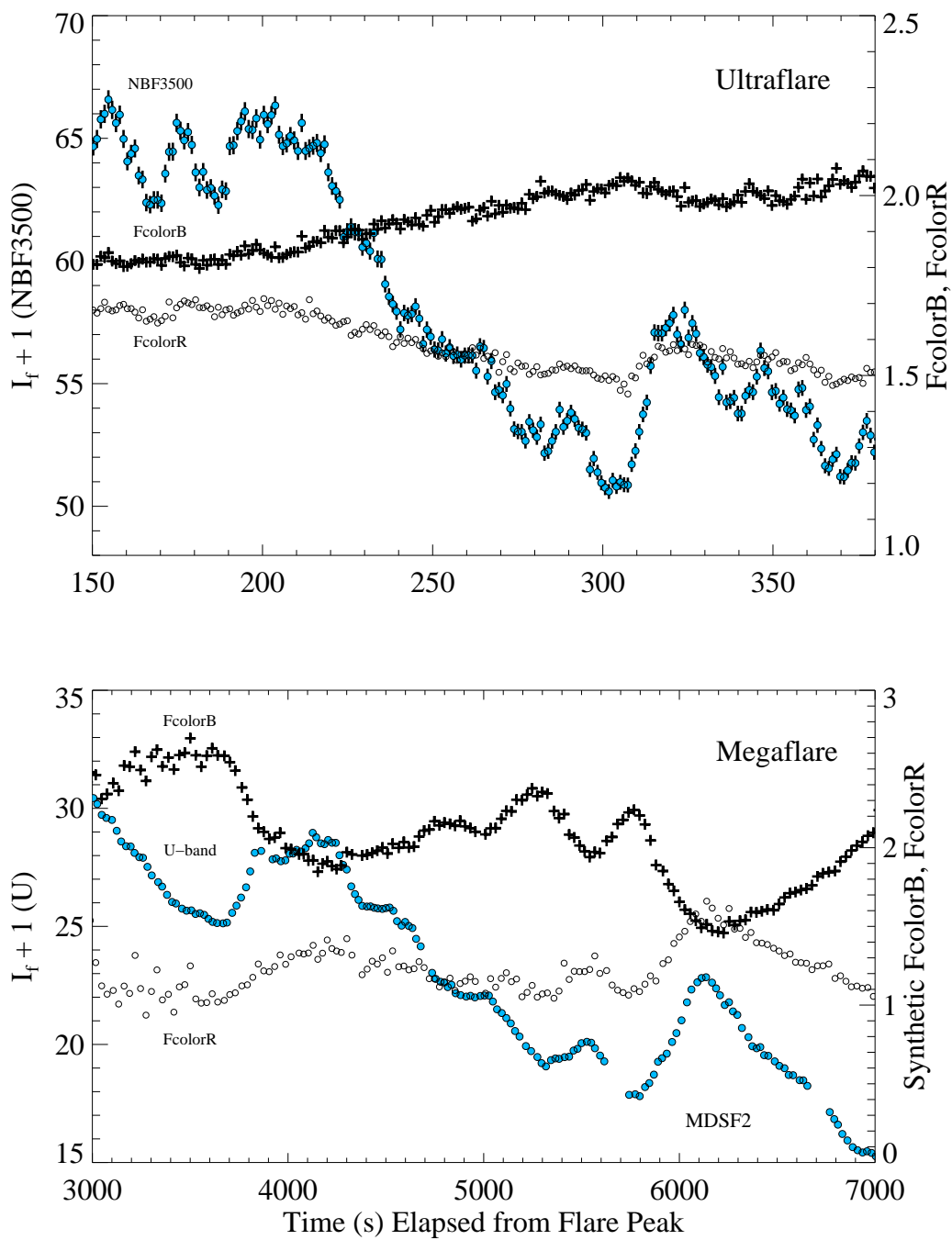}
\vspace{-10mm}
\caption{ Light curves in NBF3500 (top) and the $U$-band (bottom) of a similar sequence of secondary flares during the gradual decay phases of the 
Ultraflare (top) and Megaflare (bottom), both in the dMe star YZ CMi.  The values of FcolorB (crosses) and FcolorR (open circles) are shown on
the right axis for each flare. The colors were synthesized from the spectra for the Megaflare. } \label{fig:ultraflare_megaflare}
\end{figure}

\clearpage

\begin{deluxetable}{cclll}
\rotate
\tabletypesize{\scriptsize}
\tablewidth{0pt}
\tablecaption{Observing Log}
\tablehead{
\colhead{Date (MJD)} &
\colhead{Monitoring time (hr)} &
\colhead{Exp time (s)} &
\colhead{Telescope} &
\colhead{\# Flare events}
}
\startdata
\textbf{YZ CMi} & & & & \\
\hline
 2012 Jan 12\,--\,13 (55938\,--\,55939) & 6.87 & 2.2-5.6 [N], 1.1-2.1 [B, R]  & WHT  & $\approx$30 \\

 2012 Jan 13\,--\,14 (55939 \,--\, 55940) & 8.23 &  1.1-5.6 [N], 1.1-2.1 [B, R] & WHT &  $\approx$30  \\
\hline
\hline

\textbf{AD Leo} & & & & \\
\hline
 2012 Jan 12\,--\,13 (55938\,--\,55939) & 0.58 & 1.3 [N],
 0.3 [B, R]  & WHT & $\approx$2 \\
\hline
 2012 Jan 13\,--\,14 (55939\,--\,55940) & 1.70 &  1.3-2.6 [N],
 0.3 [B, R] & WHT & $\approx$2 \\
\hline
\hline

\textbf{Proxima Centauri} & & & & \\
\hline
2010 May 21\,--\,22 (55337\,--\,55338) & 6.72 & 10-30 [N],
 3.0 [B, R] & NTT & $\approx$5 \\
\hline
 2010 May 23 (55339) & 3.96 & 3-6 [N], 
 1.5-3 [B, R] & NTT & $\approx$13 \\
\hline
 2010 May 24\,--\,25 (55340\,--\,55341) & 6.88 & 1.5-3 [N], 1.5 [B, R]  & NTT & $\approx$14 \\
\hline
\hline

\textbf{Gl 644 AB} & & & & \\
\hline 
 2010 May 23 (55339) & 2.09 & 1 [N], 0.2 [B, R]  & NTT  & 0 \\
\hline
2010 May 25 (55341) & 2.15 & 1-2.4 [N], 0.2-0.5 [B, R]  & NTT & $\approx$5 \\
\hline
\hline
\textbf{EQ Peg A} & & & & \\
\hline
 2008 Aug 10 & 1.5 & 0.3 [N], 0.13 [B, R] & WHT & 3 \\
\enddata
\tablecomments{The range of exposure times employed on each night are labeled by the corresponding filter ([N]: NUV arm (NBF3500), [B]: blue arm (NBF4170), [R]: red arm (RC\#1 or H$\alpha$; see text)). 
  The exposure times of the blue and red arms were always the same, whereas the exposure time in the NUV arm was often twice as long. }
\label{table:obslog}
\end{deluxetable}

\begin{deluxetable}{lllllll}
\tabletypesize{\scriptsize}
\tablewidth{0pt}
\tablecaption{Quiescent Properties of Target Stars}
\tablehead{
\colhead{Star} &
\colhead{Spectral Type} &
\colhead{$I_{\rm{o},\lambda=3500}$} &
\colhead{$I_{\rm{o},\lambda=4170}$} &
\colhead{$I_{\rm{o},\lambda=6010}$} &
\colhead{QcolorB}  &
\colhead{QcolorR} 
}
\startdata
YZ CMi (Gl 285) & dM4.5e & 1.37 & 3.19 & 14.2 & 0.43 & 0.22 \\
AD Leo (Gl 388) & dM3e & 7.57 & 21.93 & 86.6 & 0.35 & 0.25 \\ 
EQ Peg A (Gl 896 A) & dM3.5e & \nodata & \nodata & \nodata & 0.40 & \nodata  \\ 
Proxima Centauri (Gl 551) & dM5.5e & 0.6 &  1.5  & 12.7 & 0.38 & 0.12 \\
Gl 644AB & dM3e  & \nodata & \nodata & \nodata & 0.35 & 0.25 \\ 
\enddata
\tablecomments{Quiescent specific fluxes ($I_{\rm{o}}$) are in units of $10^{-14}$ erg s$^{-1}$ cm$^{-2}$ \AA$^{-1}$. For the Gl 644AB 
fluxes, we use the AD Leo fluxes scaled by the magnitude difference between the two stars.  QcolorB and QcolorR are the ratio of continuum fluxes in quiescence (see text). Only QcolorB is given for EQ Peg A because the red arm data were obtained in an H$\alpha$ filter and are not used for analysis here. }
\label{table:stars}
\end{deluxetable}

\begin{deluxetable}{cccc}
\tabletypesize{\scriptsize}
\tablewidth{0pt}
\tablecaption{Timescales of F1 and F2}
\tablehead{
\colhead{Flare} &
\colhead{$t_{1/2}$ NBF3500 (s)} &
\colhead{$t_{1/2}$ NBF4170 (s)} &
\colhead{$t_{1/2}$ RC\#1 (s)} }
\startdata
F1 & 34 & 16 & 18 \\
F2 & 14 & 12 & 17 \\
\enddata
\tablecomments{The $t_{1/2}$ values give the full width at half maximum of the light
curves.}
\label{table:thalf}
\end{deluxetable}

\begin{deluxetable}{lllllll}
\tabletypesize{\scriptsize}
\tablewidth{0pt}
\tablecaption{Color Comparison between Spectra and Photometry for F1 and F2}
\tablehead{
\colhead{Flare} &
\colhead{$\chi_{\rm{flare}}$ (DIS)} &
\colhead{C4170/C6010 (DIS)} &
\colhead{FcolorB (DIS)} &
\colhead{FcolorR (DIS)} &
\colhead{FcolorB (UC)} &
\colhead{FcolorR (UC)} 
}

\startdata
F1(i) & 2.2 & 2.1 & 2.3 & 2.1 & 2.34 $\pm0.13$ & 2.16 $\pm0.21$ \\
F1(ii) & 3.0 & 1.5 & 2.8 & 1.4 & -- & -- \\
F2(i) & 1.8 & 2.0 & 1.8 & 2.1 & 1.86 $\pm0.10$ & 2.31 $\pm0.17$ \\
F2(ii-v) & 3.0 & 1.2 & 2.9 & 1.3 & $\approx$2.8\,--\,3.3 & $\approx$1.3 \\ 
\hline
F11 (dpl)&\nodata   & \nodata  & \nodata & \nodata & 11.0 & 0.80 \\
F12 (dpl)&\nodata   &\nodata   & \nodata & \nodata & 7.30 & 0.90 \\
F13 (dpl)&\nodata   & \nodata  & \nodata & \nodata & 2.10 & 1.95 \\
F13 ($\delta=3$) & 1.76 & 2.17 & \nodata & \nodata & 1.84 (1.81) & 2.18 (2.25) \\
\enddata
\tablecomments{The roman numerals (i)-(v) refer to the (APO/DIS) spectra as labeled in Figure \ref{fig:summary2}. The model spectra
are obtained at $t=2.2$~s in the double power law (dpl) simulations from K15 and at $t=2.2$~s for the $\delta=3$ F13 model (this work); these model values are obtained from the flux spectrum calculated in RADYN, and the continuum ratios are calculated after subtracting the pre-flare spectrum.
The F13 models are to be compared to the ULTRACAM (UC) flare color indices at the peak of F2.  For the $\delta=3$ F13 simulation, the
values in parentheses 
 are obtained from the model spectrum in Figure \ref{fig:f13}, which include the opacities from Landau-Zener transitions. }
\label{table:calib}
\end{deluxetable}

\begin{deluxetable}{llclc}
\tabletypesize{\scriptsize}
\tablewidth{0pt}
\tablecaption{Spectral fitting results}
\tablehead{
\colhead{Flare Spectrum} &
\colhead{$T_{\mathrm{BB}}$ (K)} &
\colhead{$X_{\mathrm{BB}}$ $\times10^{-5}$} &
\colhead{$T_{\mathrm{FcolorR}}$ (K)} &
\colhead{$X_{\mathrm{FcolorR}}$ $\times10^{-5}$ } }
\startdata
F1(i) & 12,100 & 2.7 & 10,400 & 4.4 \\
F1(ii) & 8600 & 6.5 & 7200 & 14 \\
F2(i) & 12,100 & 4.4 & 10,200 & 7.7 \\
F2(ii-v) & 9300 & 4.1 & 6900 & 15  \\
\enddata
\tablecomments{Results from fitting the spectra
in Figure \ref{fig:spectra} to a blackbody function.  The value of  $T_{\mathrm{BB}}$ was obtained
from fitting a Planck function to the blue-optical zone wavelength windows, designated as the ``BW'' regions in Table 4 of K13.
The value of $T_{\mathrm{FcolorR}}$ was obtained from convolving the spectra with the ULTRACAM filters and fitting to a Planck function.
The roman numerals (i)-(v) refer to the spectra in Figure \ref{fig:spectra}. }
\label{table:bbtable}
\end{deluxetable}

\begin{deluxetable}{lllllllllllll}
\rotate
\tabletypesize{\scriptsize}
\tablewidth{8.5in}
\tablecaption{The ULTRACAM Flare Sample}
\tablehead{
\colhead{ID} &
\colhead{Star} &
\colhead{Peak Time} &
\colhead{$I_f (3500)$} &
\colhead{$I_f (4170)$} &
\colhead{$I_f (6010)$} &
\colhead{FcolorB (err)} &
\colhead{FcolorR (err)}  &
\colhead{$t_{1/2} $} &
\colhead{$\mathcal{I}$  }&
\colhead{$t_{\rm{rise}}$}&
\colhead{$t_{\rm{fast\ decay}}$ }&
\colhead{$\mathcal{I}'$} \\
\colhead{} & \colhead{} & \colhead{(UT)} & \colhead{} & \colhead{} &\colhead{} & \colhead{} & \colhead{} & \colhead{(s)} & \colhead{} & \colhead{(s)} & \colhead{(s)} & \colhead{} }
\startdata
IF1    & YZ CMi      & 13-Jan-2012 22:33:54  &     29.9  &     6.59  &     0.73  &     1.95   (    0.10) &     2.00   (    0.02) &     42.7  &     42.0  &       31   &      44  &     59.0  \\
IF2    & Prox Cen    & 25-May-2010 04:32:11  &      2.9  &     0.44  &     0.04  &     2.46   (    0.23) &     1.46   (    0.22) &      6.1  &     28.5  &        3   &       5  &     42.8  \\
IF3    & YZ CMi      & 13-Jan-2012 22:44:31  &    105.0  &    27.71  &     2.82  &     1.63   (    0.08) &     2.21   (    0.01) &    337.4  &     18.7  &      271   &    1050  &     72.3  \\
IF4 (F2)    & YZ CMi      & 14-Jan-2012 04:32:01  &      3.7  &     0.86  &     0.09  &     1.82   (    0.10) &     2.32   (    0.17) &     13.5  &     16.2  &        8   &      16  &     32.5  \\
IF5    & YZ CMi      & 12-Jan-2012 21:12:36  &      5.3  &     0.71  &     0.10  &     3.22   (    0.17) &     1.64   (    0.03) &     24.6  &     13.0  &       26   &      17  &      8.5  \\
IF6    & EQ Peg A    & 10-Aug-2008 05:48:16  &      3.8  &     0.73  &     0.05  &     2.10   (    0.07) &   \nodata &     28.0  &      8.2  &       27   &      19  &      5.8  \\
IF7    & Prox Cen    & 25-May-2010 07:33:52  &      2.6  &     0.33  &     0.03  &     2.92   (    0.32) &     1.46   (    0.23) &     19.7  &      7.8  &        6   &       9  &     11.7  \\
IF8    & YZ CMi      & 14-Jan-2012 04:03:36  &      1.3  &     0.23  &     0.03  &     2.47   (    0.23) &     1.88   (    0.28) &     10.7  &      7.3  &        7   &       7  &      7.3  \\
IF9    & Prox Cen    & 21-May-2010 23:25:51  &      6.0  &     0.74  &     0.06  &     3.05   (    0.17) &     1.67   (    0.10) &     54.3  &      6.6  &       15   &       5  &      2.2  \\
IF10   & Prox Cen    & 25-May-2010 01:31:02  &      4.0  &     0.21  &     0.03  &     7.11   (    0.74) &     0.95   (    0.10) &     46.9  &      5.1  &       30   &      13  &      2.1  \\
IF11 (F1)  & YZ CMi      & 14-Jan-2012 02:59:25  &      2.4  &     0.43  &     0.05  &     2.34   (    0.13) &     2.22   (    0.15) &     33.9  &      4.2  &       13   &       8  &      2.5  \\
IF12   & YZ CMi      & 12-Jan-2012 21:49:33  &      2.2  &     0.24  &     0.04  &     3.98   (    0.23) &     1.44   (    0.06) &     42.5  &      3.2  &       61   &      23  &      1.2  \\
IF13   & YZ CMi      & 13-Jan-2012 02:36:28  &      0.5  &     0.08  &     0.01  &     2.61   (    0.33) &     \nodata &     11.7  &      2.8  &        7   &       7  &      2.8  \\
HF1    & Gl 644 AB   & 25-May-2010 05:08:33  &      0.5  &     0.07  &     0.00  &     2.65   (    0.18) &     \nodata &     18.3  &      1.8  &       13   &      10  &      1.3  \\
HF2    & YZ CMi      & 13-Jan-2012 00:25:30  &      0.9  &     0.12  &     0.01  &     3.26   (    0.32) &     1.63   (    0.16) &     37.7  &      1.5  &       31   &      35  &      1.7  \\
HF3    & EQ Peg A    & 10-Aug-2008 05:53:58  &      2.9  &     0.52  &     0.05  &     2.17   (    0.06) &    \nodata &    153.3  &      1.1  &       24   &     160  &      7.5  \\
HF4    & AD Leo      & 13-Jan-2012 05:29:47  &      0.4  &     0.03  &     0.00  &     3.67   (    0.34) &    \nodata &     29.6  &      0.8  &       58   &      20  &      0.3  \\
HF5    & EQ Peg A    & 10-Aug-2008 05:35:03  &      1.1  &     0.17  &     0.03  &     2.76   (    0.10) &   \nodata &    102.9  &      0.7  &      130   &     240  &      1.2  \\
GF1    & YZ CMi      & 13-Jan-2012 02:49:41  &      1.7  &     0.26  &     0.05  &     2.97   (    0.09) &     1.34   (    0.09) &    279.0  &      0.4  &       49   &     300  &      2.3  \\
GF2    & YZ CMi      & 14-Jan-2012 04:09:17  &      0.6  &     0.03  &     0.01  &     7.94   (    1.09) &     0.83   (    0.18) &     98.7  &      0.4  &        \nodata   &       \nodata  & \nodata  \\

\enddata
\tablecomments{The values of $I_f$, FcolorB, and FcolorR are shown at
  the peak flare times; the values of FcolorR have been obtained
  around the peak using a weighted mean.   The values for the IF3
  event are obtained without subtracting the emission from the IF1
  event.   The $t_{1/2}$ and $\mathcal{I}$ values are given for 
the NBF3500 filter, but the $t_{\rm{rise}}$ and $t_{\rm{fast-decay}}$
are determined from either the NBF4170 or NBF3500 filters (see text).  The IF13, HF1, and HF4 events do not have a detection in RC\#1 that has sufficient signal-to-noise for a robust value of FcolorR.  An H$\alpha$ filter was employed for the red arm during the flares on EQ Peg A (IF6, HF3, and HF5), and no FcolorR values are given. The 
$\mathcal{I}^\prime$ is  defined as $\mathcal{I} / (t_{\rm{rise}} / t_{\rm{fast\ decay}})$ (Section \ref{sec:bursty}); GF2 does not have a well defined fast decay phase.   }
\label{table:flare_properties}
\end{deluxetable}

\clearpage

\bibliographystyle{apj}
\bibliography{ucflares_v15}

\begin{thebibliography}{}
\expandafter\ifx\csname natexlab\endcsname\relax\def\natexlab#1{#1}\fi

\bibitem[{{Abbett} \& {Hawley}(1999)}]{Abbett1999}
{Abbett}, W.~P., \& {Hawley}, S.~L. 1999, \apj, 521, 906

\bibitem[{{Allred} {et~al.}(2015){Allred}, {Kowalski}, \&
  {Carlsson}}]{Allred2015}
{Allred}, J., {Kowalski}, A., \& {Carlsson}, M. 2015, in AAS/AGU Triennial
  Earth-Sun Summit, Vol.~1, AAS/AGU Triennial Earth-Sun Summit, 30207

\bibitem[{{Allred} {et~al.}(2005){Allred}, {Hawley}, {Abbett}, \&
  {Carlsson}}]{Allred2005}
{Allred}, J.~C., {Hawley}, S.~L., {Abbett}, W.~P., \& {Carlsson}, M. 2005,
  \apj, 630, 573

\bibitem[{{Allred} {et~al.}(2006){Allred}, {Hawley}, {Abbett}, \&
  {Carlsson}}]{Allred2006}
---. 2006, \apj, 644, 484

\bibitem[{{Anfinogentov} {et~al.}(2013){Anfinogentov}, {Nakariakov},
  {Mathioudakis}, {Van Doorsselaere}, \& {Kowalski}}]{Anf2013}
{Anfinogentov}, S., {Nakariakov}, V.~M., {Mathioudakis}, M., {Van
  Doorsselaere}, T., \& {Kowalski}, A.~F. 2013, \apj, 773, 156

\bibitem[{{Ayres}(2015)}]{Ayres2015}
{Ayres}, T.~R. 2015, \aj, 150, 7

\bibitem[{{Brown} {et~al.}(2012){Brown}, {Kowalski}, {Mathioudakis}, {Hooper},
  {Hawley}, {Osten}, \& {Wisniewski}}]{BrownAAS}
{Brown}, B., {Kowalski}, A.~F., {Mathioudakis}, M., {et~al.} 2012, in American
  Astronomical Society Meeting Abstracts, Vol. 220, American Astronomical
  Society Meeting Abstracts \#220, 204.52

\bibitem[{{Carlsson} \& {Stein}(1992)}]{Carlsson1992}
{Carlsson}, M., \& {Stein}, R.~F. 1992, \apjl, 397, L59

\bibitem[{{Carlsson} \& {Stein}(1994)}]{Carlsson1994}
{Carlsson}, M., \& {Stein}, R.~F. 1994, in Chromospheric Dynamics, ed.
  {Carlsson, ~M.} (Oslo: University), 47--+

\bibitem[{{Carlsson} \& {Stein}(1995)}]{Carlsson1995}
---. 1995, \apjl, 440, L29

\bibitem[{{Carlsson} \& {Stein}(1997)}]{Carlsson1997}
---. 1997, \apj, 481, 500

\bibitem[{{Carlsson} \& {Stein}(2002)}]{Carlsson2002}
---. 2002, \apj, 572, 626

\bibitem[{{Christian} {et~al.}(2003){Christian}, {Mathioudakis},
  {Jevremovi{\'c}}, {Dupuis}, {Vennes}, \& {Kawka}}]{Christian2003}
{Christian}, D.~J., {Mathioudakis}, M., {Jevremovi{\'c}}, D., {et~al.} 2003,
  \apjl, 593, L105

\bibitem[{{Cram} \& {Woods}(1982)}]{CramWoods1982}
{Cram}, L.~E., \& {Woods}, D.~T. 1982, \apj, 257, 269

\bibitem[{{Dappen} {et~al.}(1987){Dappen}, {Anderson}, \&
  {Mihalas}}]{Dappen1987}
{Dappen}, W., {Anderson}, L., \& {Mihalas}, D. 1987, \apj, 319, 195

\bibitem[{{Davenport} {et~al.}(2012){Davenport}, {Becker}, {Kowalski},
  {Hawley}, {Schmidt}, {Hilton}, {Sesar}, \& {Cutri}}]{Davenport2012}
{Davenport}, J.~R.~A., {Becker}, A.~C., {Kowalski}, A.~F., {et~al.} 2012, \apj,
  748, 58

\bibitem[{{Davenport} {et~al.}(2014){Davenport}, {Hawley}, {Hebb},
  {Wisniewski}, {Kowalski}, {Johnson}, {Malatesta}, {Peraza}, {Keil},
  {Silverberg}, {Jansen}, {Scheffler}, {Berdis}, {Larsen}, \&
  {Hilton}}]{Davenport2014}
{Davenport}, J.~R.~A., {Hawley}, S.~L., {Hebb}, L., {et~al.} 2014, \apj, 797,
  122

\bibitem[{{Dhillon} {et~al.}(2007){Dhillon}, {Marsh}, {Stevenson}, {Atkinson},
  {Kerry}, {Peacocke}, {Vick}, {Beard}, {Ives}, {Lunney}, {McLay}, {Tierney},
  {Kelly}, {Littlefair}, {Nicholson}, {Pashley}, {Harlaftis}, \&
  {O'Brien}}]{Dhillon2007}
{Dhillon}, V.~S., {Marsh}, T.~R., {Stevenson}, M.~J., {et~al.} 2007, \mnras,
  378, 825

\bibitem[{{Dhillon} {et~al.}(2014){Dhillon}, {Marsh}, {Atkinson}, {Bezawada},
  {Bours}, {Copperwheat}, {Gamble}, {Hardy}, {Hickman}, {Irawati}, {Ives},
  {Kerry}, {Leckngam}, {Littlefair}, {McLay}, {O'Brien}, {Peacocke},
  {Poshyachinda}, {Richichi}, {Soonthornthum}, \& {Vick}}]{Ultraspec}
{Dhillon}, V.~S., {Marsh}, T.~R., {Atkinson}, D.~C., {et~al.} 2014, \mnras,
  444, 4009

\bibitem[{{Donati-Falchi} {et~al.}(1985){Donati-Falchi}, {Falciani}, \&
  {Smaldone}}]{DonatiFalchi1985}
{Donati-Falchi}, A., {Falciani}, R., \& {Smaldone}, L.~A. 1985, \aap, 152, 165

\bibitem[{{Doyle} {et~al.}(1988){Doyle}, {Butler}, {Bryne}, \& {van den
  Oord}}]{Doyle1988}
{Doyle}, J.~G., {Butler}, C.~J., {Bryne}, P.~B., \& {van den Oord}, G.~H.~J.
  1988, \aap, 193, 229

\bibitem[{{Emslie}(1978)}]{Emslie1978}
{Emslie}, A.~G. 1978, \apj, 224, 241

\bibitem[{{Fletcher} \& {Hudson}(2002)}]{Fletcher2002}
{Fletcher}, L., \& {Hudson}, H.~S. 2002, \solphys, 210, 307

\bibitem[{{Fuhrmeister} {et~al.}(2008){Fuhrmeister}, {Liefke}, {Schmitt}, \&
  {Reiners}}]{Fuhrmeister2008}
{Fuhrmeister}, B., {Liefke}, C., {Schmitt}, J.~H.~M.~M., \& {Reiners}, A. 2008,
  \aap, 487, 293

\bibitem[{{Gershberg}(1972)}]{Gershberg1972}
{Gershberg}, R.~E. 1972, \apss, 19, 75

\bibitem[{{Graham} \& {Cauzzi}(2015)}]{Graham2015}
{Graham}, D.~R., \& {Cauzzi}, G. 2015, \apjl, 807, L22

\bibitem[{{Grigis} \& {Benz}(2004)}]{Grigis2004}
{Grigis}, P.~C., \& {Benz}, A.~O. 2004, \aap, 426, 1093

\bibitem[{{Hara} {et~al.}(2009){Hara}, {Watanabe}, {Bone}, {Culhane}, {van
  Driel-Gesztelyi}, \& {Young}}]{Hara2009}
{Hara}, H., {Watanabe}, T., {Bone}, L.~A., {et~al.} 2009, in Astronomical
  Society of the Pacific Conference Series, Vol. 415, The Second Hinode Science
  Meeting: Beyond Discovery-Toward Understanding, ed. B.~{Lites}, M.~{Cheung},
  T.~{Magara}, J.~{Mariska}, \& K.~{Reeves}, 459

\bibitem[{{Hawley} \& {Fisher}(1992)}]{HawleyFisher1992}
{Hawley}, S.~L., \& {Fisher}, G.~H. 1992, \apjs, 78, 565

\bibitem[{{Hawley} \& {Pettersen}(1991)}]{HawleyPettersen1991}
{Hawley}, S.~L., \& {Pettersen}, B.~R. 1991, \apj, 378, 725

\bibitem[{{Hawley} {et~al.}(1995){Hawley}, {Fisher}, {Simon}, {Cully},
  {Deustua}, {Jablonski}, {Johns-Krull}, {Pettersen}, {Smith}, {Spiesman}, \&
  {Valenti}}]{Hawley1995}
{Hawley}, S.~L., {Fisher}, G.~H., {Simon}, T., {et~al.} 1995, \apj, 453, 464

\bibitem[{{Hawley} {et~al.}(2003){Hawley}, {Allred}, {Johns-Krull}, {Fisher},
  {Abbett}, {Alekseev}, {Avgoloupis}, {Deustua}, {Gunn}, {Seiradakis}, {Sirk},
  \& {Valenti}}]{Hawley2003}
{Hawley}, S.~L., {Allred}, J.~C., {Johns-Krull}, C.~M., {et~al.} 2003, \apj,
  597, 535

\bibitem[{{Hilton} {et~al.}(2011){Hilton}, {Hawley}, {Kowalski}, \&
  {Holtzman}}]{Hilton2011}
{Hilton}, E.~J., {Hawley}, S.~L., {Kowalski}, A.~F., \& {Holtzman}, J. 2011, in
  Astronomical Society of the Pacific Conference Series, Vol. 448, Astronomical
  Society of the Pacific Conference Series, ed. C.~{Johns-Krull}, M.~K.
  {Browning}, \& A.~A. {West}, 197

\bibitem[{{Holman}(2012)}]{Holman2012}
{Holman}, G.~D. 2012, \apj, 745, 52

\bibitem[{{Holman} {et~al.}(2011){Holman}, {Aschwanden}, {Aurass}, {Battaglia},
  {Grigis}, {Kontar}, {Liu}, {Saint-Hilaire}, \& {Zharkova}}]{Holman2011}
{Holman}, G.~D., {Aschwanden}, M.~J., {Aurass}, H., {et~al.} 2011, \ssr, 159,
  107

\bibitem[{{Holtzman} {et~al.}(2010){Holtzman}, {Harrison}, \&
  {Coughlin}}]{Holtzman}
{Holtzman}, J.~A., {Harrison}, T.~E., \& {Coughlin}, J.~L. 2010, Advances in
  Astronomy, 2010, 46

\bibitem[{{Houdebine}(1992)}]{Houdebine1992}
{Houdebine}, E.~R. 1992, Irish Astronomical Journal, 20, 213

\bibitem[{{Hummer} \& {Mihalas}(1988)}]{HM88}
{Hummer}, D.~G., \& {Mihalas}, D. 1988, \apj, 331, 794

\bibitem[{{Inglis} {et~al.}(2015){Inglis}, {Ireland}, \&
  {Dominique}}]{Inglis2015}
{Inglis}, A.~R., {Ireland}, J., \& {Dominique}, M. 2015, \apj, 798, 108

\bibitem[{{Jess} {et~al.}(2010){Jess}, {Mathioudakis}, {Christian}, {Keenan},
  {Ryans}, \& {Crockett}}]{Jess2010}
{Jess}, D.~B., {Mathioudakis}, M., {Christian}, D.~J., {et~al.} 2010, \solphys,
  261, 363

\bibitem[{{Kerr} \& {Fletcher}(2014)}]{Kerr2014}
{Kerr}, G.~S., \& {Fletcher}, L. 2014, \apj, 783, 98

\bibitem[{{Kleint} {et~al.}(2016){Kleint}, {Heinzel}, {Judge}, \&
  {Krucker}}]{Kleint2016}
{Kleint}, L., {Heinzel}, P., {Judge}, P., \& {Krucker}, S. 2016, \apj, 816, 88

\bibitem[{{Kosovichev} \& {Zharkova}(2001)}]{Kosovichev2001}
{Kosovichev}, A.~G., \& {Zharkova}, V.~V. 2001, \apjl, 550, L105

\bibitem[{Kowalski {et~al.}(2016)Kowalski, Mathioudakis, Hawley, Wisniewski,
  Dhillon, Marsh, Hilton, \& Brown}]{kowalski_2016_zenodo}
Kowalski, A., Mathioudakis, M., Hawley, S., {et~al.} 2016, {M Dwarf Flare
  Continuum Variations on One-Second Timescales: Calibrating and Modeling of
  ULTRACAM Flare Color Indices}, doi:10.5281/zenodo.45878

\bibitem[{{Kowalski}(2012)}]{KowalskiThesis}
{Kowalski}, A.~F. 2012, PhD thesis, University of Washington

\bibitem[{{Kowalski}(2015)}]{Kowalski2015IAUS}
---. 2015, ArXiv e-prints, arXiv:1511.05085

\bibitem[{{Kowalski} {et~al.}(2015){Kowalski}, {Hawley}, {Carlsson}, {Allred},
  {Uitenbroek}, {Osten}, \& {Holman}}]{Kowalski2015}
{Kowalski}, A.~F., {Hawley}, S.~L., {Carlsson}, M., {et~al.} 2015, \solphys,
  arXiv:1503.07057

\bibitem[{{Kowalski} {et~al.}(2010){Kowalski}, {Hawley}, {Holtzman},
  {Wisniewski}, \& {Hilton}}]{Kowalski2010}
{Kowalski}, A.~F., {Hawley}, S.~L., {Holtzman}, J.~A., {Wisniewski}, J.~P., \&
  {Hilton}, E.~J. 2010, \apjl, 714, L98

\bibitem[{{Kowalski} {et~al.}(2011{\natexlab{a}}){Kowalski}, {Hawley},
  {Holtzman}, {Wisniewski}, \& {Hilton}}]{Kowalski2011b}
{Kowalski}, A.~F., {Hawley}, S.~L., {Holtzman}, J.~A., {Wisniewski}, J.~P., \&
  {Hilton}, E.~J. 2011{\natexlab{a}}, in IAU Symposium, Vol. 273, IAU
  Symposium, ed. D.~{Prasad Choudhary} \& K.~G. {Strassmeier}, 261--264

\bibitem[{{Kowalski} {et~al.}(2012){Kowalski}, {Hawley}, {Holtzman},
  {Wisniewski}, \& {Hilton}}]{Kowalski2012}
---. 2012, \solphys, 277, 21

\bibitem[{{Kowalski} {et~al.}(2013){Kowalski}, {Hawley}, {Wisniewski}, {Osten},
  {Hilton}, {Holtzman}, {Schmidt}, \& {Davenport}}]{Kowalski2013}
{Kowalski}, A.~F., {Hawley}, S.~L., {Wisniewski}, J.~P., {et~al.} 2013, \apjs,
  207, 15

\bibitem[{{Kowalski} {et~al.}(2011{\natexlab{b}}){Kowalski}, {Mathioudakis},
  {Hawley}, {Hilton}, {Dhillon}, {Marsh}, \& {Copperwheat}}]{Kowalski2011}
{Kowalski}, A.~F., {Mathioudakis}, M., {Hawley}, S.~L., {et~al.}
  2011{\natexlab{b}}, in Astronomical Society of the Pacific Conference Series,
  Vol. 448, 16th Cambridge Workshop on Cool Stars, Stellar Systems, and the
  Sun, ed. C.~{Johns-Krull}, M.~K. {Browning}, \& A.~A. {West}, 1157

\bibitem[{{Kunkel}(1970)}]{Kunkel1970}
{Kunkel}, W.~E. 1970, \apj, 161, 503

\bibitem[{{Lacy} {et~al.}(1976){Lacy}, {Moffett}, \& {Evans}}]{Lacy1976}
{Lacy}, C.~H., {Moffett}, T.~J., \& {Evans}, D.~S. 1976, \apjs, 30, 85

\bibitem[{{Lin} \& {Schwartz}(1987)}]{Lin1987}
{Lin}, R.~P., \& {Schwartz}, R.~A. 1987, \apj, 312, 462

\bibitem[{{Liu} \& {Fletcher}(2009)}]{LiuFletcher2009}
{Liu}, S., \& {Fletcher}, L. 2009, \apjl, 701, L34

\bibitem[{{Lovkaya}(2013)}]{Lovkaya2013}
{Lovkaya}, M.~N. 2013, Astronomy Reports, 57, 603

\bibitem[{{McKenzie}(2013)}]{McKenzie2013}
{McKenzie}, D.~E. 2013, \apj, 766, 39

\bibitem[{{Moffett}(1974)}]{Moffett1974}
{Moffett}, T.~J. 1974, \apjs, 29, 1

\bibitem[{{Nakariakov} {et~al.}(2006){Nakariakov}, {Foullon}, {Verwichte}, \&
  {Young}}]{Nakariakov2006}
{Nakariakov}, V.~M., {Foullon}, C., {Verwichte}, E., \& {Young}, N.~P. 2006,
  \aap, 452, 343

\bibitem[{{Osten} {et~al.}(2005){Osten}, {Hawley}, {Allred}, {Johns-Krull}, \&
  {Roark}}]{Osten2005}
{Osten}, R.~A., {Hawley}, S.~L., {Allred}, J.~C., {Johns-Krull}, C.~M., \&
  {Roark}, C. 2005, \apj, 621, 398

\bibitem[{{Osten} \& {Wolk}(2015)}]{Osten2015}
{Osten}, R.~A., \& {Wolk}, S.~J. 2015, ArXiv e-prints, arXiv:1506.04994

\bibitem[{{Petrosian} \& {Liu}(2004)}]{Petrosian2004}
{Petrosian}, V., \& {Liu}, S. 2004, \apj, 610, 550

\bibitem[{{Qiu} {et~al.}(2010){Qiu}, {Liu}, {Hill}, \& {Kazachenko}}]{Qiu2010}
{Qiu}, J., {Liu}, W., {Hill}, N., \& {Kazachenko}, M. 2010, \apj, 725, 319

\bibitem[{{Schmitt} {et~al.}(2008){Schmitt}, {Reale}, {Liefke}, {Wolter},
  {Fuhrmeister}, {Reiners}, \& {Peres}}]{Schmitt2008}
{Schmitt}, J.~H.~M.~M., {Reale}, F., {Liefke}, C., {et~al.} 2008, \aap, 481,
  799

\bibitem[{{Sirianni} {et~al.}(2005){Sirianni}, {Jee}, {Ben{\'{\i}}tez},
  {Blakeslee}, {Martel}, {Meurer}, {Clampin}, {De Marchi}, {Ford}, {Gilliland},
  {Hartig}, {Illingworth}, {Mack}, \& {McCann}}]{Sirianni2005}
{Sirianni}, M., {Jee}, M.~J., {Ben{\'{\i}}tez}, N., {et~al.} 2005, \pasp, 117,
  1049

\bibitem[{{Su} {et~al.}(2011){Su}, {Holman}, \& {Dennis}}]{Su2011}
{Su}, Y., {Holman}, G.~D., \& {Dennis}, B.~R. 2011, \apj, 731, 106

\bibitem[{{Tian} {et~al.}(2015){Tian}, {Young}, {Reeves}, {Chen}, {Liu}, \&
  {McKillop}}]{Tian2015}
{Tian}, H., {Young}, P.~R., {Reeves}, K.~K., {et~al.} 2015, \apj, 811, 139

\bibitem[{{Tremblay} \& {Bergeron}(2009)}]{Tremblay2009}
{Tremblay}, P.-E., \& {Bergeron}, P. 2009, \apj, 696, 1755

\bibitem[{{Tulloch} \& {Dhillon}(2011)}]{qucam}
{Tulloch}, S.~M., \& {Dhillon}, V.~S. 2011, \mnras, 411, 211

\bibitem[{{Uitenbroek}(2001)}]{Uitenbroek2001}
{Uitenbroek}, H. 2001, \apj, 557, 389

\bibitem[{{Wang}(2009)}]{Wang2009}
{Wang}, L. 2009, \apj, 694, 247

\bibitem[{{Warmuth} {et~al.}(2009){Warmuth}, {Holman}, {Dennis}, {Mann},
  {Aurass}, \& {Milligan}}]{Warmuth2009}
{Warmuth}, A., {Holman}, G.~D., {Dennis}, B.~R., {et~al.} 2009, \apj, 699, 917

\bibitem[{{Zarro} \& {Zirin}(1985)}]{Zarro1985}
{Zarro}, D.~M., \& {Zirin}, H. 1985, \aap, 148, 240

\bibitem[{{Zharkova} \& {Gordovskyy}(2006)}]{Zharkova2006}
{Zharkova}, V.~V., \& {Gordovskyy}, M. 2006, \apj, 651, 553

\bibitem[{{Zhilyaev} {et~al.}(2007){Zhilyaev}, {Romanyuk}, {Svyatogorov},
  {Verlyuk}, {Kaminsky}, {Andreev}, {Sergeev}, {Gershberg}, {Lovkaya},
  {Avgoloupis}, {Seiradakis}, {Contadakis}, {Antov}, {Konstantinova-Antova}, \&
  {Bogdanovski}}]{Zhilyaev2007}
{Zhilyaev}, B.~E., {Romanyuk}, Y.~O., {Svyatogorov}, O.~A., {et~al.} 2007,
  \aap, 465, 235

\end{thebibliography}

\end{document}